\documentclass[a4paper,fleqn]{cas-sc}

\usepackage[authoryear,longnamesfirst]{natbib}
\usepackage{tabularx}
\usepackage{placeins}

\def\tsc#1{\csdef{#1}{\textsc{\lowercase{#1}}\xspace}}
\tsc{WGM}
\tsc{QE}

\begin{document}
\let\WriteBookmarks\relax
\def\floatpagepagefraction{1}
\def\textpagefraction{.001}

\shorttitle{Cross-Evidence Authorization State Learning}   

\shortauthors{Siqi Lin et al.}

\title [mode = title]{Cross-Evidence Authorization State Learning for Runtime Detection of Broken Object-Level Authorization in Black-Box Web Applications}  



%


\author[1,2]{Siqi Lin}[
    orcid=0009-0006-2510-9784
]

\cormark[1]


\ead{lsq@moefire.net}




\affiliation[1]{
    organization={Xiamen Moefire Technology Co., Ltd.},
    city={Xiamen},
    postcode={361000},
    state={Fujian},
    country={China}
}

\affiliation[2]{
    organization={School of ZhongXin International Engineering, Shenyang Jianzhu University},
    city={Shenyang},
    postcode={110168},
    state={Liaoning},
    country={China}
}

\author[3]{Qianheng Cheng}[
    orcid=0009-0006-8649-4751
]

\ead{3122056505@qq.com}

\affiliation[3]{
    organization={School of Information Engineering, Changji University},
    city={Changji},
    postcode={831100},
    state={Xinjiang},
    country={China}
}

\author[4]{Yiping Mao} [
    orcid=0009-0000-6455-0413
]

\ead{2539273248@qq.com}

\affiliation[4]{
    organization={Digital Technology School of Sias, Sias University},
    city={Xinzheng},
    postcode={451150},
    state={Henan},
    country={China}
}







\cortext[1]{Corresponding author}



\begin{abstract}
Broken object-level authorization (BOLA) remains difficult to detect in black-box Web applications because external runtime monitors generally lack access to source code, authorization policies, and object-ownership records. This paper proposes a cross-evidence authorization-state learning method that reconstructs subject--resource--object--action relations from request--response transactions. The method represents authorization relations with confidence-aware states, aggregates positive and negative evidence while accounting for provenance independence, and applies contamination-gated promotion, evidence quarantine and rollback, expiration, and revocation. A Python prototype is evaluated using a deterministic synthetic benchmark comprising 64,400 requests across five independent random seeds. In this benchmark, the proposed method achieves 100.00\% recall, an F1-score of 56.63\%, and a false-positive rate of 33.65\%, while the measured unauthorized-relation pollution rate is 0.00\%. Ablation experiments show that removing provenance independence or evidence quarantine increases the pollution rate to 100.00\%. The experiments provide an initial evaluation under controlled synthetic conditions.
\end{abstract}


\begin{highlights}
\item A cross-evidence model learns object-level authorization states from runtime traffic.
\item Independent evidence confirmation reduces authorization-state poisoning.
\item Contamination-gated updates support rollback, revocation, and expiration.
\item Action-specific relations improve runtime detection of BOLA attacks.
\end{highlights}


\maketitle

\section{Introduction}\label{sec:introduction}

\subsection{Background and motivation}\label{subsec:background}

Modern Web applications and application programming interfaces (APIs) expose business resources through explicit object identifiers, such as user identifiers, contract numbers, order identifiers, document identifiers, device identifiers, and universally unique identifiers. These identifiers are commonly transmitted in URI path segments, query parameters, form fields, or structured request bodies. Although this resource-oriented design facilitates interface composition and service integration, it also creates a persistent security risk when the server fails to verify whether an authenticated requester is authorized to perform the requested operation on the referenced object.

The Open Worldwide Application Security Project (OWASP) defines this weakness as Broken Object Level Authorization (BOLA) and ranks it as the first risk in the OWASP API Security Top 10 \citep{owasp2023bola}. A BOLA vulnerability may allow an authenticated attacker to read, modify, delete, export, or otherwise operate on resources for which the requester lacks the required authorization, merely by manipulating an object identifier in an otherwise legitimate request. Unlike injection attacks or malformed protocol messages, BOLA requests frequently contain valid authentication credentials, supported HTTP methods, syntactically correct parameters, and structurally normal payloads. Consequently, the decisive security question is not whether a request is authenticated or syntactically anomalous, but whether a particular subject is authorized to perform a particular action on a particular object at the time of the request.

Conventional access-control models express authorization through roles, attributes, relationships, or usage-control conditions \citep{sandhu1996rbac,hu2014abac,fong2011rebac,park2004ucon}. These models normally operate within the protected application and assume access to trusted policies, attributes, ownership relations, or organizational data. An external runtime protection component operates under substantially different conditions: it may observe request--response transactions but have no access to application source code, database ownership records, policy engines, or complete business authorization configurations. This situation is common in legacy applications, third-party systems, rapidly evolving API environments, and multi-service deployments in which authorization logic is distributed across several enforcement points.

Prior work introduced a non-intrusive traffic analysis framework that uses request--response correlation to organize heterogeneous runtime evidence and support explainable authorization-risk decisions without modifying application code \citep{lin2026framework}. Its binary object-evidence representation, however, cannot fully express evidence confidence, source independence, operation-specific authorization, or changes in authorization over time. A more reliable runtime mechanism must therefore infer authorization states from multiple observable evidence sources while preserving uncertainty and preventing untrusted transactions from contaminating subsequent decisions.

\subsection{Research gap and approach}\label{subsec:research-gap}

Existing approaches have four principal limitations in the target black-box runtime setting. First, active black-box testing does not maintain persistent authorization states during ordinary runtime traffic. Second, a successful response cannot serve as independent authorization evidence when the protected application may itself contain the authorization defect being detected. Third, external traffic monitors lack the trusted policies, attributes, and ownership information assumed by conventional access-control models. Fourth, authorization relations may change because of object transfer, permission revocation, temporary-access expiration, or evolving application behavior.

This study therefore addresses the following problem: given a stream of request--response transactions from a Web application whose source code, internal authorization policy, and database ownership relations are unavailable, how can an external runtime component learn and maintain reliable subject--resource--object--action authorization states while resisting evidence contamination and adapting to authorization changes?

To address this problem, we propose a cross-evidence authorization state learning method. The method maintains action-specific authorization relations from request--response transactions, distinguishes provisional and confirmed evidence, constrains evidence independence, suspends state promotion when incoming transactions trigger contamination indicators, and supports rollback, revocation, expiration, and temporal confidence decay.

The method operates at a reverse proxy, API gateway, ingress controller, or an equivalent observation point after TLS termination. It does not replace application-side authorization. Instead, it constructs a partial runtime authorization view for external BOLA detection and explainable security decisions.

\subsection{Contributions}\label{subsec:contributions}

The main contributions of this work are as follows:

\begin{enumerate}
    \item
    We formulate traffic-side object-level authorization as a dynamic subject--resource--object--action relation with explicit confidence, lifecycle state, provenance, and temporal validity. The formulation distinguishes object visibility from permission to perform state-changing operations.

    \item
    We propose a cross-evidence authorization confirmation mechanism that evaluates evidence credibility, polarity, freshness, and independence. The mechanism prevents a single successful response or correlated observation from serving as self-validating authorization evidence.

    \item
    We implement the method as a non-intrusive runtime component and evaluate it in terms of authorization-state accuracy, BOLA detection performance, cold-start behavior, contamination resistance, authorization-change responsiveness, cross-application applicability, and runtime overhead.
\end{enumerate}

\section{Background and related work}\label{sec:related-work}

\subsection{Object-level authorization and access-control models}\label{subsec:authorization-models}

Authentication establishes the identity associated with a request, whereas authorization determines whether the identified subject may perform a specified action on a protected resource. BOLA occurs when an application accepts a client-controlled object identifier but fails to enforce the corresponding subject--object authorization constraint \citep{owasp2023bola}.

Object-level authorization may depend on ownership, organizational membership, delegation, resource hierarchy, object status, operation type, or environmental conditions. Permission to read an object does not necessarily imply permission to modify, delete, export, or approve it.

RBAC associates permissions with organizational roles \citep{sandhu1996rbac}, whereas ABAC evaluates subject, object, operation, and environmental attributes \citep{hu2014abac}. ReBAC represents authorization through relationships among users and resources \citep{fong2011rebac}, and UCON supports continuous decisions and mutable contextual conditions \citep{park2004ucon}.

These models assume that the authorization mechanism can access trusted roles, attributes, relationships, or policies. A traffic-side learner observes only a partial projection of authorization-relevant state through requests, responses, object references, and temporal sequences. It must therefore preserve uncertainty, provenance, and temporal validity rather than convert every observation into an authoritative policy rule.

\subsection{Black-box Web and API security testing}\label{subsec:black-box-testing}

Black-box scanners interact with Web applications through externally visible interfaces and avoid language-specific source-code requirements. Previous evaluations showed that automated scanners can identify several classes of input-validation vulnerabilities but remain limited in maintaining authenticated state, traversing complex workflows, and detecting application-specific logic flaws \citep{bau2010blackbox,doupe2010johnny}.

Stateful API testing improves workflow coverage by modeling dependencies among operations. RESTler infers producer--consumer dependencies from API specifications and generates stateful request sequences \citep{atlidakis2019restler}. RESTest similarly derives valid and invalid test cases from OpenAPI descriptions and parameter constraints \citep{martinlopez2021restest}.

These techniques primarily support active vulnerability discovery. They require generated test traffic and often depend on an API specification. They do not continuously maintain authorization states from naturally occurring runtime transactions. Comparative evaluations also report unresolved challenges in authentication handling, semantically constrained input generation, deep-state exploration, and inter-operation dependency satisfaction \citep{zhang2023openproblems}.

\subsection{Runtime and authorization-specific analysis}\label{subsec:runtime-authorization-analysis}

Runtime monitoring techniques demonstrate that application-logic violations require stateful contextual analysis. Swaddler learns relationships among internal application-state variables and identifies executions that violate the learned model \citep{cova2007swaddler}. Felmetsger et al. infer likely invariants from normal executions to identify logic vulnerabilities that cannot be represented through generic attack signatures \citep{felmetsger2010logic}.

These approaches generally rely on internal program state, instrumented traces, or execution-level observations. A traffic-side component has less semantic visibility and must infer authorization evidence from protocol-visible identifiers, operation semantics, business outcomes, and historical transactions.

Authorization-specific methods provide more direct evidence. AuthScope identifies authorization-related protocol fields, substitutes their values, and compares responses across controlled executions \citep{zuo2017authscope}. BolaRay combines static program analysis and database-relation analysis to infer object-level authorization models in database-backed applications \citep{huang2024bolaray}. Formal specification-driven methods can also transform OpenAPI descriptions into executable or analyzable models for detecting BOLA-related events \citep{santosfilho2025openapi}.

These methods provide strong results under their respective assumptions, but require active request manipulation, internal program or database visibility, or accurate API specifications. They are not designed to maintain persistent authorization states from passive runtime traffic.

\subsection{Summary and research positioning}\label{subsec:research-positioning}

A distinct challenge of online authorization learning is that observed transactions can modify the state used for future decisions. If every successful response is treated as positive evidence, a successful unauthorized request may contaminate the learned state. Public search results, aggregated collections, and repeated correlated observations may create similar contamination. Previously valid evidence may also become stale after object transfer, permission revocation, or temporary-access expiration.

The proposed method differs from existing approaches in five respects. It operates on passive or proxy-side request--response transactions, maintains persistent action-specific authorization states, separates provisional from confirmed evidence, constrains evidence independence, and supports contamination control and authorization lifecycle management.

Table~\ref{tab:related-work-comparison} summarizes the main differences between representative approaches.

\begin{table}[t]
\centering
\caption{Comparison of representative approaches to Web and API authorization analysis. Paired values indicate internal visibility / active request generation and action-specific relation support / contamination control, respectively.}
\label{tab:related-work-comparison}
\scriptsize
\setlength{\tabcolsep}{4pt}
\begin{tabularx}{\textwidth}{
    >{\raggedright\arraybackslash}p{2.55cm}
    >{\centering\arraybackslash}p{2.35cm}
    >{\raggedright\arraybackslash}p{2.15cm}
    >{\raggedright\arraybackslash}p{3.0cm}
    >{\raggedright\arraybackslash}X}
\toprule
Approach &
Visibility / probing &
Persistent state &
Action specificity / contamination &
Operating context \\
\midrule
Generic black-box scanners \citep{bau2010blackbox,doupe2010johnny} &
No / Yes & No & No / No & Security testing \\

Stateful API fuzzers \citep{atlidakis2019restler,martinlopez2021restest} &
No / Yes & No & Limited / No & Specification-based API testing \\

Internal runtime monitoring \citep{cova2007swaddler,felmetsger2010logic} &
Yes / No & Behavioral state &
No / Not authorization-specific & Instrumented applications \\

Differential authorization testing \citep{zuo2017authscope} &
No / Yes & Limited & Limited / No & Controlled client testing \\

Database-aware BOLA analysis \citep{huang2024bolaray} &
Yes / No & Inferred authorization model &
Yes / Not traffic-learning-oriented & Static analysis \\

OpenAPI formal modeling \citep{santosfilho2025openapi} &
Specification required / No & Formal execution state &
Partially / Limited & Log and model analysis \\

Proposed method &
No / No & Yes & Yes / Yes &
Passive or proxy-side runtime monitoring \\
\bottomrule
\end{tabularx}
\end{table}

\section{System model and problem formulation}\label{sec:system-model}

\subsection{Observation and deployment model}\label{subsec:observation-model}

The proposed method operates as an external runtime component between
Web clients and a protected application or API service. The component
observes application-layer requests and responses at a reverse proxy,
API gateway, ingress controller, or an equivalent traffic observation
point. It does not require modifications to the protected
application's source code, database schema, or internal authorization
logic.

Let $Q_i$ denote the $i$th HTTP request and $P_i$ denote its
corresponding response. The request is observed at time $t_i^{q}$, and
the corresponding response is observed at time $t_i^{p}$. After the
response is returned, the component correlates $Q_i$ and $P_i$ into a
transaction
\begin{equation}
T_i =
\left\langle
\eta_i,u_i,e_i,\mathcal{O}_i,a_i,y_i,
t_i^{q},t_i^{p},\kappa_i
\right\rangle ,
\label{eq:transaction}
\end{equation}

where $\eta_i$ is the tenant or application context, $u_i$ is the
observed access subject, $e_i$ is the normalized endpoint,
$\mathcal{O}_i$ is the set of candidate object references, $a_i$ is
the requested action, $y_i$ is the observed response outcome,
$t_i^{q}$ is the request-arrival time, $t_i^{p}$ is the
response-observation time, and $\kappa_i\in[0,1]$ is the
request--response correlation confidence. The candidate object set is
defined as
\begin{equation}
\mathcal{O}_i
=
\left\{
\left\langle r_{ij},o_{ij}\right\rangle
\right\}_{j=1}^{m_i},
\label{eq:candidate-object-set}
\end{equation}

where $r_{ij}$ is the resource type, $o_{ij}$ is the corresponding
object identifier, and $m_i$ is the number of candidate object
references extracted from transaction $T_i$.

The response outcome $y_i$ contains both protocol-level and application-level information when available. Protocol-level information includes the HTTP status code and response headers, whereas application-level information may include a business success indicator, an application-specific result code, and a structured response summary. A successful protocol response is not assumed to prove that the corresponding operation was authorized.

For inline deployments, the decision for $Q_i$ is made before the request is forwarded. It therefore uses only the current request context and authorization evidence established by earlier transactions. If the request is forwarded, $P_i$ is subsequently observed and used to update the state for future requests. The response to the current request cannot retrospectively alter the decision already made for that request. This temporal separation prevents the current transaction from validating itself.

In passive deployments, the same transaction model can be constructed from mirrored traffic, gateway logs, or request traces. Passive deployment supports state learning, auditing, and retrospective analysis but does not perform inline enforcement. In both deployment modes, application-layer visibility is required for complete extraction of object identifiers and business outcomes. When plaintext application-layer content is unavailable at the observation point, the method can process only the metadata exposed by the available traffic or logging interface.

\subsection{Entities and notation}\label{subsec:entities-notation}

Table~\ref{tab:notation} summarizes the principal entities and symbols used throughout the paper.

\begin{table}[t]
\centering
\caption{Principal notation used in the system and authorization-state models.}
\label{tab:notation}
\footnotesize
\begin{tabularx}{\linewidth}{
    >{\raggedright\arraybackslash}p{2.15cm}
    >{\raggedright\arraybackslash}X}
\toprule
Symbol & Meaning \\
\midrule

$Q_i$, $P_i$
&
The $i$th HTTP request and its corresponding response.
\\

$T_i$
&
The correlated transaction formed from $Q_i$ and $P_i$ after the
response is observed.
\\

$\eta$, $u$, $e$
&
Tenant or application context, observed access subject, and normalized
endpoint, respectively.
\\

$\mathcal{O}_i$
&
Set of candidate object references extracted from transaction $T_i$;
each reference contains a resource type and an object identifier.
\\

$r$, $o$, $a$
&
Resource type, business-object identifier, and requested action class,
respectively.
\\

$y_i$
&
Observed response outcome, including available protocol-level and
application-level result information.
\\

$t_i^{q}$, $t_i^{p}$
&
Request-arrival time and response-observation time of transaction
$T_i$, respectively.
\\

$\kappa_i$
&
Request--response correlation confidence for transaction $T_i$, where
$\kappa_i\in[0,1]$.
\\

$\mathcal{T}_{\leq t}$
&
Set of complete transactions whose responses have been observed no
later than time $t$.
\\

$\mathcal{H}_{t}$
&
Historical request and transaction context available before time $t$
for online authorization-risk decisions.
\\

$\mathcal{E}_{t}$
&
Authorization-relevant evidence admitted to the learning process up to
time $t$.
\\

$\mathcal{A}_{t}$
&
Authorization-state mapping maintained by the runtime component at
time $t$.
\\

$s$, $c$
&
Lifecycle state and confidence value of an authorization relation,
where $c\in[0,1]$.
\\

$\Pi$, $\tau$
&
Evidence provenance and time of the most recent relation update,
respectively.
\\

$d_i$
&
Object-level authorization-risk decision produced for request $Q_i$.
\\

$K$
&
Authorization-relation key consisting of tenant, subject, resource,
object, and action.
\\

$\varepsilon_k$
&
The $k$th authorization-evidence item.
\\

$c^{+}$, $c^{-}$
&
Positive and negative evidence support associated with an authorization
relation.
\\

$\pi_k$
&
Provenance descriptor of evidence item $\varepsilon_k$.
\\

$\mathcal{K}_i$
&
Set of object-level authorization-relation keys extracted from request
$Q_i$.
\\

$\rho_i(K)$, $\rho_i$
&
Risk value of relation key $K$ and aggregated object-level risk of
request $Q_i$, respectively.
\\

$\Delta\mathcal{E}_i$
&
Set of authorization-evidence items extracted from transaction $T_i$.
\\

$b_k$
&
Positive-evidence promotion-gate result for evidence item
$\varepsilon_k$.
\\

$v_t(K)$
&
Monotonically increasing version number of relation record $K$ at time
$t$.
\\

$\theta_{\mathrm{review}}$, $\theta_{\mathrm{block}}$
&
Decision thresholds for review and blocking.
\\

$\mathbf{h}_k$, $\chi_k$
&
Contamination-indicator vector and aggregate contamination score of
evidence item $\varepsilon_k$.
\\

$H_k$
&
Quarantine record associated with a held evidence item.
\\

$R_t(K)$
&
Indicator of an active trusted revocation record for relation key $K$.
\\

$\mathcal{D}(G)$
&
Set of authorization relations whose evidence depends on provenance
group $G$.
\\

$V_{\ell}$, $U_{\ell}$
&
Object-level revocation and reauthorization events, respectively.
\\

$\Delta_{\mathrm{life}}(K)$
&
Maximum configured evidence lifetime of authorization relation $K$.
\\
\bottomrule
\end{tabularx}
\end{table}

The subject identifier $u$ should be stable within the observation period and sufficiently precise to distinguish different authenticated subjects. When only a low-confidence temporary identifier is available, the resulting transactions may contribute to anomaly analysis but should not independently establish high-confidence authorization relations.

The endpoint identifier $e$ combines the HTTP method with a normalized URI path. Dynamic path segments, including numeric identifiers, UUIDs, and application-generated tokens, are replaced with placeholders. Resource type $r$ represents the semantic class of an object rather than a single endpoint. Consequently, several endpoints may refer to the same resource type even when they use different paths or parameter names.

The candidate-object set $\mathcal{O}_i$ may contain zero, one, or multiple object references. Each reference is represented by a resource--object pair $\langle r_{ij},o_{ij}\rangle$. A transaction may reference multiple objects in batch operations or nested structured payloads. Candidate references are not immediately treated as authorization evidence; they must first be evaluated according to their extraction source, transaction context, and evidence admissibility.

\subsection{Runtime authorization relation}\label{subsec:authorization-relation-model}

The runtime authorization state for a tenant, subject, resource, object, and action is represented as
\begin{equation}
\mathcal{A}_{t}(\eta,u,r,o,a)
=
\left\langle
s,c,\Pi,\tau
\right\rangle ,
\label{eq:authorization-relation}
\end{equation}

where $s$ is the lifecycle state, $c$ is the confidence value, $\Pi$ contains the provenance of the supporting or contradicting evidence, and $\tau$ is the most recent update time.

The authorization key contains an explicit action dimension. The following relations are therefore distinct:
\begin{equation}
\mathcal{A}_{t}(\eta,u,r,o,\mathrm{read})
\neq
\mathcal{A}_{t}(\eta,u,r,o,\mathrm{delete}).
\label{eq:action-separation}
\end{equation}

This separation is necessary because observing or reading an object does not imply authorization to modify, delete, export, or approve it. When an operation cannot be mapped reliably to a predefined action class, it is assigned to an application-specific or generic \textit{other} class rather than being merged with a more permissive operation.

The lifecycle state belongs to the set
\begin{equation}
\mathcal{S}
=
\{
\mathrm{Unknown},
\mathrm{Candidate},
\mathrm{Confirmed},
\mathrm{Conflicted},
\mathrm{Revoked},
\mathrm{Expired}
\}.
\label{eq:state-set}
\end{equation}

An \textit{Unknown} relation has no sufficient admissible evidence. A \textit{Candidate} relation is supported by provisional positive evidence but has not satisfied the confirmation requirements. A \textit{Confirmed} relation is supported by sufficiently credible and independent positive evidence. A \textit{Conflicted} relation contains materially inconsistent evidence, such as positive observations contradicted by reliable negative evidence or evidence whose provenance indicates possible contamination. A \textit{Revoked} relation has been invalidated by an explicit trusted revocation signal or sufficiently strong object-level negative evidence. An \textit{Expired} relation has lost temporal validity because its supporting evidence has exceeded the configured lifetime.

These states represent the external component's current evidence-based view rather than the authoritative policy of the protected application. In particular, an \textit{Unknown} state does
not by itself prove that an operation is unauthorized, and a \textit{Confirmed} state does not represent permanent permission. A relation may subsequently become \textit{Conflicted}, \textit{Revoked}, or \textit{Expired} when contradictory evidence, trusted object-level revocation information, or temporal invalidation is observed.

The provenance component $\Pi$ records information required to evaluate evidence dependence and support later state correction. It may include the originating transaction, endpoint, evidence source, session context, observation time, and correlation group. Evidence items derived from the same transaction or response are linked through their provenance and are not automatically treated as independent confirmation.

\subsection{Threat model and assumptions}\label{subsec:threat-model}

\paragraph{Adversary capabilities.}
The adversary possesses a valid authenticated session for the protected application and can issue syntactically valid HTTP requests through the same interfaces available to legitimate users. The adversary can modify object identifiers in URI paths, query parameters, form fields, and structured request bodies. The adversary can also vary request ordering and timing, use low-rate access patterns, and invoke different operations on the same or related objects.

The adversary may exploit an application endpoint that returns a successful response despite an incorrect object-level authorization decision. The adversary may also attempt to influence the learned authorization state by repeatedly accessing unauthorized objects, causing objects to appear in apparently successful transactions, or using public and aggregated collections as misleading authorization evidence.

\paragraph{Trusted components.}
The traffic observation component, its state store, and its audit records are assumed to be protected from direct modification by the adversary. The mapping between an authenticated session and the observed subject identifier is assumed to be available from the protected application or its access gateway. The supplied subject binding is treated as trusted input and is not learned or reinterpreted by the proposed method. Request and response contents observed after TLS termination are assumed not to be altered between the observation point and the learning component.

\paragraph{Application assumptions.}
The protected application exposes one or more business objects through identifiers visible in application-layer transactions. The same object may be referenced through different endpoints or parameter names, and the application may return public, shared, aggregated, or subject-specific collections. No assumption is made that the application provides complete ownership metadata or an explicit object-level authorization policy.

\paragraph{Scope.}
The objective is to identify object-level authorization inconsistencies under valid authenticated sessions. The method does not evaluate the correctness of the authentication mechanism itself and does not claim to reconstruct the complete internal access-control policy. Denial-of-service attacks, transport-layer attacks, client compromise, and direct tampering with the trusted observation infrastructure are outside the scope of the model.

\subsection{Problem definition}\label{subsec:problem-definition}

Let
\begin{equation}
\mathcal{T}_{\leq t}
=
\left\{
T_i
\mid
t_i^{p}\leq t
\right\},
\label{eq:transaction-stream}
\end{equation}

denote the transaction stream observed up to time $t$. The objective is to maintain an authorization-state store $\mathcal{A}_{t}$ from $\mathcal{T}_{\leq t}$ without access to the protected application's source code, ownership database, or authoritative authorization policy.

The online process contains two temporally separated functions. Before the response to request $Q_i$ is available, the decision function computes
\begin{equation}
d_i
=
f\left(
Q_i,
\mathcal{A}_{(t_i^{q})^-},
\mathcal{H}_{(t_i^{q})^-}
\right),
\label{eq:online-decision}
\end{equation}

where $d_i$ is the object-level authorization-risk decision, $\mathcal{A}_{(t_i^{q})^-}$ is the authorization-state mapping available immediately before $Q_i$ is evaluated, and $\mathcal{H}_{(t_i^{q})^-}$ is the historical request and transaction context available at that instant.

After the corresponding response is observed and transaction $T_i$ is formed, the state-update function computes
\begin{equation}
\mathcal{A}_{t_i^{p}}
=
g\left(
T_i,
\mathcal{A}_{(t_i^{p})^-},
\mathcal{E}_{(t_i^{p})^-}
\right),
\label{eq:state-update}
\end{equation}

Here, $(t_i^{q})^-$ denotes the instant immediately before request $Q_i$ is evaluated, and $(t_i^{p})^-$ denotes the instant immediately before the corresponding transaction updates the authorization state.

The update function must not assume that a successful response is authoritative evidence of authorization. Instead, it must determine whether the transaction contributes positive, negative, ambiguous, or inadmissible evidence and whether that evidence is sufficiently independent from previously observed items.

The problem is therefore not equivalent to binary request classification. It requires the joint maintenance of a dynamic authorization state and the use of that state in subsequent online decisions. A suitable solution should satisfy the following design requirements.

\begin{enumerate}
    \item
    \textbf{Non-intrusive operation.}
    The method should operate without application source-code modification or direct access to the internal authorization database.

    \item
    \textbf{Action-sensitive representation.}
    Authorization evidence for one operation should not automatically authorize other operations on the same object.

    \item
    \textbf{Uncertainty preservation.}
    Limited or ambiguous evidence should remain provisional rather than being converted immediately into a confirmed relation.

    \item
    \textbf{Evidence-independence control.}
    Correlated observations originating from the same transaction, response, or tightly coupled context should not be counted as independent confirmation.

    \item
    \textbf{Contamination resistance.}
    Transactions that trigger contamination indicators should not promote authorization states without sufficient independent corroboration, and provisional updates should remain traceable and correctable.

    \item
    \textbf{Authorization lifecycle support.}
    The state model should support confidence decay, contradiction, revocation, expiration, and changes in object ownership or access conditions.

    \item
    \textbf{Online efficiency.}
    State lookup and update operations should be sufficiently efficient for runtime deployment at a proxy or gateway.
\end{enumerate}

Section~\ref{sec:authorization-learning} presents the evidence model and authorization state-transition mechanism used to satisfy these requirements.

\section{Cross-evidence authorization state model}\label{sec:authorization-learning}

\subsection{Authorization evidence representation}\label{subsec:evidence-representation}

For a tenant, subject, resource, object, and action, the corresponding authorization-relation key is defined as

\begin{equation}
K =
\left\langle
\eta,u,r,o,a
\right\rangle .
\label{eq:relation-key}
\end{equation}

The subject identifier $u$ is treated as an externally supplied and stable binding within the observation period. The authorization learner does not modify or reinterpret this binding. Its objective is limited to determining whether observable runtime evidence supports the relation between subject $u$, resource type $r$, object $o$, and action $a$.

An authorization-evidence item associated with relation key $K_k$ is represented as

\begin{equation}
\varepsilon_k
=
\left\langle
K_k,z_k,p_k,q_k,t_k,\pi_k
\right\rangle ,
\label{eq:evidence-item}
\end{equation}

where $z_k$ denotes the evidence-source class, $p_k$ is the evidence polarity, $q_k\in[0,1]$ is the base credibility, $t_k$ is the observation time, and $\pi_k$ is the provenance descriptor.

The evidence polarity is defined as

\begin{equation}
p_k
\in
\{-1,0,+1\},
\label{eq:evidence-polarity}
\end{equation}

where $+1$ denotes positive authorization evidence, $-1$ denotes negative authorization evidence, and $0$ denotes ambiguous evidence. Ambiguous evidence is retained for auditing and possible subsequent corroboration, but does not directly increase either positive or negative support.

The runtime evidence classes considered by the model are summarized in Table~\ref{tab:evidence-classes}. Their interpretation depends only on observable object-level transaction semantics and does not require access to the protected application's internal authorization policy.

\begin{table}[t]
\centering
\caption{Runtime evidence classes used by the authorization-state model.}
\label{tab:evidence-classes}
\footnotesize
\begin{tabularx}{\linewidth}{
    >{\raggedright\arraybackslash}p{2.5cm}
    >{\raggedright\arraybackslash}p{4.0cm}
    >{\raggedright\arraybackslash}X}
\toprule
Evidence class &
Observable pattern &
Default interpretation \\
\midrule

Object creation or assignment
&
An object identifier is returned by a successful object-creation or object-assignment transaction.
&
Positive candidate evidence for compatible object--action relations.
\\

Subject-scoped collection
&
An object occurs in a collection whose observable transaction semantics indicate that the collection is scoped to the current subject.
&
Positive candidate evidence, subject to source-specificity and action compatibility.
\\

Direct object operation
&
An object-specific endpoint returns an apparent business-success outcome.
&
Provisional evidence only; the response cannot independently confirm authorization.
\\

Object-specific denial
&
The application explicitly rejects an operation on a particular object through a protocol-level or business-level authorization result.
&
Negative evidence when the denial semantics are sufficiently specific.
\\

Object-level revocation
&
An observable transaction indicates removal, transfer, expiration, or withdrawal of access to a specific object.
&
Strong negative evidence for the affected object--action relation.
\\

Public or aggregate exposure
&
An object occurs in a public search result, recommendation response, cross-subject aggregate, or unscoped collection.
&
Ambiguous evidence that cannot directly confirm subject-specific authorization.
\\

\bottomrule
\end{tabularx}
\end{table}

The evidence interpretation is deliberately conservative. In particular, a successful object-specific response is not treated as self-validating authorization evidence because the protected endpoint may itself contain an authorization defect. Similarly, the appearance of an object in a response supports authorization only when the observable response semantics indicate that the collection is scoped to the corresponding subject and compatible with the target action.

\subsection{Evidence credibility and admissibility}\label{subsec:evidence-credibility}

The credibility of an evidence item depends on the reliability of the transaction correlation, object extraction, evidence source, and action mapping. Its base credibility is calculated as

\begin{equation}
q_k
=
\kappa_k^{\alpha_{\kappa}}
\xi_k^{\alpha_{\xi}}
\sigma_k^{\alpha_{\sigma}}
\gamma_k^{\alpha_{\gamma}},
\label{eq:base-credibility}
\end{equation}

subject to

\begin{equation}
\alpha_{\kappa}
+
\alpha_{\xi}
+
\alpha_{\sigma}
+
\alpha_{\gamma}
=
1,
\qquad
\alpha_{\kappa},
\alpha_{\xi},
\alpha_{\sigma},
\alpha_{\gamma}
\geq 0,
\label{eq:credibility-weights}
\end{equation}

where $\kappa_k$ is the request--response correlation confidence, $\xi_k$ is the object-extraction and resource-alignment confidence, $\sigma_k$ is the source-specificity coefficient, and $\gamma_k$ is the action-compatibility coefficient. All four factors are normalized to $[0,1]$.

The correlation term $\kappa_k$ represents the reliability of the mapping between a request and its corresponding response. The extraction term $\xi_k$ represents the confidence that an extracted value identifies a business object and has been mapped to the correct resource type. The source-specificity coefficient $\sigma_k$ distinguishes subject-scoped evidence from public, aggregate, or otherwise ambiguous observations. The action-compatibility coefficient $\gamma_k$ prevents evidence associated with one operation from being automatically transferred to another operation.

The weighted geometric form in Eq.~\eqref{eq:base-credibility} ensures that a low-quality factor cannot be fully compensated for by unrelated high-quality factors. The exponents allow the relative importance of the four factors to be calibrated without changing the evidence model.

The temporal weight of evidence item $\varepsilon_k$ at time $t$ is defined as

\begin{equation}
\omega_k(t)
=
I_k q_k
\exp
\left[
-\lambda_{z_k}(t-t_k)
\right],
\qquad
t\geq t_k,
\label{eq:temporal-evidence-weight}
\end{equation}

where $\lambda_{z_k}\geq0$ is the decay coefficient associated with source class $z_k$, and $I_k\in\{0,1\}$ is the evidence-admissibility indicator.

Evidence is admitted to the authorization-state model only when the transaction and extraction results satisfy the minimum quality requirements:

\begin{equation}
I_k =
\begin{cases}
1,
&
\kappa_k\geq\theta_{\kappa},
\;
\xi_k\geq\theta_{\xi},
\;
K_k\text{ is complete},
\;
\gamma_k>0,
\\[2mm]
0,
&
\text{otherwise},
\end{cases}
\label{eq:evidence-admissibility}
\end{equation}

where $\theta_{\kappa}$ and $\theta_{\xi}$ are the minimum correlation and extraction-confidence thresholds.

An inadmissible evidence item can remain in the audit record but cannot change the authorization-support profile. This separation prevents incomplete transactions, unreliable object extraction, and semantically incompatible observations from influencing subsequent authorization decisions.

\subsection{Action-specific evidence projection}\label{subsec:action-projection}

Let $\hat{a}_k$ denote the action directly associated with evidence item $\varepsilon_k$, and let $a$ denote the target action of relation key $K$. Their compatibility is defined by

\begin{equation}
\gamma_k
=
\Gamma(\hat{a}_k,a),
\qquad
\Gamma:
\mathcal{C}_{\mathrm{act}}
\times
\mathcal{C}_{\mathrm{act}}
\rightarrow
[0,1],
\label{eq:action-compatibility}
\end{equation}

where $\mathcal{C}_{\mathrm{act}}$ is the set of normalized action classes.

The default compatibility rule is

\begin{equation}
\Gamma(\hat{a},a)
=
\begin{cases}
1, & \hat{a}=a,\\
0, & \hat{a}\neq a.
\end{cases}
\label{eq:default-action-compatibility}
\end{equation}

A nonzero cross-action value may be configured only when the application semantics provide explicit and reproducible evidence that an observation associated with one action supports another action. The default rule therefore prevents evidence for a read operation from confirming update, delete, export, or approval authorization.

The endpoint identifier is not included in relation key $K$ because different endpoints may operate on the same resource--object pair. Evidence observed through distinct endpoints may support the same relation only when their object references are aligned to the same $\langle r,o\rangle$ pair. Endpoint information remains part of the provenance descriptor and is used to evaluate evidence dependence and alignment reliability.

This design permits controlled cross-endpoint aggregation while preventing unrelated endpoints or similarly shaped identifiers from being merged solely because they contain the same literal value.

\subsection{Provenance-aware evidence independence}\label{subsec:evidence-independence}

Repeated observations are not necessarily independent. Several candidate identifiers may originate from the same response, and the same object may occur repeatedly within one collection or a tightly coupled transaction sequence. Counting these observations separately would allow one derivation source to accumulate disproportionate authorization support.

The provenance descriptor of evidence item $\varepsilon_k$ is represented as

\begin{equation}
\pi_k
=
\left\langle
\mathrm{tid}_k,
\mathrm{rid}_k,
e_k,
\mathrm{src}_k,
\mathrm{grp}_k
\right\rangle ,
\label{eq:provenance-descriptor}
\end{equation}

where $\mathrm{tid}_k$ is the transaction identifier, $\mathrm{rid}_k$ is the response or record identifier, $e_k$ is the normalized endpoint, $\mathrm{src}_k$ is the extraction source, and $\mathrm{grp}_k$ is the provenance-group identifier.

Let $g(\pi_k)$ map a provenance descriptor to its derivation root. Two evidence items are treated as provenance-dependent when

\begin{equation}
\varepsilon_i
\sim
\varepsilon_j
\quad\Longleftrightarrow\quad
g(\pi_i)=g(\pi_j).
\label{eq:provenance-equivalence}
\end{equation}

At minimum, evidence extracted from the same transaction, the same response, or the same collection instance is assigned to the same provenance group. Application-specific grouping rules may additionally associate observations produced by a single multi-step operation when they originate from the same derivation root.

Let $\mathcal{G}_{K,t}$ denote the set of provenance groups containing admissible evidence for relation key $K$ by time $t$. For each group $G\in\mathcal{G}_{K,t}$, the representative positive and negative weights are

\begin{align}
\overline{\omega}_{G}^{+}(t)
&=
\max_{\varepsilon_k\in G,\;p_k=+1}
\omega_k(t),
\label{eq:group-positive-weight}
\\
\overline{\omega}_{G}^{-}(t)
&=
\max_{\varepsilon_k\in G,\;p_k=-1}
\omega_k(t).
\label{eq:group-negative-weight}
\end{align}

If a provenance group contains no evidence of the required polarity, the corresponding representative weight is defined as zero.

Using only the maximum weight from each provenance group prevents repeated or correlated observations from being counted as multiple independent confirmations. Evidence from distinct derivation roots can still accumulate support when it originates from provenance-independent transactions.

\subsection{Positive and negative support aggregation}\label{subsec:support-aggregation}

For relation key $K$, positive and negative support at time $t$ are aggregated across provenance groups as

\begin{align}
c_t^{+}(K)
&=
1-
\prod_{G\in\mathcal{G}_{K,t}}
\left(
1-\overline{\omega}_{G}^{+}(t)
\right),
\label{eq:positive-support}
\\
c_t^{-}(K)
&=
1-
\prod_{G\in\mathcal{G}_{K,t}}
\left(
1-\overline{\omega}_{G}^{-}(t)
\right).
\label{eq:negative-support}
\end{align}

The evidence-support profile is therefore

\begin{equation}
\mathbf{c}_t(K)
=
\left\langle
c_t^{+}(K),
c_t^{-}(K)
\right\rangle .
\label{eq:support-profile}
\end{equation}

The multiplicative-complement aggregation has three relevant properties. First, each additional provenance-independent group can increase support without causing the result to exceed one. Second, repeated evidence from the same derivation root cannot accumulate without bound because only the maximum group-level weight is used. Third, positive and negative support remain separate rather than being collapsed into a scalar whose opposing components may cancel.

The numbers of independently corroborating positive and negative groups are defined as

\begin{align}
n_t^{+}(K)
&=
\sum_{G\in\mathcal{G}_{K,t}}
\mathbb{I}
\left[
\overline{\omega}_{G}^{+}(t)
\geq\theta_g
\right],
\label{eq:positive-group-count}
\\
n_t^{-}(K)
&=
\sum_{G\in\mathcal{G}_{K,t}}
\mathbb{I}
\left[
\overline{\omega}_{G}^{-}(t)
\geq\theta_g
\right],
\label{eq:negative-group-count}
\end{align}

where $\theta_g$ is the minimum group-level corroboration threshold and $\mathbb{I}[\cdot]$ is the indicator function.

The independent-group counts are maintained in addition to the aggregated support. Consequently, one high-credibility transaction can establish strong candidate evidence but cannot independently satisfy a confirmation rule that requires multiple provenance-independent groups.

\subsection{Authorization-state interpretation}\label{subsec:state-interpretation}

The lifecycle state of relation key $K$ is interpreted using its positive support, negative support, number of independent provenance groups, and explicit lifecycle evidence.

An \textit{Unknown} relation has insufficient admissible evidence. A \textit{Candidate} relation has provisional positive support but has not satisfied the confidence and independence requirements for confirmation. A \textit{Confirmed} relation is supported by sufficiently credible positive evidence from multiple independent provenance groups.

A \textit{Conflicted} relation contains materially inconsistent evidence. This state applies when positive authorization observations are contradicted by reliable negative evidence, or when the provenance of previously accumulated evidence indicates possible contamination. The \textit{Conflicted} state characterizes the evidence associated with a specific subject--resource--object--action relation and does not characterize the subject or transaction source itself.

A \textit{Revoked} relation has been invalidated by an explicit trusted object-level revocation signal or by sufficiently strong independent negative evidence after the relation was established. An \textit{Expired} relation has lost temporal validity because its supporting evidence has decayed or exceeded the configured lifetime.

The minimum confirmation condition is

\begin{equation}
c_t^{+}(K)
\geq
\theta_{\mathrm{conf}}
\quad\land\quad
n_t^{+}(K)
\geq
m_{\mathrm{conf}}
\quad\land\quad
c_t^{-}(K)
<
\theta_{\mathrm{neg}},
\label{eq:confirmation-condition}
\end{equation}

where $\theta_{\mathrm{conf}}$ is the positive-support threshold, $m_{\mathrm{conf}}\geq2$ is the minimum number of independent provenance groups, and $\theta_{\mathrm{neg}}$ is the maximum negative support permitted for confirmation.

A relation satisfies the conflict condition when

\begin{equation}
c_t^{+}(K)
\geq
\theta_{\mathrm{pos}}
\quad\land\quad
c_t^{-}(K)
\geq
\theta_{\mathrm{conflict}},
\label{eq:conflict-condition}
\end{equation}

where $\theta_{\mathrm{pos}}$ and $\theta_{\mathrm{conflict}}$ are the minimum positive and negative support thresholds for identifying material evidence inconsistency.

A relation satisfies the evidence-based revocation condition when

\begin{equation}
c_t^{-}(K)
\geq
\theta_{\mathrm{rev}}
\quad\land\quad
n_t^{-}(K)
\geq
m_{\mathrm{rev}},
\label{eq:revocation-condition}
\end{equation}

where $\theta_{\mathrm{rev}}$ is the revocation-support threshold and $m_{\mathrm{rev}}$ is the required number of independent negative provenance groups. An explicitly trusted object-level revocation signal may trigger revocation without waiting for repeated observations.

Equations~\eqref{eq:confirmation-condition}--\eqref{eq:revocation-condition} define the evidence conditions associated with the lifecycle states. They do not determine the complete transition precedence. The ordering of conflicting conditions, rollback behavior, expiration, and concurrent state updates is defined by the online procedure in Section~\ref{sec:learning}.

\subsection{Model properties}\label{subsec:model-properties}

The cross-evidence authorization-state model maintains five principal properties.

\paragraph{No self-confirmation.}
Evidence derived from a single provenance group cannot independently establish a \textit{Confirmed} relation because $m_{\mathrm{conf}}\geq2$.

\paragraph{Action isolation.}
Evidence for action $\hat{a}$ cannot support a different action $a$ unless an explicit nonzero compatibility value $\Gamma(\hat{a},a)$ has been defined.

\paragraph{Bounded provenance contribution.}
Repeated evidence derived from the same root contributes at most the largest positive and negative weights assigned to its provenance group.

\paragraph{Conflict visibility.}
Positive and negative evidence are maintained independently. Strong contradictory evidence is therefore represented by a \textit{Conflicted} state rather than being concealed through scalar cancellation.

\paragraph{Temporal boundedness.}
Evidence is subject to source-specific temporal decay. Historical observations therefore cannot provide permanent authorization support unless they are refreshed by admissible evidence.

These properties provide the evidence and state foundation for the online update and authorization-decision procedures presented in Section~\ref{sec:learning}.

\section{Online state update and authorization decision}\label{sec:learning}

\subsection{Online processing sequence}\label{subsec:online-processing-sequence}

The online procedure separates request-time authorization-risk assessment from response-time state learning. This separation ensures that the response generated for a request cannot retrospectively validate the same request.

For each candidate object reference $\langle r_{ij},o_{ij}\rangle\in\mathcal{O}_i$ extracted from request $Q_i$, the component constructs a relation key

\begin{equation}
K_{ij}
=
\left\langle
\eta_i,
u_i,
r_{ij},
o_{ij},
a_i
\right\rangle .
\label{eq:request-relation-key}
\end{equation}

The set of relation keys associated with request $Q_i$ is

\begin{equation}
\mathcal{K}_i
=
\left\{
K_{ij}
\mid
\left\langle r_{ij},o_{ij}\right\rangle
\in
\mathcal{O}_i
\right\}.
\label{eq:request-relation-key-set}
\end{equation}

At request-arrival time $t_i^{q}$, the component retrieves the authorization states associated with $\mathcal{K}_i$ and produces an object-level authorization-risk decision. The decision uses only evidence observed before $(t_i^{q})^-$. If the request is forwarded, the corresponding response $P_i$ is subsequently correlated with $Q_i$, producing complete transaction $T_i$ at time $t_i^{p}$.

The response-time procedure extracts authorization-relevant evidence from $T_i$, evaluates its admissibility and provenance, and atomically updates the affected authorization relations. The updated state becomes available only to requests evaluated after the update has been committed.

The complete processing sequence is therefore

\begin{equation}
Q_i
\longrightarrow
\mathcal{K}_i
\longrightarrow
d_i
\longrightarrow
P_i
\longrightarrow
T_i
\longrightarrow
\Delta\mathcal{E}_i
\longrightarrow
\mathcal{A}_{t_i^{p}},
\label{eq:online-processing-sequence}
\end{equation}

where $d_i$ is the request-time decision and $\Delta\mathcal{E}_i$ is the set of evidence items extracted from transaction $T_i$.

\subsection{Request-time state retrieval}\label{subsec:request-time-retrieval}

Before evaluating request $Q_i$, the runtime component retrieves the most recent committed state for each $K\in\mathcal{K}_i$. Let

\begin{equation}
\mathcal{A}_{(t_i^{q})^-}(K)
=
\left\langle
s_i^{\mathrm{pre}}(K),
\mathbf{c}_i^{\mathrm{pre}}(K),
\Pi_i^{\mathrm{pre}}(K),
\tau_i^{\mathrm{pre}}(K)
\right\rangle .
\label{eq:pre-request-state}
\end{equation}

denote the state available immediately before request evaluation, where

\begin{equation}
\mathbf{c}_i^{-}(K)
=
\left\langle
c_i^{+}(K),
c_i^{-}(K)
\right\rangle .
\label{eq:pre-request-support-profile}
\end{equation}

Here, the superscript $-$ attached to the state variables denotes the instant before the current request is evaluated and should not be confused with negative evidence support.

Temporal decay is evaluated lazily when a relation is retrieved. For each provenance group $G$ associated with relation key $K$, its representative weight is updated to the request-arrival time as

\begin{equation}
\overline{\omega}_{G}^{p}(t_i^{q})
=
\overline{\omega}_{G}^{p}(t_G)
\exp
\left[
-\lambda_G
\left(
t_i^{q}-t_G
\right)
\right],
\qquad
p\in\{+,-\},
\label{eq:lazy-group-decay}
\end{equation}

where $t_G$ is the most recent evaluation time of provenance group $G$, and $\lambda_G$ is the decay coefficient inherited from its evidence-source class.

The positive and negative support values are then evaluated using Eqs.~\eqref{eq:positive-support} and \eqref{eq:negative-support}. Lazy evaluation avoids periodic scans of the complete authorization-state store while preserving time-dependent evidence validity.

If no state exists for relation key $K$, the component initializes

\begin{equation}
\mathcal{A}_{(t_i^{q})^-}(K)
=
\left\langle
\mathrm{Unknown},
\left\langle 0,0\right\rangle,
\varnothing,
t_i^{q}
\right\rangle .
\label{eq:unknown-state-initialization}
\end{equation}

The initialization of an \textit{Unknown} relation does not imply that the requested operation is unauthorized. It indicates only that the runtime component lacks sufficient admissible evidence for the corresponding relation.

\subsection{Object-level authorization-risk evaluation}\label{subsec:risk-evaluation}

The state of a relation is converted into an object-level risk value through a state-dependent prior and its positive and negative evidence support. For relation key $K$, the risk value at request-arrival time is defined as

\begin{equation}
\rho_i(K)
=
\operatorname{clip}_{[0,1]}
\left[
\beta\!\left(s_i^{\mathrm{pre}}(K)\right)
+
\mu_{-}c_i^{-}(K)
-
\mu_{+}c_i^{+}(K)
\right].
\label{eq:relation-risk}
\end{equation}

where $\beta(s)$ is the prior risk associated with lifecycle state $s$, $\mu_{-}\geq0$ controls the contribution of negative evidence, $\mu_{+}\geq0$ controls the risk reduction contributed by positive evidence, and

\begin{equation}
\operatorname{clip}_{[0,1]}(x)
=
\min
\left\{
1,
\max\{0,x\}
\right\}.
\label{eq:clip-function}
\end{equation}

The state priors satisfy the ordering

\begin{equation}
\beta(\mathrm{Confirmed})
<
\beta(\mathrm{Candidate})
<
\beta(\mathrm{Unknown})
\leq
\beta(\mathrm{Expired})
<
\beta(\mathrm{Conflicted})
<
\beta(\mathrm{Revoked}).
\label{eq:state-risk-order}
\end{equation}

This ordering reflects the evidential meaning of each lifecycle state. A \textit{Confirmed} relation has the lowest prior risk, whereas a \textit{Revoked} relation has the highest. A \textit{Conflicted} relation is assigned a high risk because its positive and negative evidence are materially inconsistent. An \textit{Unknown} relation is assigned an intermediate prior because the absence of evidence is not equivalent to evidence of unauthorized access.

For a request containing multiple object references, the request-level risk is conservatively aggregated as

\begin{equation}
\rho_i
=
\max_{K\in\mathcal{K}_i}
\rho_i(K).
\label{eq:request-risk}
\end{equation}

The maximum operator prevents a low-risk relation from concealing another object reference with a high-risk authorization state. This rule is particularly relevant to batch requests and structured payloads that contain several object identifiers.

The request-time decision is

\begin{equation}
d_i
=
\begin{cases}
\mathrm{Allow},
&
\rho_i < \theta_{\mathrm{review}},
\\[1mm]
\mathrm{Review},
&
\theta_{\mathrm{review}}
\leq
\rho_i
<
\theta_{\mathrm{block}},
\\[1mm]
\mathrm{Block},
&
\rho_i
\geq
\theta_{\mathrm{block}},
\end{cases}
\label{eq:request-decision}
\end{equation}

where

\begin{equation}
0
\leq
\theta_{\mathrm{review}}
<
\theta_{\mathrm{block}}
\leq
1.
\label{eq:decision-threshold-order}
\end{equation}

In passive deployment, the labels in Eq.~\eqref{eq:request-decision} represent risk categories for logging and retrospective analysis. In inline deployment, they can be mapped to forwarding, secondary verification, rate limitation, or blocking actions according to the configured enforcement policy.

The authorization-state model supplies object-level evidence to the decision function but does not replace the protected application's own authorization mechanism. A low runtime risk therefore indicates that the observed relation is sufficiently supported by external evidence; it does not constitute an authoritative access-control decision.

\subsection{Cold-start decision policy}\label{subsec:cold-start-policy}

During cold start, a large proportion of relation keys may be in the \textit{Unknown} state. Treating every unknown relation as a confirmed violation would produce excessive false positives, whereas allowing all unknown relations without observation would provide little protection.

The runtime component therefore supports three deployment policies.

\begin{table}[t]
\centering
\caption{Decision policies for relations without established
authorization evidence.}
\label{tab:cold-start-policies}
\footnotesize
\begin{tabularx}{\linewidth}{
    >{\raggedright\arraybackslash}p{2.2cm}
    >{\raggedright\arraybackslash}p{3.4cm}
    >{\raggedright\arraybackslash}X}
\toprule
Policy &
Treatment of unknown relations &
Intended use \\
\midrule

Observation
&
The request is forwarded and the resulting transaction is used for
evidence collection.
&
Initial deployment, model calibration, and retrospective validation.
\\

Review
&
The request is forwarded or challenged according to the surrounding
gateway policy and is assigned an elevated audit priority.
&
Gradual activation in environments where blocking unknown relations is
not acceptable.
\\

Enforcement
&
Unknown relations associated with designated sensitive actions may be
blocked or require additional verification.
&
Mature deployments with sufficient authorization-state coverage.
\\

\bottomrule
\end{tabularx}
\end{table}

The cold-start policy changes only the enforcement treatment of an \textit{Unknown} relation. It does not change the evidence required for promotion to \textit{Candidate} or \textit{Confirmed}. Consequently, relaxing enforcement during initial learning does not allow a single successful request to confirm its own authorization relation.

Actions with substantial state-changing effects may be assigned higher decision priors than read-only actions. Such configuration is applied to the object action $a$ and does not require inference of subject roles, organizational privilege levels, or identity attributes.

\subsection{Response-time evidence generation}\label{subsec:response-time-evidence}

After response $P_i$ is observed, transaction $T_i$ is processed by an evidence-extraction function

\begin{equation}
\Delta\mathcal{E}_i
=
\Phi(T_i),
\label{eq:evidence-extraction-function}
\end{equation}

where each

\begin{equation}
\varepsilon_k
\in
\Delta\mathcal{E}_i
\end{equation}

is represented according to Eq.~\eqref{eq:evidence-item}.

Evidence extraction includes four operations:

\begin{enumerate}
    \item
    correlating the request and response and evaluating $\kappa_k$;

    \item
    extracting candidate object references and mapping them to resource--object pairs;

    \item
    normalizing the requested action and evaluating action compatibility;

    \item
    assigning evidence polarity, source class, credibility factors, and provenance information.
\end{enumerate}

The extracted set is divided into admitted, held, and rejected evidence:

\begin{equation}
\Delta\mathcal{E}_i
=
\Delta\mathcal{E}_i^{\mathrm{adm}}
\cup
\Delta\mathcal{E}_i^{\mathrm{hold}}
\cup
\Delta\mathcal{E}_i^{\mathrm{rej}},
\label{eq:evidence-partition}
\end{equation}

with the three subsets being mutually disjoint.

Rejected evidence fails the basic admissibility requirements in Eq.~\eqref{eq:evidence-admissibility} and cannot modify authorization support. Held evidence satisfies the basic quality requirements but is temporarily prevented from promoting an authorization relation because the transaction triggers one or more contamination indicators. Admitted evidence is eligible to update the state model.

Let $b_k\in\{0,1\}$ denote the promotion-gate result for evidence item $\varepsilon_k$. Its online admission rule is

\begin{equation}
\varepsilon_k
\in
\Delta\mathcal{E}_i^{\mathrm{adm}}
\quad\Longleftrightarrow\quad
I_k=1
\land
\left(
p_k\leq0
\;\lor\;
b_k=1
\right).
\label{eq:online-evidence-admission}
\end{equation}

Positive evidence with $I_k=1$ and $b_k=0$ is placed in $\Delta\mathcal{E}_i^{\mathrm{hold}}$. It remains available for audit and later corroboration but does not increase positive authorization support. Negative evidence is not automatically discarded by the positive-evidence promotion gate because doing so could conceal object-specific denials or revocation information.

The construction of contamination indicators and the subsequent release or rollback of held evidence are presented in Section~\ref{sec:defense}. The online procedure in this section treats $b_k$ as a deterministic gate supplied by that mechanism.

\subsection{Provenance-group update}\label{subsec:provenance-group-update}

For each admitted evidence item $\varepsilon_k$, the runtime component identifies its relation key $K_k$ and provenance group $G_k=g(\pi_k)$. If the group does not exist, a new group record is created. Otherwise, the existing representative weights are first decayed to response-observation time $t_i^{p}$.

For polarity $p_k=+1$, the positive representative weight becomes

\begin{equation}
\overline{\omega}_{G_k}^{+}(t_i^{p})
=
\max
\left\{
\widehat{\omega}_{G_k}^{+}(t_i^{p}),
\omega_k(t_i^{p})
\right\},
\label{eq:incremental-positive-group-update}
\end{equation}

where

\begin{equation}
\widehat{\omega}_{G_k}^{+}(t_i^{p})
=
\overline{\omega}_{G_k}^{+}(t_G)
\exp
\left[
-\lambda_{G_k}
\left(
t_i^{p}-t_G
\right)
\right]
\label{eq:decayed-positive-group-weight}
\end{equation}

is the previous group weight decayed from its last evaluation time $t_G$.

Similarly, for polarity $p_k=-1$,

\begin{equation}
\overline{\omega}_{G_k}^{-}(t_i^{p})
=
\max
\left\{
\widehat{\omega}_{G_k}^{-}(t_i^{p}),
\omega_k(t_i^{p})
\right\}.
\label{eq:incremental-negative-group-update}
\end{equation}

Evidence with $p_k=0$ is retained in provenance but does not directly modify either representative support value.

After all admitted evidence items associated with relation key $K$ have been processed, the runtime component recalculates

\begin{equation}
\mathbf{c}_{t_i^{p}}(K)
=
\left\langle
c_{t_i^{p}}^{+}(K),
c_{t_i^{p}}^{-}(K)
\right\rangle
\label{eq:post-update-support-profile}
\end{equation}

and the corresponding independent-group counts $n_{t_i^{p}}^{+}(K)$ and $n_{t_i^{p}}^{-}(K)$.

\subsection{Lifecycle transition precedence}\label{subsec:transition-precedence}

Several lifecycle conditions may become true at the same update time. For example, a relation may retain substantial positive support while also receiving reliable negative evidence. The state-transition procedure therefore applies explicit precedence rather than evaluating conditions independently.

Let

\begin{equation}
s^{-}(K)
=
s_{(t_i^{p})^-}(K)
\end{equation}

denote the state before transaction $T_i$ is committed, and let

\begin{equation}
s^{+}(K)
=
s_{t_i^{p}}(K)
\end{equation}

denote the resulting state. The transition precedence is shown in Table~\ref{tab:transition-precedence}.

\begin{table}[t]
\centering
\caption{Precedence of online authorization-state transitions.}
\label{tab:transition-precedence}
\footnotesize
\begin{tabularx}{\linewidth}{
    >{\centering\arraybackslash}p{1.0cm}
    >{\raggedright\arraybackslash}p{4.3cm}
    >{\centering\arraybackslash}p{1.7cm}
    >{\raggedright\arraybackslash}X}
\toprule
Order &
Condition &
Resulting state &
Rationale \\
\midrule

1
&
A trusted object-level revocation signal is admitted.
&
\textit{Revoked}
&
Explicit revocation has priority over previously accumulated positive
support.
\\

2
&
The conflict condition in Eq.~\eqref{eq:conflict-condition} is
satisfied.
&
\textit{Conflicted}
&
Material positive and negative evidence must remain visible rather than
being collapsed into either authorization or denial.
\\

3
&
The evidence-based revocation condition in
Eq.~\eqref{eq:revocation-condition} is satisfied without material
positive conflict.
&
\textit{Revoked}
&
Independent negative evidence invalidates the previously supported
relation.
\\

4
&
The relation has exceeded its lifetime or its positive support has
fallen below the expiration threshold.
&
\textit{Expired}
&
Previously established evidence is no longer temporally valid.
\\

5
&
The confirmation condition in Eq.~\eqref{eq:confirmation-condition} is
satisfied.
&
\textit{Confirmed}
&
Sufficient positive support has been obtained from independent
provenance groups.
\\

6
&
Positive admissible evidence exists but confirmation requirements are
not satisfied.
&
\textit{Candidate}
&
The relation remains provisional.
\\

7
&
No admissible evidence supports another lifecycle state.
&
\textit{Unknown}
&
The component has no sufficient evidence-based relation.
\\

\bottomrule
\end{tabularx}
\end{table}

Let $\Delta_{\mathrm{life}}(K)$ denote the maximum configured lifetime of positive support for relation key $K$. The expiration condition is

\begin{equation}
t_i^{p}-\tau^{+}(K)
\geq
\Delta_{\mathrm{life}}(K)
\quad\lor\quad
c_{t_i^{p}}^{+}(K)
<
\theta_{\mathrm{exp}},
\label{eq:expiration-condition}
\end{equation}

where $\tau^{+}(K)$ is the most recent time at which admissible positive evidence refreshed relation $K$, and $\theta_{\mathrm{exp}}$ is the minimum positive support required to retain temporal validity.

The resulting state can be expressed as

\begin{equation}
s^{+}(K)
=
\mathcal{F}
\left(
s^{-}(K),
\mathbf{c}_{t_i^{p}}(K),
n_{t_i^{p}}^{+}(K),
n_{t_i^{p}}^{-}(K),
\Pi_{t_i^{p}}(K),
t_i^{p}
\right),
\label{eq:state-transition-function}
\end{equation}

where $\mathcal{F}$ applies the precedence in Table~\ref{tab:transition-precedence}.

A \textit{Revoked} relation is not restored by ordinary positive evidence alone. Its reactivation requires explicit and independently verifiable object-level reauthorization evidence. Similarly, evidence that has been held because of contamination indicators cannot promote a relation until the hold condition is resolved. These lifecycle operations are specified in Section~\ref{sec:defense}.

\subsection{Atomic state commit}\label{subsec:atomic-state-commit}

All evidence and state changes caused by one completed transaction are committed atomically for each affected relation key. For relation $K$, the committed record is

\begin{equation}
R_{t_i^{p}}(K)
=
\left\langle
s_{t_i^{p}}(K),
\mathbf{c}_{t_i^{p}}(K),
\Pi_{t_i^{p}}(K),
\tau_{t_i^{p}}(K),
v_{t_i^{p}}(K)
\right\rangle ,
\label{eq:versioned-relation-record}
\end{equation}

where $v_{t_i^{p}}(K)$ is a monotonically increasing version number.

The update procedure reads version $v^{-}(K)$, computes the new provenance, support profile, and lifecycle state, and commits the result only if the stored version remains equal to $v^{-}(K)$. Otherwise, the procedure reloads the current state and repeats the update. This compare-and-commit rule prevents lost updates when several responses affect the same relation concurrently.

For a transaction containing several distinct relation keys, updates may be committed independently because the authorization state is maintained per key. If an application operation requires all referenced objects to be updated as one indivisible business unit, the deployment may group the affected keys under the same transaction commit identifier. The grouped commit preserves transaction-level traceability without merging their authorization evidence.

\subsection{Out-of-order and delayed responses}\label{subsec:out-of-order-responses}

Concurrent requests may complete in a different order from their arrival order. The state learner therefore distinguishes request time $t_i^{q}$ from response-observation time $t_i^{p}$ and retains the original evidence observation time in each provenance record.

State updates are committed in response-observation order, but temporal weights are evaluated using the evidence timestamp $t_k$. Consequently, a delayed response cannot be treated as newly generated evidence merely because it was processed later.

For two transactions $T_i$ and $T_j$ affecting the same relation, it is possible that

\begin{equation}
t_i^{q}
<
t_j^{q}
\quad\land\quad
t_i^{p}
>
t_j^{p}.
\label{eq:out-of-order-condition}
\end{equation}

In this case, $T_j$ is committed first. When $T_i$ is subsequently processed, the versioned update procedure recomputes its contribution against the latest committed state. The evidence timestamp remains $t_i^{p}$ or the more precise application event time when such a value is available.

A stale response is not permitted to overwrite a more recent trusted revocation or reauthorization record. Lifecycle events therefore carry their own event time and precedence metadata. Detailed handling of revocation, reauthorization, and rollback ordering is presented in Section~\ref{sec:defense}.

\subsection{Online update procedure}\label{subsec:online-update-procedure}

For each completed transaction $T_i$, the response-time update procedure performs the following steps:

\begin{enumerate}
    \item
    Extract evidence set $\Delta\mathcal{E}_i=\Phi(T_i)$.

    \item
    Evaluate the basic admissibility indicator $I_k$ for every evidence item.

    \item
    Apply the positive-evidence promotion gate $b_k$ and divide the evidence into admitted, held, and rejected subsets.

    \item
    Group admitted evidence by relation key and provenance root.
    
    \item
    Retrieve the current versioned state for each affected relation key.

    \item
    Decay existing provenance-group weights to time $t_i^{p}$.

    \item
    Merge admitted positive and negative evidence into the corresponding provenance groups.

    \item
    Recalculate positive and negative support and independent-group counts.

    \item
    Evaluate lifecycle conditions according to the transition precedence in Table~\ref{tab:transition-precedence}.

    \item
    Atomically commit the updated relation record and append the evidence decision to the audit trail.
\end{enumerate}

For request-time processing, the component performs the following steps:

\begin{enumerate}
    \item
    Extract candidate object references and construct
    $\mathcal{K}_i$.

    \item
    Retrieve and temporally normalize the state of every relation key.

    \item
    Compute per-relation risk values using Eq.~\eqref{eq:relation-risk}.

    \item
    Aggregate the request-level risk using Eq.~\eqref{eq:request-risk}.
    
    \item
    Produce decision $d_i$ according to the configured decision thresholds and cold-start policy.

    \item
    Record the relation states and versions used for the decision so that the result remains auditable after later state updates.
\end{enumerate}

The request-time procedure does not update positive authorization support. This restriction preserves the temporal separation between decision and learning and prevents the current request from influencing its own authorization assessment.

\subsection{Computational complexity}\label{subsec:online-complexity}

Let $m_i=|\mathcal{K}_i|$ denote the number of object-level relation keys extracted from request $Q_i$. With a hash-indexed or key--value-based state store, retrieving one relation record has expected complexity $O(1)$. The expected request-time lookup complexity is therefore

\begin{equation}
O(m_i).
\label{eq:request-time-complexity}
\end{equation}

Let $g_K$ denote the number of active provenance groups retained for relation key $K$. Recomputing positive and negative support directly from all active groups requires

\begin{equation}
O(g_K)
\label{eq:relation-recomputation-complexity}
\end{equation}

time. Updating the representative weight of an existing provenance group requires expected $O(1)$ time when groups are hash-indexed. Consequently, the response-time complexity for transaction $T_i$ is

\begin{equation}
O
\left(
|\Delta\mathcal{E}_i|
+
\sum_{K\in\mathcal{K}_i^{\mathrm{upd}}}
g_K
\right),
\label{eq:response-time-complexity}
\end{equation}

where $\mathcal{K}_i^{\mathrm{upd}}$ is the set of relation keys updated by the transaction.

The state-store space requirement is

\begin{equation}
O
\left(
|\mathcal{A}|
+
\sum_{K\in\mathcal{A}}g_K
\right),
\label{eq:state-space-complexity}
\end{equation}

where $|\mathcal{A}|$ is the number of retained authorization relations. Provenance groups whose weights have decayed below the retention threshold and whose audit retention periods have elapsed can be compacted or removed. This bounds the active state required for online decisions while preserving longer-term audit records in separate storage.

\subsection{Decision and update properties}\label{subsec:decision-update-properties}

The online procedure satisfies five operational properties.

\paragraph{Temporal isolation.}
The decision for request $Q_i$ uses only the state committed before $t_i^{q}$. Evidence extracted from response $P_i$ can affect only subsequent requests.

\paragraph{Conservative multi-object aggregation.}
A high-risk relation cannot be hidden by other low-risk object references in the same request because request risk is aggregated using the maximum relation risk.

\paragraph{Auditable state mutation.}
Every state change retains the originating evidence, provenance group, previous version, resulting version, and transition condition.

\paragraph{Conflict-preserving updates.}
Positive evidence does not overwrite negative evidence, and negative evidence does not silently erase positive support. Material inconsistency produces a \textit{Conflicted} state.

\paragraph{Concurrent-update safety.}
Versioned atomic commits prevent later transactions from overwriting state changes that were committed after the update procedure began.

These properties connect the cross-evidence state model to runtime authorization-risk decisions. Section~\ref{sec:defense} further specifies the mechanisms used to identify contamination attempts, quarantine provisional evidence, roll back affected support, and process object-level authorization evolution.

\section{Poisoning defense and authorization evolution}\label{sec:defense}

\subsection{Threats to online authorization-state learning}\label{subsec:poisoning-threats}

An online authorization learner derives future decisions from previously observed transactions. This dependency creates a distinct attack surface: an adversary may attempt to cause unauthorized transactions to be incorporated as positive evidence and thereby reduce the risk assigned to subsequent accesses.

The protected application cannot be treated as an authoritative source of authorization evidence because the application may itself contain the object-level authorization defect under analysis. In particular, a successful response to an object-specific request does not prove that the requester was authorized to perform the operation.

The proposed defense considers five object-level contamination patterns.

\begin{enumerate}
    \item
    \textbf{Self-validation.}
    A transaction attempts to use its own successful response to establish the authorization relation required to justify the same request.
    
    \item
    \textbf{Repeated correlated evidence.}
    Multiple observations derived from the same transaction, response, collection, or derivation root are presented as if they were independent confirmation.

    \item
    \textbf{Public or aggregate exposure.}
    Objects observed in public search results, recommendations, cross-subject collections, or unscoped responses are treated incorrectly as subject-specific authorization evidence.

    \item
    \textbf{Unsupported object injection.}
    Previously unseen object references are introduced through successful direct operations without independent creation, assignment, subject-scoped collection, or other corroborating evidence.

    \item
    \textbf{Stale relation retention.}
    Previously valid authorization evidence remains effective after object transfer, access revocation, temporary-access expiration, or other observable object-level authorization changes.
\end{enumerate}

These patterns concern the evidential status of a specific subject--resource--object--action relation. The method assumes that the subject binding supplied by the application or gateway is trusted input and does not infer or modify identity attributes, role membership, or privilege levels.

\subsection{Contamination indicators}\label{subsec:contamination-indicators}

For positive evidence item $\varepsilon_k$, the contamination-indicator vector is defined as

\begin{equation}
\mathbf{h}_k
=
\left\langle
h_k^{\mathrm{self}},
h_k^{\mathrm{scope}},
h_k^{\mathrm{dep}},
h_k^{\mathrm{nov}},
h_k^{\mathrm{align}},
h_k^{\mathrm{conf}}
\right\rangle ,
\label{eq:contamination-indicator-vector}
\end{equation}

where each component belongs to $[0,1]$.

The indicator components have the following meanings.

\begin{table}[t]
\centering
\caption{Object-level contamination indicators used by the positive
evidence promotion gate.}
\label{tab:contamination-indicators}
\footnotesize
\begin{tabularx}{\linewidth}{
    >{\raggedright\arraybackslash}p{2.25cm}
    >{\raggedright\arraybackslash}X}
\toprule
Indicator & Interpretation \\
\midrule

$h_k^{\mathrm{self}}$
&
The evidence originates from the response to the same direct object-specific operation whose authorization relation would be
promoted.
\\

$h_k^{\mathrm{scope}}$
&
The response scope is public, aggregate, cross-subject, or insufficiently specific to the observed subject.
\\

$h_k^{\mathrm{dep}}$
&
The evidence is strongly dependent on previously observed evidence through the same provenance root or tightly coupled derivation chain.
\\

$h_k^{\mathrm{nov}}$
&
The object relation is newly observed and lacks prior independent creation, assignment, or subject-scoped evidence.
\\

$h_k^{\mathrm{align}}$
&
The resource type, endpoint, object identifier, or action alignment is uncertain or inconsistent with existing evidence.
\\

$h_k^{\mathrm{conf}}$
&
The evidence contradicts reliable negative evidence or an existing revocation record for the same relation.
\\

\bottomrule
\end{tabularx}
\end{table}

The self-validation indicator is defined conservatively as

\begin{equation}
h_k^{\mathrm{self}}
=
\begin{cases}
1,
&
z_k=\mathrm{DirectOperation}
\ \land\
K_k\in\mathcal{K}_i,
\\
0,
&
\text{otherwise},
\end{cases}
\label{eq:self-validation-indicator}
\end{equation}

where $\mathcal{K}_i$ is the set of relation keys evaluated for request $Q_i$. Thus, a direct successful response can be retained as provisional evidence but cannot independently validate the request that generated it.

The scope indicator is derived from the source-specificity coefficient:

\begin{equation}
h_k^{\mathrm{scope}}
=
1-\sigma_k.
\label{eq:scope-indicator}
\end{equation}

The alignment indicator is similarly defined as

\begin{equation}
h_k^{\mathrm{align}}
=
1-\xi_k.
\label{eq:alignment-indicator}
\end{equation}

The provenance-dependence indicator is

\begin{equation}
h_k^{\mathrm{dep}}
=
\max_{\varepsilon_j\in\mathcal{E}_{t_k^-}(K_k)}
\operatorname{dep}
\left(
\pi_k,\pi_j
\right),
\label{eq:dependence-indicator}
\end{equation}

where

\begin{equation}
\operatorname{dep}
\left(
\pi_k,\pi_j
\right)
\in[0,1]
\label{eq:provenance-dependence-function}
\end{equation}

measures the derivational dependence between two provenance descriptors. Evidence sharing the same provenance root has dependence one.

Let

\begin{equation}
\operatorname{prior}(K_k,t_k)
=
\mathbb{I}
\left[
n_{t_k^-}^{+}(K_k)>0
\right]
\label{eq:prior-positive-evidence}
\end{equation}

indicate whether relation $K_k$ already has an independently corroborated positive provenance group. The novelty indicator is

\begin{equation}
h_k^{\mathrm{nov}}
=
1-
\operatorname{prior}(K_k,t_k).
\label{eq:novelty-indicator}
\end{equation}

For evidence produced by an explicit object-creation or assignment transaction, $h_k^{\mathrm{nov}}$ may be set to zero because the creation or assignment event itself explains the appearance of the new object relation.

The contradiction indicator is defined as

\begin{equation}
h_k^{\mathrm{conf}}
=
\max
\left\{
c_{t_k^-}^{-}(K_k),
R_{t_k^-}(K_k)
\right\},
\label{eq:contradiction-indicator}
\end{equation}

where $R_t(K)\in\{0,1\}$ indicates whether a trusted object-level revocation record is active for relation key $K$.

\subsection{Contamination score and promotion gate}\label{subsec:contamination-gate}

The aggregate contamination score of positive evidence item $\varepsilon_k$ is

\begin{equation}
\chi_k
=
1-
\prod_{j\in\mathcal{J}}
\left(
1-w_j h_k^{j}
\right),
\label{eq:contamination-score}
\end{equation}

where

\begin{equation}
\mathcal{J}
=
\{
\mathrm{self},
\mathrm{scope},
\mathrm{dep},
\mathrm{nov},
\mathrm{align},
\mathrm{conf}
\},
\end{equation}

and $w_j\in[0,1]$ controls the contribution of indicator $j$.

The multiplicative-complement form preserves the score in $[0,1]$ and allows several moderate indicators to produce a high aggregate contamination score. A single indicator with value zero does not remove the effect of the remaining indicators.

The positive-evidence promotion gate is

\begin{equation}
b_k
=
\begin{cases}
1,
&
\chi_k<\theta_{\mathrm{hold}}
\ \land\
h_k^{\mathrm{self}}<1
\ \land\
R_{t_k^-}(K_k)=0,
\\[1mm]
0,
&
\text{otherwise},
\end{cases}
\label{eq:promotion-gate}
\end{equation}

where $\theta_{\mathrm{hold}}$ is the maximum contamination score permitted for immediate admission.

Equation~\eqref{eq:promotion-gate} gives absolute priority to two conditions. First, evidence produced by the current direct operation cannot independently promote its own authorization relation. Second, ordinary positive observations cannot override an active trusted revocation record.

A positive evidence item satisfying $I_k=1$ but $b_k=0$ is placed in the held set defined in Eq.~\eqref{eq:evidence-partition}. Holding is different from rejection: the evidence remains traceable and may later be released if independent corroboration resolves the contamination indicators.

Negative and ambiguous evidence are not interpreted through the same positive promotion gate. Negative evidence is evaluated according to its specificity, credibility, provenance independence, and revocation semantics. Ambiguous evidence remains non-promotional.

\subsection{Held-evidence quarantine}\label{subsec:held-evidence-quarantine}

A held evidence record is represented as

\begin{equation}
H_k
=
\left\langle
\varepsilon_k,
\chi_k,
\mathbf{h}_k,
t_k^{\mathrm{hold}},
t_k^{\mathrm{exp}},
\zeta_k
\right\rangle ,
\label{eq:held-evidence-record}
\end{equation}

where $t_k^{\mathrm{hold}}$ is the quarantine time, $t_k^{\mathrm{exp}}$ is the hold-expiration time, and $\zeta_k$ is the release status.

The release status belongs to

\begin{equation}
\zeta_k
\in
\{
\mathrm{Pending},
\mathrm{Released},
\mathrm{Rejected},
\mathrm{Expired}
\}.
\label{eq:hold-status-set}
\end{equation}

Held positive evidence does not contribute to $c_t^{+}(K_k)$ or $n_t^{+}(K_k)$ while its status is $\mathrm{Pending}$. It may be released only when at least one of the following object-level corroboration conditions is satisfied:

\begin{enumerate}
    \item
    a provenance-independent object-creation or assignment event supports the same relation;

    \item
    the same object--action relation appears in a provenance-independent subject-scoped source;

    \item
    multiple independent evidence groups satisfy the configured release credibility and support thresholds;

    \item
    a trusted object-level reauthorization event explicitly restores the relation.
\end{enumerate}

Let

\begin{equation}
\mathcal{G}_{k}^{\mathrm{new}}(t)
=
\left\{
G\in\mathcal{G}_{K_k,t}
\mid
G\not\sim g(\pi_k),
\;
t_G>t_k
\right\}
\label{eq:new-independent-groups}
\end{equation}

denote the provenance-independent groups observed after evidence item $\varepsilon_k$ was held.

The evidence may be released when

\begin{equation}
\left|
\left\{
G\in\mathcal{G}_{k}^{\mathrm{new}}(t)
\mid
\overline{\omega}_{G}^{+}(t)
\geq
\theta_{\mathrm{rel}}
\right\}
\right|
\geq
m_{\mathrm{rel}}
\label{eq:held-evidence-release-condition}
\end{equation}

and

\begin{equation}
c_t^{-}(K_k)
<
\theta_{\mathrm{neg}},
\qquad
R_t(K_k)=0,
\label{eq:held-release-negative-condition}
\end{equation}

where $\theta_{\mathrm{rel}}$ is the minimum credibility of a release group and $m_{\mathrm{rel}}$ is the required number of independent groups.

When released, the original evidence timestamp $t_k$ is retained. Therefore, delayed release does not make old evidence appear newly observed and does not bypass temporal decay.

A pending evidence item expires when

\begin{equation}
t
\geq
t_k^{\mathrm{exp}}
\label{eq:held-evidence-expiration}
\end{equation}

without satisfying the release condition. Expired held evidence remains in the audit record but cannot subsequently contribute to authorization support unless it is observed again as a new evidence item.

\subsection{Provenance-based rollback}\label{subsec:evidence-rollback}

Evidence that was initially admitted may later be determined to be contaminated. For example, an endpoint previously interpreted as subject-scoped may be identified as public, or several apparently independent observations may later be linked to the same derivation root.

Each admitted evidence item and provenance group therefore has an activity flag

\begin{equation}
a_k^{\mathrm{ev}}
\in
\{0,1\},
\qquad
a_G^{\mathrm{grp}}
\in
\{0,1\}.
\label{eq:evidence-activity-flags}
\end{equation}

Inactive evidence remains auditable but is excluded from support aggregation. The active provenance-group set is

\begin{equation}
\mathcal{G}_{K,t}^{\mathrm{act}}
=
\left\{
G\in\mathcal{G}_{K,t}
\mid
a_G^{\mathrm{grp}}=1
\right\}.
\label{eq:active-provenance-groups}
\end{equation}

When provenance root $G^{*}$ is invalidated, rollback sets

\begin{equation}
a_{G^{*}}^{\mathrm{grp}}
\leftarrow
0
\label{eq:group-invalidation}
\end{equation}

and recomputes

\begin{align}
c_t^{+}(K)
&=
1-
\prod_{G\in\mathcal{G}_{K,t}^{\mathrm{act}}}
\left(
1-\overline{\omega}_{G}^{+}(t)
\right),
\label{eq:rollback-positive-support}
\\
c_t^{-}(K)
&=
1-
\prod_{G\in\mathcal{G}_{K,t}^{\mathrm{act}}}
\left(
1-\overline{\omega}_{G}^{-}(t)
\right).
\label{eq:rollback-negative-support}
\end{align}

The lifecycle state is then re-evaluated according to the transition precedence in Table~\ref{tab:transition-precedence}.

Let

\begin{equation}
\mathcal{D}(G^{*})
=
\left\{
K
\mid
G^{*}\in\mathcal{G}_{K,t}
\right\}
\label{eq:rollback-dependency-set}
\end{equation}

denote the set of authorization relations whose support depends on provenance group $G^{*}$. Rollback is applied to every relation in $\mathcal{D}(G^{*})$.

A relation previously in the \textit{Confirmed} state may become \textit{Candidate}, \textit{Unknown}, or \textit{Conflicted} after rollback. It does not remain confirmed merely because it was confirmed before the contaminated evidence was identified.

A rollback record contains

\begin{equation}
B_{\ell}
=
\left\langle
G^{*},
\mathcal{D}(G^{*}),
t_{\ell},
\mathrm{reason}_{\ell},
\mathrm{before}_{\ell},
\mathrm{after}_{\ell}
\right\rangle ,
\label{eq:rollback-record}
\end{equation}

which preserves the invalidated provenance root, affected relations, rollback time, reason, and state versions before and after correction.

\subsection{Explicit revocation}\label{subsec:explicit-revocation}

An explicit revocation event is represented as

\begin{equation}
V_{\ell}
=
\left\langle
K_{\ell},
t_{\ell}^{\mathrm{event}},
q_{\ell}^{\mathrm{rev}},
\pi_{\ell}^{\mathrm{rev}}
\right\rangle ,
\label{eq:revocation-event}
\end{equation}

where $K_{\ell}$ is the affected authorization relation, $t_{\ell}^{\mathrm{event}}$ is the event time, $q_{\ell}^{\mathrm{rev}}\in[0,1]$ is the revocation credibility, and $\pi_{\ell}^{\mathrm{rev}}$ is the revocation provenance.

A revocation event is treated as trusted when

\begin{equation}
q_{\ell}^{\mathrm{rev}}
\geq
\theta_{\mathrm{rev}}^{\mathrm{trusted}}
\label{eq:trusted-revocation-condition}
\end{equation}

and its observable semantics identify the affected resource, object, subject, and action with sufficient specificity.

When admitted, the event sets

\begin{equation}
R_{t_{\ell}^{\mathrm{event}}}(K_{\ell})
=
1
\label{eq:activate-revocation-record}
\end{equation}

and transitions the corresponding relation to \textit{Revoked}. Previously accumulated positive support is retained for audit but is marked as superseded by the revocation event.

Ordinary successful accesses observed after revocation are held rather than admitted as positive support. This rule prevents an application that continues to allow unauthorized operations from automatically undoing the external revocation state.

Revocation may be action-specific. For example, observable evidence may invalidate update or delete authorization without invalidating read authorization. The affected action set must therefore be determined from the revocation semantics rather than expanded automatically to all actions associated with the object.

\subsection{Evidence-based revocation}\label{subsec:evidence-based-revocation}

An explicit revocation signal may not always be externally visible. The model therefore also supports evidence-based revocation through independent negative observations.

Relation $K$ satisfies the evidence-based revocation condition when

\begin{equation}
c_t^{-}(K)
\geq
\theta_{\mathrm{rev}}
\quad\land\quad
n_t^{-}(K)
\geq
m_{\mathrm{rev}}
\quad\land\quad
c_t^{+}(K)
<
\theta_{\mathrm{pos}},
\label{eq:extended-evidence-revocation}
\end{equation}

or when the corresponding condition in Eq.~\eqref{eq:revocation-condition} is met according to the configured transition precedence.

When reliable positive and negative evidence remain simultaneously strong, the resulting state is \textit{Conflicted} rather than \textit{Revoked}. This preserves observable disagreement and avoids silently discarding established positive evidence.

Evidence-based revocation is intentionally more conservative than trusted explicit revocation because an object-specific denial can result from transient application errors, incomplete workflows, or temporary business conditions. It therefore requires independent negative provenance groups and configurable persistence.

\subsection{Object transfer and reassignment}\label{subsec:object-transfer}

An object transfer changes the subject-specific authorization relation associated with an object. Let $u_{\mathrm{old}}$ denote the previously supported subject and $u_{\mathrm{new}}$ denote the subject appearing in a reliable reassignment transaction.

The old and new relation keys are

\begin{align}
K_{\mathrm{old}}
&=
\left\langle
\eta,
u_{\mathrm{old}},
r,
o,
a
\right\rangle ,
\label{eq:old-transfer-key}
\\
K_{\mathrm{new}}
&=
\left\langle
\eta,
u_{\mathrm{new}},
r,
o,
a
\right\rangle .
\label{eq:new-transfer-key}
\end{align}

A sufficiently credible transfer event activates a revocation or expiration transition for $K_{\mathrm{old}}$. It may simultaneously produce positive candidate evidence for $K_{\mathrm{new}}$.

However, the transfer event does not automatically establish all actions for the new subject. Evidence is projected only to actions explicitly supported by the observable transfer semantics and the action-compatibility function in Eq.~\eqref{eq:action-compatibility}.

For example, evidence that an object is assigned to a new subject may support object visibility or a designated management action but does not automatically confirm export, deletion, or approval authorization. Each affected object--action relation remains subject to the ordinary confirmation and independence requirements.

The model does not infer organizational role changes or privilege hierarchies from the transfer. It updates only the directly observable object-level relations associated with the transferred object.

\subsection{Expiration and temporal invalidation}\label{subsec:authorization-expiration}

Authorization evidence may become stale even when no explicit revocation event is observed. For each relation key $K$, let $\tau^{+}(K)$ denote the most recent time at which admissible positive evidence refreshed the relation.

The relation reaches temporal expiration when

\begin{equation}
t-\tau^{+}(K)
\geq
\Delta_{\mathrm{life}}(K)
\label{eq:lifetime-expiration}
\end{equation}

or when evidence decay reduces its positive support below the expiration threshold:

\begin{equation}
c_t^{+}(K)
<
\theta_{\mathrm{exp}}.
\label{eq:support-expiration}
\end{equation}

A relation in the \textit{Expired} state is not equivalent to a revoked relation. Expiration indicates that the runtime component no longer has sufficiently recent evidence, whereas revocation represents strong evidence that the previously supported relation should no longer be used.

New admissible evidence may return an expired relation to \textit{Candidate} and later to \textit{Confirmed}. By contrast, an active revocation record requires explicit reauthorization before ordinary positive evidence can restore the relation.

Different evidence sources may use different lifetimes and decay coefficients. Object-creation or assignment evidence may retain validity longer than a single direct operation, while temporary-access evidence may require a short lifetime.

\subsection{Object-level reauthorization}\label{subsec:reauthorization}

A revoked relation can be restored only through an object-level reauthorization event whose provenance is independent of the evidence that caused the previous revocation.

A reauthorization event is represented as

\begin{equation}
U_{\ell}
=
\left\langle
K_{\ell},
t_{\ell}^{\mathrm{event}},
q_{\ell}^{\mathrm{reauth}},
\pi_{\ell}^{\mathrm{reauth}}
\right\rangle .
\label{eq:reauthorization-event}
\end{equation}

The event is admissible when

\begin{equation}
q_{\ell}^{\mathrm{reauth}}
\geq
\theta_{\mathrm{reauth}}
\quad\land\quad
\pi_{\ell}^{\mathrm{reauth}}
\not\sim
\pi_{\ell}^{\mathrm{rev}},
\label{eq:reauthorization-condition}
\end{equation}

where $\pi_{\ell}^{\mathrm{rev}}$ is the provenance of the active revocation record.

Upon valid reauthorization, the revocation record is closed:

\begin{equation}
R_t(K_{\ell})
\leftarrow
0.
\label{eq:close-revocation-record}
\end{equation}

The relation then enters \textit{Candidate} unless the reauthorization event itself satisfies a configured trusted confirmation rule. Ordinary successful object operations observed after revocation do not qualify as reauthorization because they may be manifestations of the same authorization defect.

Reauthorization remains action-specific. Restoring one action does not implicitly restore other actions on the same object.

\subsection{Ordering of lifecycle events}\label{subsec:lifecycle-event-ordering}

Revocation, transfer, expiration, reauthorization, and ordinary transaction evidence may be observed out of order. Each lifecycle event therefore includes an event time and a processing time.

Let $t_{\ell}^{\mathrm{event}}$ denote the time at which lifecycle event $\ell$ occurred and $t_{\ell}^{\mathrm{proc}}$ denote the time at which it was processed. Event ordering is determined primarily by $t_{\ell}^{\mathrm{event}}$.

A lifecycle event is stale with respect to relation key $K$ when

\begin{equation}
t_{\ell}^{\mathrm{event}}
<
\tau_{\mathrm{life}}(K)
\label{eq:stale-lifecycle-event}
\end{equation}

and it does not carry higher-priority trusted semantics than the current lifecycle record. Here, $\tau_{\mathrm{life}}(K)$ is the event time of the most recent accepted revocation, transfer, or reauthorization event.

A stale ordinary positive observation cannot overwrite a newer revocation. Similarly, a delayed revocation event whose event time precedes a later trusted reauthorization is retained for audit but does not reactivate the revoked state.

When event times are unavailable or unreliable, the method uses processing order conservatively and assigns the resulting relation to \textit{Conflicted} when incompatible high-credibility lifecycle events cannot be ordered reliably.

\subsection{Bounded contamination influence}\label{subsec:bounded-contamination}

The provenance and promotion mechanisms limit the effect of contaminating transactions.

\paragraph{Single-transaction bound.}
A single transaction contributes to at most one representative positive and one representative negative weight per relation and provenance group. Repeating identifiers within the same response therefore cannot increase the independent-group count.

\paragraph{Self-validation exclusion.}
A direct operation response associated with the current request has $h_k^{\mathrm{self}}=1$ and cannot pass the immediate promotion gate. It may remain held, but it cannot confirm the same relation without later independent corroboration.

\paragraph{Public-source exclusion.}
Evidence from a source with insufficient subject specificity has a high scope indicator and cannot independently confirm subject-specific authorization.

\paragraph{Revocation priority.}
An active trusted revocation record prevents ordinary positive transactions from restoring the relation.

\paragraph{Rollback completeness.}
Invalidating provenance root $G^{*}$ causes every relation in $\mathcal{D}(G^{*})$ to be recomputed without the affected group.

These bounds do not guarantee that all authorization evidence is correct. They ensure that commonly observable contamination patterns cannot silently and irreversibly promote authorization states through unlimited repeated observations.

\subsection{Defense processing procedure}\label{subsec:defense-processing-procedure}

For each positive evidence item $\varepsilon_k$, the defense procedure performs the following operations:

\begin{enumerate}
    \item
    compute contamination indicators $\mathbf{h}_k$;

    \item
    compute aggregate contamination score $\chi_k$;

    \item
    evaluate the promotion gate $b_k$;

    \item
    admit, hold, or reject the evidence according to Eqs.~\eqref{eq:evidence-admissibility}, \eqref{eq:online-evidence-admission}, and \eqref{eq:promotion-gate};

    \item
    create a quarantine record for held evidence;

    \item
    periodically or event-triggeredly evaluate independent corroboration for pending held evidence;

    \item
    release, reject, or expire held evidence according to its corroboration and lifetime;

    \item
    invalidate affected provenance groups when contamination is confirmed and recompute dependent authorization relations.
\end{enumerate}

For each lifecycle event, the procedure performs the following operations:

\begin{enumerate}
    \item
    identify the affected resource--object--action relations;

    \item
    evaluate event credibility, specificity, provenance, and event
    time;

    \item
    compare the event with the current lifecycle record;

    \item
    apply revocation, transfer, expiration, or reauthorization semantics;

    \item
    atomically commit the resulting relation states and append the lifecycle event to the audit trail.
\end{enumerate}

The defense mechanisms preserve the same relation key used throughout the authorization-state model. They do not introduce separate identity, role, or organizational privilege states.

\subsection{Defense properties}\label{subsec:defense-properties}

The proposed defense provides six principal properties.

\paragraph{Non-self-validating learning.}
A successful direct operation cannot immediately confirm the authorization relation required for that operation.

\paragraph{Dependence-aware corroboration.}
Repeated observations sharing the same derivation root do not satisfy independent evidence requirements.

\paragraph{Quarantined uncertainty.}
Potentially useful but contamination-prone evidence is held rather than being treated immediately as either trusted support or permanent rejection.

\paragraph{Reversible learning.}
Admitted evidence remains connected to its provenance, allowing affected authorization states to be recomputed when a source is invalidated.

\paragraph{Lifecycle precedence.}
Trusted revocation, reauthorization, transfer, and expiration events are processed according to explicit temporal and semantic precedence.

\paragraph{Action-bounded evolution.}
Changes to one object-level action do not automatically modify other actions associated with the same object.

These properties complete the state-learning and defense model. The following section describes the prototype implementation and the runtime mechanisms used to realize the proposed procedures.

\section{Prototype Implementation}\label{sec:implementation}

\subsection{Prototype Architecture}\label{subsec:prototype-architecture}

A lightweight prototype was implemented to validate the proposed cross-evidence authorization state learning method. The prototype is written in Python and adopts a single-process reverse-proxy architecture. Its implementation objective is to realize the request-time risk assessment and response-time evidence-learning workflow described in Sections~\ref{sec:system-model}--\ref{sec:defense}, rather than to reproduce a complete production-grade Web application firewall.

The prototype contains five logical components: request parsing, request--response processing, cross-evidence learning, authorization state management, and runtime risk detection. These components are implemented as modules within one Python application and share an in-memory authorization-state store. FastAPI provides the HTTP interface, while HTTPX forwards requests to the upstream application. The implementation does not require Redis, a relational database, or independently deployed services.

Table~\ref{tab:prototype-components} summarizes the implementation mapping.

\begin{table}[t]
\centering
\caption{Main components of the Python prototype.}
\label{tab:prototype-components}
\footnotesize
\begin{tabularx}{\linewidth}{
    >{\raggedright\arraybackslash}p{3.0cm}
    >{\raggedright\arraybackslash}p{4.0cm}
    >{\raggedright\arraybackslash}X}
\toprule
Component & Implementation & Responsibility \\
\midrule
Request parser
&
FastAPI request handlers and configurable extraction rules
&
Extracts the trusted subject context, tenant, resource identifiers,
endpoint, and action class.
\\

Transaction processor
&
HTTPX asynchronous client and transaction identifiers
&
Forwards requests, receives upstream responses, and correlates each
response with its originating request.
\\

Cross-evidence learner
&
Python classes and dictionary-based provenance groups
&
Evaluates evidence quality, polarity, provenance dependence, temporal
weight, and positive-evidence promotion conditions.
\\

Authorization state manager
&
In-memory dictionaries and thread synchronization
&
Maintains relation records, lifecycle states, held evidence, revocation
records, and rollback dependencies.
\\

Runtime risk detector
&
State-dependent risk calculation
&
Aggregates object-level risks and produces Allow, Review, or Block
decisions.
\\

\bottomrule
\end{tabularx}
\end{table}

The prototype was executed with Python 3.13.5, FastAPI 0.128.2, and HTTPX 0.28.1. The authorization state is maintained in process memory using Python dictionaries. Evidence and state-transition events are exported as JSON Lines (JSONL), while request-level labels and prediction results are exported as CSV files for subsequent analysis. This arrangement keeps the experimental implementation reproducible without introducing an external storage service.

The proxy is positioned at an observation point where application-layer HTTP requests and responses are available. The target application remains unchanged. In the controlled implementation, an upstream authentication component supplies the subject context through the \texttt{X-Authenticated-User} header, and the tenant context is supplied through \texttt{X-Tenant}. These headers are treated as trusted context only when inserted by the observation environment. A real deployment must remove client-supplied copies and inject the values after authentication; otherwise, the subject binding assumption in Section~\ref{subsec:threat-model} would not hold.

The processing sequence follows the temporal separation defined in Section~\ref{sec:learning}. The proxy first extracts relation keys and evaluates the state available before the current request. If the request is forwarded, the upstream response is subsequently correlated with the request and processed by the evidence learner. Evidence obtained from the response is therefore unavailable to the decision that preceded that response.

\subsection{Request--Response Processing}\label{subsec:request-response-processing}

The request parser extracts the information required to construct the relation keys in Eq.~\eqref{eq:request-relation-key-set}. In the current prototype, document identifiers can be obtained from configured URI path patterns, query parameters, and JSON request-body fields. The configured field names include \texttt{object\_id}, \texttt{document\_id}, and \texttt{resource\_id}; batch identifiers can be supplied through \texttt{object\_ids} and \texttt{document\_ids}. Generic fields such as \texttt{id} are not interpreted as object identifiers without an application-specific mapping, since they may represent users, sessions, or unrelated entities.

The current adapter uses \textit{document} as its resource class. Additional resource classes can be introduced through route and field mappings without changing the authorization-state model. The HTTP method is mapped to an action class using the default mapping shown in Table~\ref{tab:action-mapping}. Application-specific operations, such as export or approval, require explicit endpoint mappings and are not automatically merged with read or update operations.

\begin{table}[t]
\centering
\caption{Default HTTP method to action mapping in the prototype.}
\label{tab:action-mapping}
\footnotesize
\begin{tabular}{ll}
\toprule
HTTP method & Action class \\
\midrule
GET & \textit{read} \\
POST & \textit{create} \\
PUT, PATCH & \textit{update} \\
DELETE & \textit{delete} \\
Other configured methods & \textit{other} \\
\bottomrule
\end{tabular}
\end{table}

For each incoming request, the proxy generates a unique transaction identifier. The identifier is retained during upstream forwarding and response processing, thereby providing an explicit correlation reference for the resulting transaction. When a request contains multiple object identifiers, the parser constructs one relation key for each identifier while retaining the common transaction context.

Before forwarding, the proxy retrieves the current state of each relation key and invokes the risk detector. A request classified as \textit{Block} is rejected locally with an HTTP 403 response. A request classified as \textit{Review} is forwarded in the initial monitoring configuration and is marked with the runtime decision and risk value in the response headers. An \textit{Allow} request is forwarded without additional intervention. This distinction permits the same learning component to be evaluated in observation and enforcement configurations.

After receiving the upstream response, the transaction processor extracts the available protocol-level information and applies configured response-semantic rules. A successful HTTP status code is not considered sufficient evidence of authorization. In particular, a successful direct object operation is represented as provisional positive evidence and is passed to the contamination gate as a \textit{direct operation} observation.

The current HTTP adapter implements two principal response extraction paths. First, successful responses to object-specific operations generate direct-operation evidence for the corresponding subject--resource--object--action keys. Second, a configured subject-scoped collection endpoint, represented by \texttt{/api/me/documents} in the controlled implementation, generates collection evidence for the returned object identifiers. The latter is associated with the subject context of the request and is assigned a separate provenance root for each returned object. Object-assignment, creation, explicit revocation, and reauthorization events use the same evidence interface and can be supplied by application-specific response mappings or controlled experimental event generators.

Responses generated locally by the proxy are excluded from evidence extraction. This prevents a locally generated blocking response from being interpreted as an observation of the target application's authorization behavior. Similarly, transactions for which the subject binding, object extraction, or request--response correlation is insufficient are retained for diagnostic purposes but do not promote authorization states.

\subsection{Cross-Evidence Learning Engine}\label{subsec:cross-evidence-engine}

The cross-evidence learning engine implements the evidence representation and aggregation procedures defined in Section~\ref{sec:authorization-learning}. Each evidence item is represented by a Python data object containing the tenant, subject, resource type, object identifier, action, source class, polarity, credibility value, observation timestamp, and provenance root. Additional flags identify direct operations, public-scope observations, explicit revocation, and explicit reauthorization.

Evidence quality is first calculated using the geometric aggregation in Eq.~\eqref{eq:base-credibility}. The prototype uses equal exponents of 0.25 for the four credibility factors; correlation and extraction confidence are supplied by the transaction correlator and parser, while source specificity and action compatibility are assigned by the configured evidence and action mappings. The minimum admissibility threshold is set to 0.60. Evidence below this threshold is rejected from support aggregation but remains identifiable in the experimental audit trail. For admitted evidence, the learner calculates positive and negative support separately and groups observations by their provenance root.

Within each provenance group, the maximum effective positive weight and maximum effective negative weight are retained. Repeated observations from the same derivation root therefore do not increase the independent-group count. The support values are recalculated using the multiplicative-complement aggregation in Eqs.~\eqref{eq:positive-support} and~\eqref{eq:negative-support}. Temporal decay is applied when group weights are evaluated, using the evidence timestamp and the configured decay coefficient.

Positive evidence is passed through the contamination gate before it can modify the active support profile. The gate evaluates self-validation, source scope, provenance dependence, relation novelty, alignment confidence, and conflict with an active revocation record. The indicator weights used by the prototype are 0.35, 0.25, 0.15, 0.10, 0.10, and 0.05, respectively. Their aggregate score is calculated according to Eq.~\eqref{eq:contamination-score}, and the immediate-promotion threshold is set to 0.50.

Two conditions are enforced independently of the aggregate score. A successful direct object operation cannot immediately promote the authorization relation that was evaluated by that operation, and public or aggregate observations cannot independently establish subject-specific authorization. Such evidence is retained in a quarantine collection with its original provenance and contamination indicators.

Held positive evidence can be released only after the corresponding relation has obtained sufficient corroboration from at least two admitted, provenance-independent positive groups, the positive-support threshold has been reached, the relation remains in the \textit{Confirmed} state at the release timestamp, and no active revocation, expiration, or material negative conflict remains. Public-scope evidence is not released as subject-specific evidence. When held evidence is released, its original observation timestamp is preserved so that quarantine does not reset its temporal age.

Negative evidence is evaluated separately. An object-specific denial can increase negative support, while a trusted explicit revocation event can invalidate the affected relation immediately when its credibility is at least 0.90 and its object-level scope is sufficiently specific. This separation prevents the positive-evidence promotion gate from suppressing potentially important negative authorization evidence.

The learner also implements provenance-based rollback. Each admitted group is indexed by its provenance root, and a reverse lookup identifies the relation records that depend on a given root. When a root is invalidated, the corresponding group is deactivated and all affected positive and negative support values are recalculated. Historical audit events are retained, allowing the resulting state correction to be traced to the evidence that was removed.

\subsection{Authorization State Management}\label{subsec:authorization-state-management}

The authorization state manager stores each relation under the composite key
\begin{equation}
K=\langle\eta,u,r,o,a\rangle,
\end{equation}
consistent with Eq.~\eqref{eq:relation-key}. A Python dictionary maps each key to a relation record. Each record contains its lifecycle state, provenance-group dictionary, held-evidence collection, revocation timestamp, latest positive-evidence timestamp, and monotonically increasing version number. A separate audit list records evidence admission, rejection, quarantine, release, revocation, and rollback events.

Each provenance-group record stores the representative positive and negative weights, their most recent evaluation timestamps, and an activity flag. The support profile is calculated on demand rather than by periodically scanning the entire state store. This lazy evaluation avoids a background expiration task and allows the current confidence values to be obtained when a relation is queried or updated.

The state-transition implementation follows the precedence defined in Table~\ref{tab:transition-precedence}. The prototype uses a positive-support threshold of 0.75 and requires at least two independent positive groups, each with an effective weight of at least 0.60, before a relation can enter the \textit{Confirmed} state. A relation becomes \textit{Conflicted} when positive support is at least 0.35 and negative support reaches 0.55. Evidence-based revocation requires negative support of at least 0.80 from at least two independent negative groups. An explicit trusted revocation event takes precedence over ordinary evidence accumulation.

The relation lifetime is configured as 100,000 timestamp units in the controlled implementation, and the default temporal decay coefficient is \(10^{-6}\). These values are implementation defaults for the reported controlled experiment, not universal parameters for all Web applications. Their operational values should be calibrated against the transaction rate, expected authorization lifetime, and evidence-source characteristics of the target application.

When a trusted revocation is accepted, the previously accumulated positive support is superseded while the historical evidence events remain in the audit record. Ordinary successful operations cannot restore the revoked relation. A subsequent trusted reauthorization event closes the active revocation record and starts a new candidate relation; the old positive support is not automatically reinstated. Further independent evidence is required for confirmation.

State mutations are protected by a process-level reentrant lock. This prevents concurrent request handlers from interleaving modifications to the same in-memory structures. The version number is updated when lifecycle transitions, revocation, or rollback operations change the relation record. This synchronization is sufficient for the single-process experimental prototype; distributed consistency and multi-instance state replication are outside its scope.

The state store is intentionally volatile. Restarting the prototype clears the active in-memory authorization state. JSONL files preserve the experimental evidence and state-transition history, but the current prototype does not implement automatic restoration of a live state store from those files. Each controlled experiment therefore initializes its state explicitly and records the initialization events to ensure repeatability.

\subsection{Runtime Risk Detection}\label{subsec:runtime-risk-detection}

The runtime risk detector uses the lifecycle state and current positive and negative support values to calculate the risk associated with each relation key. The implementation follows the state-dependent risk model in Eq.~\eqref{eq:relation-risk}. The default state priors are 0.55 for \textit{Unknown}, 0.50 for \textit{Candidate}, 0.08 for \textit{Confirmed}, 0.82 for \textit{Conflicted}, 0.98 for \textit{Revoked}, and 0.65 for \textit{Expired}. Positive and negative support adjust the risk value according to the state-specific calculation implemented by the detector.

The positive- and negative-support coefficients are set to \(\mu_{+}=0.05\) and \(\mu_{-}=0.20\), respectively. The review and blocking thresholds are set to 0.40 and 0.75. A relation with insufficient evidence therefore remains subject to review rather than being automatically treated as a confirmed authorization violation. A conflicted or revoked relation receives a high risk value, whereas a confirmed relation receives a low risk value unless sufficiently strong negative evidence changes its state.

For requests containing multiple object references, the detector aggregates the individual relation risks using the maximum operator in Eq.~\eqref{eq:request-risk}. This prevents a low-risk object reference from masking another object reference with a higher authorization risk. Requests without an extractable object reference are not assigned an artificial object-level relation and are passed through the object-level detector without an object-specific decision.

The detector produces three outcomes according to Eq.~\eqref{eq:request-decision}. \textit{Allow} requests are forwarded, \textit{Review} requests are forwarded with an elevated audit indication in the monitoring configuration, and \textit{Block} requests are rejected before reaching the upstream application. The decision is computed exclusively from the state available before the current request is processed. The response generated by the target application is used only for subsequent evidence learning.

For reproducibility, the experimental runner records the generated transaction identifiers, ground-truth labels, predictions from each baseline, and the proposed method's evidence and state-transition events. Request-level data and predictions are exported to CSV, while the proposed method's audit events are exported to JSONL. These outputs provide the implementation-level records required for the experimental comparisons in the following sections.

\section{Experimental Methodology}\label{sec:experimental-methodology}

This section specifies the controlled evaluation protocol used to examine the proposed cross-evidence authorization-state learning method. The evaluation is designed around the streaming request--evidence sequence implemented in the preceding prototype-implementation section. It focuses on object-level authorization decisions, the reliability of evidence used to establish authorization relations, and the effect of provenance dependence, evidence quarantine, temporal validity, and authorization lifecycle events. The current preprint uses a deterministic synthetic benchmark generated by the supplied Python code. It does not treat synthetic observations as production traffic or as a substitute for independent application-level validation.

\subsection{Evaluation Objectives and Research Questions}\label{subsec:exp-objectives}

The experiments address the following research questions:

\begin{itemize}
    \item \textbf{RQ1:} How accurately can the proposed method distinguish legitimate object-level requests from unauthorized BOLA requests when authorization evidence varies in source count, provenance independence, quality, and age?
    \item \textbf{RQ2:} How does provenance-aware cross-evidence learning compare with URL-rule, frequency-based, binary allow-set, and single-channel success-based baselines?
    \item \textbf{RQ3:} To what extent do provenance-independence checks and quarantine prevent successful but unauthorized object operations from contaminating learned authorization state?
    \item \textbf{RQ4:} How does the method behave under delegated access, cold-start conditions, stale or low-quality evidence, public-scope observations, explicit revocation, reauthorization, action escalation, and cross-tenant object references?
    \item \textbf{RQ5:} Does the configured object-identifier extractor correctly handle supported URI, query, and JSON formats while rejecting ambiguous or malformed inputs?
\end{itemize}

The first four questions concern the authorization-state learner and its runtime decisions. RQ5 isolates the input-extraction component because an incorrect object identifier can produce an invalid relation key before the learning mechanism is invoked.

\subsection{Benchmark Environment and Synthetic Data Generation}\label{subsec:exp-benchmark}

The benchmark is generated by \texttt{experiment\_v2.py} using Python's deterministic pseudo-random generator. Five independent seeds, \texttt{20261030}--\texttt{20261034}, are used. Each seed creates a separate authorization graph and event stream. The benchmark contains three tenants, 100 subjects per tenant, and five documents per subject. Local document identifiers are intentionally reused across tenant namespaces so that the tenant component of the relation key is necessary for correct isolation. Each authorization relation is evaluated through four chronologically ordered requests.

Table~\ref{tab:exp-benchmark-scale} reports the per-seed scale and the aggregate size across all five runs. The totals are read directly from the generated run summary rather than entered manually.

\begin{table}[t]
\centering
\caption{Scale of the deterministic synthetic benchmark.}
\label{tab:exp-benchmark-scale}
\footnotesize
\begin{tabularx}{\linewidth}{>{\raggedright\arraybackslash}Xrr}
\toprule
Quantity & Per-run setting / mean & Across five runs \\
\midrule
Tenants & 3 & 3 in each run \\
Users per tenant & 100 & 100 per tenant \\
Documents per user & 5 & 5 per user \\
Requests per relation & 4 & 4 per relation \\
Authorization relations & 3,220 & 16,100 \\
Initially authorized relations & 2,700 & 13,500 \\
Static unauthorized relations & 520 & 2,600 \\
Evaluation requests & 12,880 & 64,400 \\
Legitimate requests & 10,560 & 52,800 \\
Unauthorized requests & 2,320 & 11,600 \\
Evidence events & 5,371.6 & 26,858 \\
\bottomrule
\end{tabularx}
\end{table}

For each tenant, the simulator creates owner read relations for all subject--document pairs and owner update relations for the configured subset of documents. It also creates delegated read relations for a deterministic subset of documents. Delegated update or delete permission is not implied by a delegated read grant. Unauthorized relations are generated separately and include horizontal object access, action escalation, cross-tenant object references, and enumeration behavior. The simulator's authorization graph is retained as ground truth and is not passed to the runtime detector.

The simulator assigns HTTP 200 to every generated object request. This setting represents an intentionally vulnerable object endpoint that does not enforce the relevant object-level authorization check. It prevents response status alone from separating legitimate and unauthorized requests and allows the experiment to test whether the learner incorrectly treats a successful operation as proof of authorization.

\subsection{Evidence Profiles and Scenario Construction}\label{subsec:exp-scenarios}

Owner relations are assigned one of eight evidence profiles. The profile weights are fixed in the generator and held constant across seeds; the actual assignment is randomized by each seed. Table~\ref{tab:exp-owner-profiles} lists the configured proportions.

\begin{table}[t]
\centering
\caption{Configured evidence profiles for owner relations.}
\label{tab:exp-owner-profiles}
\footnotesize
\begin{tabular}{lr}
\toprule
Evidence profile & Sampling weight \\
\midrule
Two independent sources & 50\% \\
Single source & 10\% \\
Correlated sources & 10\% \\
Cold start & 10\% \\
Noisy, then confirmed & 7\% \\
Stale evidence & 5\% \\
Public scope & 3\% \\
Mid-stream revocation & 5\% \\
\bottomrule
\end{tabular}
\end{table}

The profiles distinguish independent positive sources from correlated observations, single-source evidence, delayed evidence, noisy evidence, stale evidence, public-scope observations, and relations revoked during the event stream. The attack scenarios are generated independently of the owner profiles. Table~\ref{tab:exp-scenario-matrix} lists the resulting scenario categories and their mean request composition per seed. Legitimate and unauthorized counts are derived from the scenario-level evaluation output; the positive class denotes an unauthorized request.

\begin{table*}[t]
\centering
\caption{Scenario matrix and per-seed request composition.}
\label{tab:exp-scenario-matrix}
\scriptsize
\begin{tabularx}{\textwidth}{>{\raggedright\arraybackslash}p{3.0cm}rrrX}
\toprule
Scenario & Requests & Legitimate & Unauthorized & Experimental condition \\
\midrule
Cold start & 951.2 & 951.2 & 0 & The first requests arrive before the supporting evidence is observed. \\
Correlated sources & 971.2 & 971.2 & 0 & Two observations share one provenance root and are not independent. \\
Cross-tenant object reference & 240 & 0 & 240 & A request uses a locally repeated object identifier under another tenant. \\
Delegated two-source access & 1,200 & 1,200 & 0 & Delegated read access is supported by a grant and a scoped collection result. \\
High-volume enumeration & 640 & 0 & 640 & An attacker touches multiple distinct objects at elevated volume. \\
Horizontal BOLA & 480 & 0 & 480 & A subject requests another subject's object within the same tenant. \\
Low-frequency enumeration & 240 & 0 & 240 & An attacker touches a small number of distinct objects. \\
Noisy, then confirmed & 665.6 & 665.6 & 0 & Low-quality evidence is followed by fresh independent corroboration. \\
Post-revocation access & 240 & 0 & 240 & An access request occurs after the trusted revocation event. \\
Public scope & 301.6 & 301.6 & 0 & A public-scope observation must not establish subject-specific access. \\
Mid-stream revocation & 240 & 240 & 0 & An initially authorized relation is explicitly revoked during the stream. \\
Single source & 957.6 & 957.6 & 0 & Only one positive provenance group is available. \\
Stale evidence & 515.2 & 515.2 & 0 & Previously valid evidence is older than the configured relation lifetime. \\
Two independent sources & 4,757.6 & 4,757.6 & 0 & Owner relation supported by two provenance-independent sources. \\
Vertical action escalation & 480 & 0 & 480 & A subject attempts an action not granted for the object. \\
\bottomrule
\end{tabularx}
\end{table*}

Horizontal BOLA is represented by a subject requesting another subject's document within the same tenant. Vertical or action-level escalation is represented by a subject attempting an action, such as deletion, that is not present in the relation's authorization key. Cross-tenant cases deliberately reuse local object identifiers in different tenant namespaces. Low-frequency enumeration uses attackers that touch only a small number of distinct objects, whereas high-volume enumeration uses attackers that touch more objects. These cases separate authorization-state violations from simple request-volume anomalies.

\subsection{Ground Truth, Event Ordering, and Leakage Prevention}\label{subsec:exp-ground-truth}

Ground-truth authorization relations are written to a separate \texttt{ground\_truth\_relations.csv} file for each seed. The ground truth is derived from the simulator's explicit authorization graph, including tenant, subject, resource, object identifier, action, and whether the relation is authorized at the beginning of the stream. A relation that is initially authorized but later revoked is labeled as unauthorized only for requests occurring after the revocation event. Static unauthorized relations are tracked separately for the evidence-pollution measurement.

The runtime detector receives a \texttt{RequestObservation} object containing only the request-time fields required for a decision. The object deliberately excludes the ground-truth label, scenario name, and attack-family field. These fields are retained in the evaluation records and joined with predictions only after the detector has made its decision. Evidence-stream records likewise contain no ground-truth or scenario columns. Request identifiers and subject identifiers use opaque formats that do not encode the label or attack category. The validator also checks that no external positive evidence is generated for a statically unauthorized relation.

Events are sorted by timestamp. Before each request decision, the detector ingests only evidence whose timestamp is less than or equal to the request timestamp. The detector then evaluates the request using the current authorization state. The simulated response is observed only after that decision, so evidence derived from the current successful operation cannot retroactively justify the same request. This ordering models the online operation of the proxy and avoids future-evidence leakage.

No conventional random train--test split is used. The method is an online state learner rather than a batch classifier trained on a labeled feature matrix. Each seed creates a fresh graph, evidence stream, and detector state; the five seeds are independent repeated controlled runs used to estimate variability under the simulator's random profile assignments. Ground-truth labels are used only for post-decision scoring and validation.

\subsection{Compared Methods and Ablation Settings}\label{subsec:exp-baselines}

Seven methods are evaluated under the same request ordering and the same per-seed data. The first four provide rule-based, frequency-based, and simplified state-learning baselines. The next two are ablations that remove one of the proposed method's central safeguards. The final method is the complete cross-evidence implementation. Table~\ref{tab:exp-methods} summarizes their operational differences.

\begin{table*}[t]
\centering
\caption{Compared detection methods and ablation settings.}
\label{tab:exp-methods}
\footnotesize
\begin{tabularx}{\textwidth}{>{\raggedright\arraybackslash}p{4.1cm}X}
\toprule
Method & Evaluation configuration \\
\midrule
URL path rule & Marks requests matching the configured object route without learning authorization state. \\
Distinct-object frequency & Uses the number of distinct objects observed for each tenant--subject pair; threshold set to five. \\
Binary AllowSet & Maintains a binary allow set and accepts successful object responses as positive authorization evidence. \\
Single-channel success learner & Promotes a relation after repeated successful object responses, without cross-source corroboration. \\
Cross-evidence without provenance independence & Ablation that counts positive evidence events without grouping correlated provenance roots. \\
Cross-evidence without quarantine & Ablation that disables quarantine for direct-operation evidence. \\
Proposed cross-evidence method & Uses provenance-aware support aggregation, evidence quarantine, temporal decay, lifecycle transitions, and revocation handling. \\
\bottomrule
\end{tabularx}
\end{table*}

All methods process the same chronologically ordered requests within each seed. They are reset between seeds. For methods that learn from response outcomes, the response observation is supplied only after the current decision, consistent with the event ordering described above. The frequency detector uses the fixed distinct-object threshold of five. The ablation methods preserve the remaining implementation settings and disable only the named mechanism, allowing the effect of provenance independence and quarantine to be examined separately.

\subsection{Evaluation Metrics}\label{subsec:exp-metrics}

The positive class is an unauthorized BOLA request; legitimate requests form the negative class. A \textit{Review} or \textit{Block} decision is counted as a risk alert for binary classification, while an \textit{Allow} decision is counted as a non-alert. This convention measures whether a request is surfaced for intervention, but it does not equate a review decision with a confirmed attack. Operational decision rates are therefore reported separately.

Precision, recall, F1-score, false-positive rate, false-negative rate, and accuracy are calculated as
\begin{align}
\mathrm{Precision} &= \frac{TP}{TP+FP}, \\
\mathrm{Recall} &= \frac{TP}{TP+FN}, \\
F_1 &= \frac{2\,\mathrm{Precision}\,\mathrm{Recall}}
{\mathrm{Precision}+\mathrm{Recall}}, \\
\mathrm{FPR} &= \frac{FP}{FP+TN}, \\
\mathrm{FNR} &= \frac{FN}{FN+TP}, \\
\mathrm{Accuracy} &= \frac{TP+TN}{TP+FP+TN+FN}.
\end{align}
The Matthews correlation coefficient (MCC) is also reported to account for all four confusion-matrix cells:
\begin{equation}
\mathrm{MCC}=\frac{TP\cdot TN-FP\cdot FN}
{\sqrt{(TP+FP)(TP+FN)(TN+FP)(TN+FN)}}.
\end{equation}

To distinguish analyst workload from enforcement behavior, the evaluation reports the proportion of requests assigned \textit{Review}, \textit{Block}, and \textit{Allow}. Scenario-level results are calculated separately for each scenario and seed. For scenarios containing only legitimate requests, FPR is the relevant error measure; for attack-only scenarios, recall is the relevant detection measure. Both are reported for mixed scenarios.

Evidence contamination is evaluated using the unauthorized-relation pollution rate:
\begin{equation}
\mathrm{PollutionRate}=\frac{N_{\mathrm{promoted,unauthorized}}}
{N_{\mathrm{static,unauthorized}}}.
\end{equation}
Here, the numerator is the number of statically unauthorized relation keys that reach the \textit{Confirmed} state, and the denominator is the number of relations marked unauthorized at the beginning of the run. Relations that become unauthorized only after an explicit mid-stream revocation are excluded from this denominator and evaluated through the post-revocation request scenario.

\subsection{Fixed Parameters and Evidence-Quality Configuration}\label{subsec:exp-parameters}

All state thresholds are held constant across the five seeds. The values below are read from the prototype constructor and the benchmark configuration. They are not optimized against the evaluation labels on a per-seed basis. The frequency threshold is included for the frequency baseline; the remaining values govern evidence admission, state transitions, temporal validity, quarantine, and runtime decisions.

\begin{table}[t]
\centering
\caption{Fixed benchmark and authorization-state parameters.}
\label{tab:exp-parameters}
\footnotesize
\begin{tabularx}{\linewidth}{>{\raggedright\arraybackslash}p{3.4cm}rX}
\toprule
Parameter & Value & Role \\
\midrule
Frequency threshold & 5 & Distinct-object frequency baseline threshold. \\
Minimum evidence quality & 0.6 & Minimum admissible evidence quality. \\
Confirmation threshold & 0.75 & Minimum positive support for confirmation. \\
Negative conflict threshold & 0.55 & Negative support threshold for a conflict state. \\
Revocation threshold & 0.8 & Negative support threshold for evidence-based revocation. \\
Provenance-group threshold & 0.6 & Minimum effective weight for an independent provenance group. \\
Minimum confirmation groups & 2 & Minimum number of independent positive groups. \\
Quarantine threshold & 0.5 & Contamination score above which positive evidence is held. \\
Review threshold & 0.4 & Risk threshold for a Review decision. \\
Block threshold & 0.75 & Risk threshold for a Block decision. \\
Temporal decay rate & 1e-06 & Exponential temporal-decay coefficient. \\
Relation lifetime & 100000 & Configured relation lifetime in simulator timestamp units. \\
\bottomrule
\end{tabularx}
\end{table}

Evidence credibility is computed as a weighted geometric aggregation of correlation confidence, extraction confidence, source specificity, and action compatibility. Equal exponents of 0.25 are used for the four factors in the controlled generator. The source-specific factors are assigned by the simulator according to the evidence type; low-quality evidence is rejected below the minimum admissibility threshold. These values define the benchmark's evidence semantics and should not be interpreted as universally calibrated probabilities for arbitrary applications.

\subsection{Parser and Mechanism Validation}\label{subsec:exp-validation}

The configured object-identifier extractor is evaluated separately from the authorization classifier. The parser benchmark contains 3,300 test cases, including positive cases for URI paths, query parameters, JSON scalar fields, and JSON batch fields, together with negative cases for generic identifiers, malformed JSON, missing object references, and duplicate identifiers. Table~\ref{tab:exp-parser-cases} reports the generated test composition.

\begin{table}[t]
\centering
\caption{Object-identifier parser test cases.}
\label{tab:exp-parser-cases}
\scriptsize
\begin{tabularx}{\linewidth}{>{\raggedright\arraybackslash}p{3.1cm}rrX}
\toprule
Case type & Samples & Expected identifiers & Test purpose \\
\midrule
Duplicate-ID deduplication & 200 & 200 & Repeated identifiers from path and body are deduplicated. \\
Generic ID rejection & 200 & 0 & A generic user or record id must not be misclassified as an object id. \\
JSON batch field & 500 & 1,000 & Multiple identifiers in a configured JSON array field. \\
JSON scalar field & 500 & 500 & Configured scalar identifier in a top-level JSON field. \\
Malformed JSON rejection & 200 & 0 & Malformed JSON must not create an extracted identifier. \\
No object reference & 200 & 0 & A request without an object reference must return an empty extraction. \\
URI path & 1,000 & 1,000 & Object identifier in a configured URI path. \\
Query parameter & 500 & 500 & Object identifier in a configured query parameter. \\
\bottomrule
\end{tabularx}
\end{table}

The mechanism regression suite contains 10 deterministic tests. Table~\ref{tab:exp-mechanism-tests} lists the behaviors covered by the suite. The dataset validator additionally checks request-ID uniqueness, opaque identifier formats, ground-truth isolation from runtime observations, the absence of unauthorized positive evidence, repeated local object identifiers across tenants, HTTP response consistency, chronological request order, and evidence-stream field isolation. The generated validation summary is reported in Table~\ref{tab:exp-integrity-summary}.

\begin{table*}[t]
\centering
\caption{Mechanism-level regression tests.}
\label{tab:exp-mechanism-tests}
\footnotesize
\begin{tabularx}{\textwidth}{>{\raggedright\arraybackslash}p{4.0cm}X}
\toprule
Test & Expected behavior \\
\midrule
Provenance independence & Repeated observations from one root cannot satisfy the independent-source requirement. \\
Direct-operation quarantine & A successful direct operation cannot validate its own authorization relation. \\
Quarantine release after corroboration & Held evidence is released only after fresh independent corroboration. \\
Stale evidence is not refreshed & A new held operation does not refresh expired positive support. \\
Explicit revocation & A trusted revocation takes precedence and produces a Block decision. \\
Reauthorization requires confirmation & Reauthorization does not restore superseded positive support. \\
Tenant and action isolation & Evidence for one tenant/action does not confirm another tenant/action key. \\
Public-scope evidence remains held & Public-scope evidence remains held from subject-specific promotion. \\
Provenance rollback & Invalidating provenance roots reverses dependent support and state transitions. \\
Low-quality evidence rejection & Evidence below the admissibility threshold is rejected. \\
\bottomrule
\end{tabularx}
\end{table*}

\begin{table}[t]
\centering
\caption{Dataset-integrity checks across independent seeds.}
\label{tab:exp-integrity-summary}
\footnotesize
\begin{tabularx}{\linewidth}{Xr}
\toprule
Integrity check & Passed runs \\
\midrule
All target responses use HTTP 200 & 5/5 \\
Chronological request order & 5/5 \\
Evidence stream excludes ground truth & 5/5 \\
Runtime observation excludes labels & 5/5 \\
No positive evidence for unauthorized relations & 5/5 \\
Opaque principal identifiers & 5/5 \\
Opaque request identifiers & 5/5 \\
Repeated local object IDs across tenants & 5/5 \\
Unique request identifiers & 5/5 \\
All checks & 45/45 \\
\bottomrule
\end{tabularx}
\end{table}

\subsection{Reproducibility and Scope}\label{subsec:exp-reproducibility}

The complete benchmark is executed from the experiment package with the command \texttt{python run\_all\_v2.py}. The pipeline generates the five request and evidence streams, ground-truth relation files, predictions for all compared methods, per-seed and aggregate metrics, pollution measurements, parser results, mechanism-test records, data-integrity checks, figures, LaTeX result tables, and this section's LaTeX source. The main outputs are stored under \texttt{results\_v2/}; each seed has a separate subdirectory containing \texttt{requests.csv}, \texttt{evidence\_stream.csv}, \texttt{ground\_truth\_relations.csv}, \texttt{predictions.csv}, scenario metrics, and the proposed method's JSONL audit trace.

The recorded software environment is Python 3.13.5, FastAPI 0.128.2, and HTTPX 0.28.1. The fixed seeds are 20261030, 20261031, 20261032, 20261033, 20261034. Re-running the pipeline with the same source code and dependency versions regenerates the deterministic data and classification outputs. Platform-specific execution time is not treated as a cross-machine performance claim in this chapter.

The benchmark is synthetic and controlled. Its ground truth is generated from the simulator's authorization graph rather than independently annotated production traffic. Scenario proportions and evidence-source semantics are simulator assumptions, and the benchmark does not establish generalization to arbitrary Web applications. The current preprint does not include OWASP crAPI, VAmPI, or other independently deployed third-party applications; those experiments are deferred to a subsequent revision and must use independently verified test-account permission matrices. The results should therefore be interpreted as controlled mechanism and baseline evaluations, not as a multi-application field study.

\section{Results and Analysis}\label{sec:results}

\subsection{Experimental Overview}\label{subsec:results-overview}

The proposed cross-evidence authorization-state learning method was evaluated using the deterministic synthetic benchmark described in Section~\ref{sec:experimental-methodology}. Five independent runs were conducted with random seeds ranging from 20261030 to 20261034. All compared methods processed the same request sequences under identical benchmark conditions.

Across the five runs, the benchmark contained 64,400 object-level requests, including 52,800 legitimate requests and 11,600 unauthorized requests. A total of 26,858 evidence events and 16,100 authorization relation records were generated. Each run used an independently initialized authorization graph and evidence stream.

The evaluation covered seven methods: URL path rule, distinct-object frequency, binary AllowSet, single-channel success learner, cross-evidence without provenance independence, cross-evidence without quarantine, and the proposed cross-evidence method.

All reported aggregate metrics are calculated as the mean and sample standard deviation across the five runs. The classification results follow the convention defined in Section~\ref{subsec:exp-metrics}: both \textit{Review} and \textit{Block} are treated as risk alerts, whereas \textit{Allow} is treated as a non-alert.

\subsection{Overall Detection Performance}\label{subsec:overall-performance}

Table~\ref{tab:overall-performance} presents the overall detection performance of the seven methods.

\begin{table*}[t]
\centering
\caption{Overall BOLA detection performance across five independent synthetic runs. Values are mean $\pm$ standard deviation (\%). MCC is reported on its original scale.}
\label{tab:overall-performance}
\scriptsize
\setlength{\tabcolsep}{2.5pt}
\renewcommand{\arraystretch}{1.12}
\begin{tabularx}{\textwidth}{
    >{\raggedright\arraybackslash}p{3.35cm}
    *{7}{>{\centering\arraybackslash}X}}
\toprule
Method
& Precision
& Recall
& F1
& FPR
& FNR
& Accuracy
& MCC \\
\midrule
URL path rule
& 18.01 $\pm$ 0.17
& 100.00 $\pm$ 0.00
& 30.53 $\pm$ 0.25
& 100.00 $\pm$ 0.00
& 0.00 $\pm$ 0.00
& 18.01 $\pm$ 0.17
& 0.000 $\pm$ 0.000 \\

Distinct-object frequency
& 43.75 $\pm$ 0.00
& 30.17 $\pm$ 0.29
& 35.72 $\pm$ 0.20
& 8.52 $\pm$ 0.02
& 69.83 $\pm$ 0.29
& 80.43 $\pm$ 0.17
& 0.252 $\pm$ 0.003 \\

Binary AllowSet
& 65.28 $\pm$ 1.07
& 32.75 $\pm$ 0.65
& 43.62 $\pm$ 0.77
& 3.83 $\pm$ 0.12
& 67.25 $\pm$ 0.65
& 84.75 $\pm$ 0.10
& 0.388 $\pm$ 0.008 \\

Single-channel success learner
& 19.16 $\pm$ 0.27
& 55.17 $\pm$ 0.43
& 28.44 $\pm$ 0.36
& 51.14 $\pm$ 0.11
& 44.83 $\pm$ 0.43
& 50.00 $\pm$ 0.00
& 0.031 $\pm$ 0.003 \\

Cross-evidence without provenance independence
& 53.27 $\pm$ 1.15
& 55.17 $\pm$ 0.43
& 54.20 $\pm$ 0.75
& 10.64 $\pm$ 0.39
& 44.83 $\pm$ 0.43
& 83.20 $\pm$ 0.33
& 0.439 $\pm$ 0.009 \\

Cross-evidence without quarantine
& 44.91 $\pm$ 0.91
& 55.17 $\pm$ 0.43
& 49.51 $\pm$ 0.70
& 14.87 $\pm$ 0.34
& 44.83 $\pm$ 0.43
& 79.73 $\pm$ 0.29
& 0.373 $\pm$ 0.008 \\

Proposed cross-evidence method
& 39.50 $\pm$ 0.70
& 100.00 $\pm$ 0.00
& 56.63 $\pm$ 0.72
& 33.65 $\pm$ 0.69
& 0.00 $\pm$ 0.00
& 72.41 $\pm$ 0.60
& 0.512 $\pm$ 0.007 \\
\bottomrule
\end{tabularx}
\end{table*}

The proposed method achieved a recall of 100.00\% and an F1-score of 56.63\%, representing the highest F1-score among the seven evaluated methods. Its MCC reached 0.512, also exceeding the values obtained by the baseline and ablation methods.

Compared with the cross-evidence method without provenance independence, the proposed method improved recall from 55.17\% to 100.00\% and F1-score from 54.20\% to 56.63\%. MCC increased from 0.439 to 0.512. These results indicate that explicitly distinguishing independent evidence sources improves the ability to identify unauthorized object-level requests under the evaluated conditions.

The comparison with the method without quarantine shows a similar pattern. The complete method achieved an F1-score of 56.63\%, compared with 49.51\% for the corresponding ablation. Recall increased by 44.83 percentage points.

However, the proposed method also exhibited a relatively high false-positive rate of 33.65\%. In contrast, Binary AllowSet achieved a lower false-positive rate of 3.83\%, while its recall was only 32.75\%. The distinct-object frequency baseline also produced a relatively low false-positive rate of 8.52\%, but detected only 30.17\% of unauthorized requests.

This comparison demonstrates that the proposed method prioritizes identifying potentially unauthorized requests over minimizing all risk alerts. Its performance advantage is therefore not uniform across every metric. In particular, the improvement in recall is accompanied by a greater number of legitimate requests requiring further examination.

\begin{figure}[t]
\centering
\includegraphics[width=0.96\textwidth]{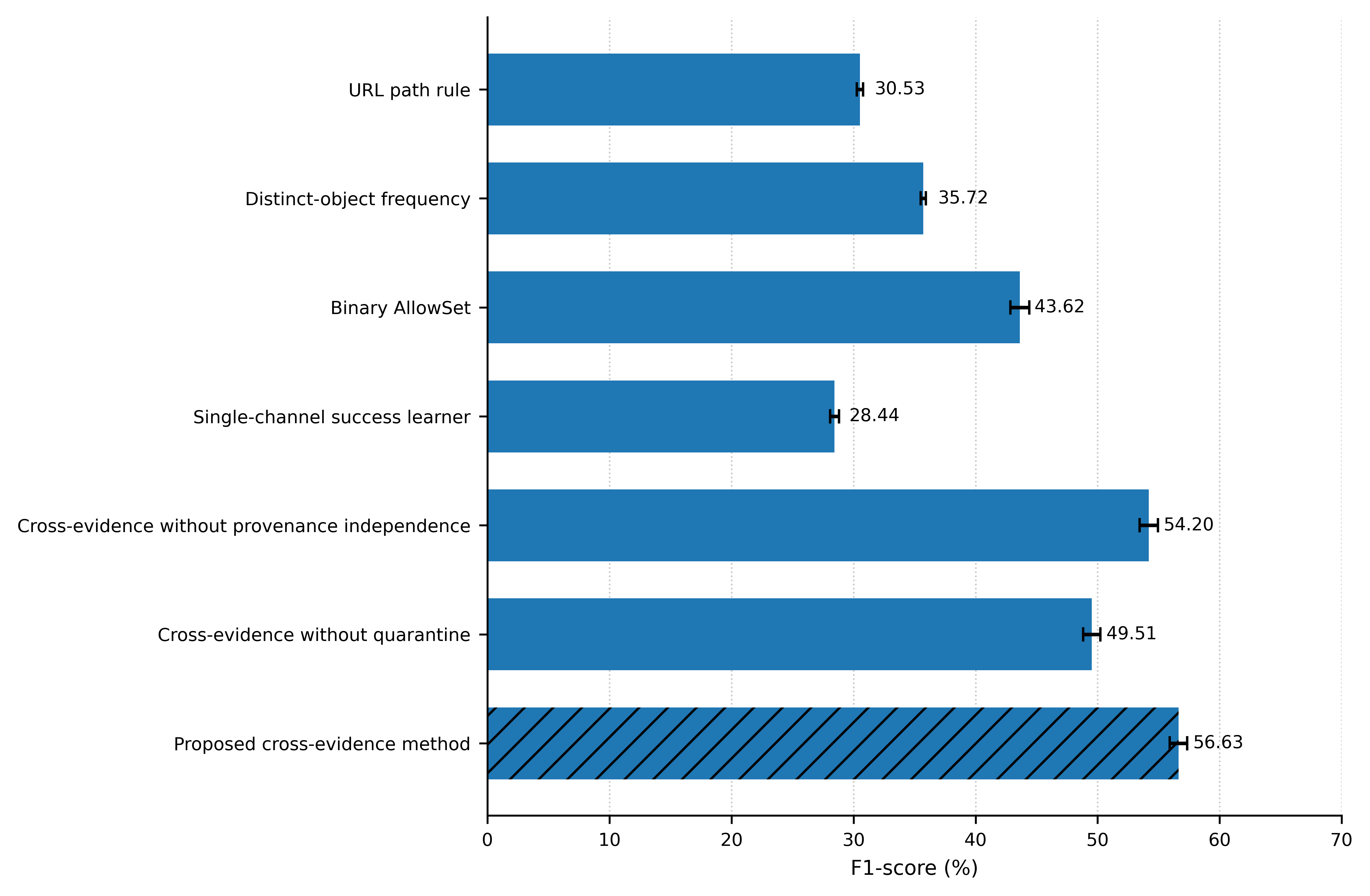}
\caption{F1-score comparison across seven detection methods. Bars show the mean across five independent seeds; error bars denote the sample standard deviation.}
\label{fig:f1-comparison}
\end{figure}

\begin{figure}[t]
\centering
\includegraphics[width=0.92\textwidth]{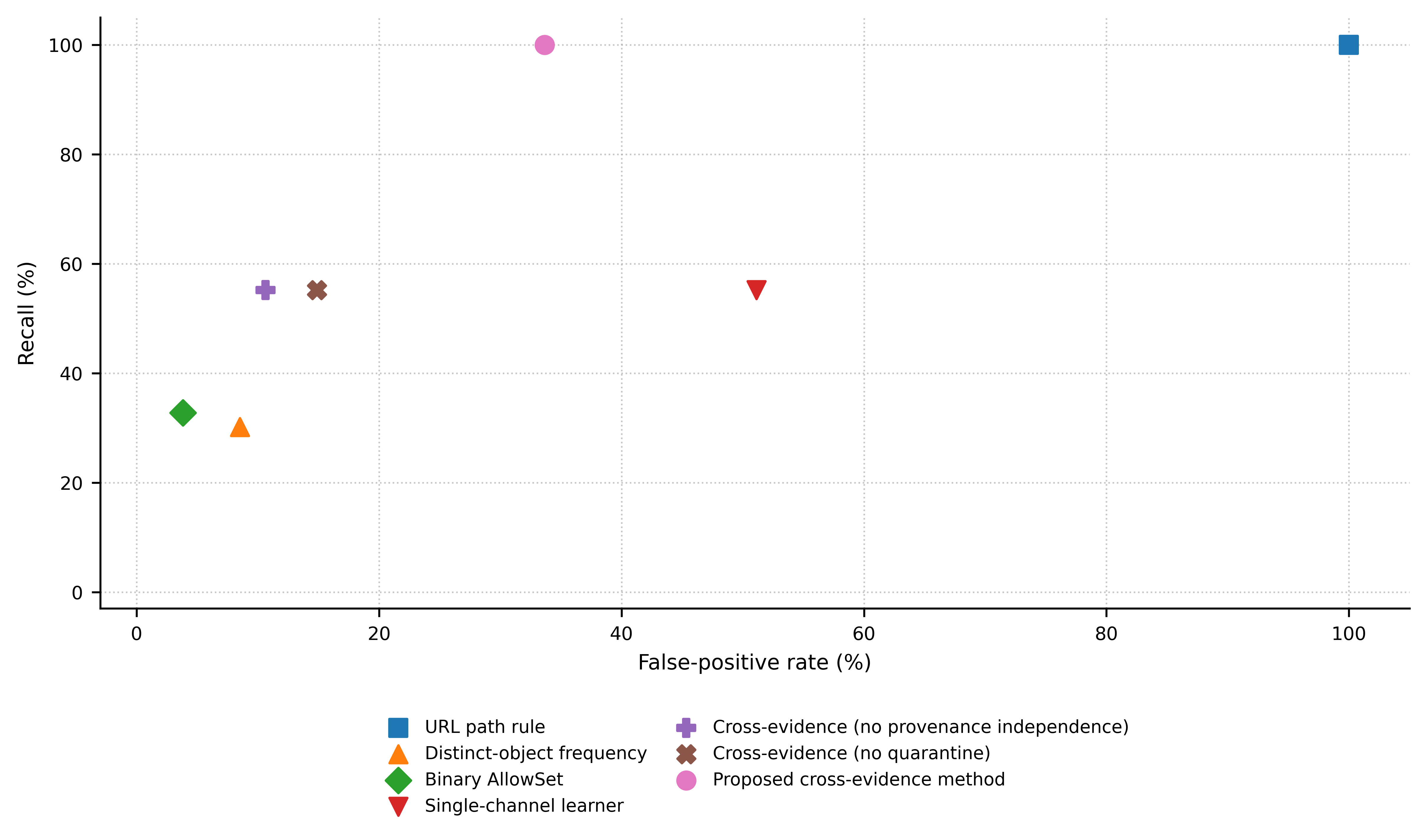}
\caption{Recall--FPR operating points for the seven evaluated methods. Each point represents the mean metric values across five independent seeds.}
\label{fig:recall-fpr-tradeoff}
\end{figure}

Figure~\ref{fig:f1-comparison} summarizes the F1-score comparison across all evaluated methods. Figure~\ref{fig:recall-fpr-tradeoff} further illustrates the operating-point trade-off between recall and false-positive rate. The proposed method occupies the full-recall operating point, whereas Binary AllowSet achieves a lower false-positive rate at the cost of substantially lower recall.

\FloatBarrier

\subsection{Operational Decision Distribution}\label{subsec:decision-distribution}

Binary classification metrics do not distinguish between requests requiring manual examination and requests that are directly blocked. To further investigate the operational behavior of the proposed method, the proportions of \textit{Allow}, \textit{Review}, and \textit{Block} decisions are examined separately.

Table~\ref{tab:decision-distribution} presents the decision distribution across all evaluated methods.

\begin{table*}[t]
\centering
\caption{Operational decision distribution across five independent runs. Values are mean $\pm$ standard deviation (\%).}
\label{tab:decision-distribution}
\small
\begin{tabular}{lrrr}
\toprule
Method & Review rate & Block rate & Allow rate \\
\midrule
URL path rule
& 100.00 $\pm$ 0.00
& 0.00 $\pm$ 0.00
& 0.00 $\pm$ 0.00 \\

Distinct-object frequency
& 12.42 $\pm$ 0.00
& 0.00 $\pm$ 0.00
& 87.58 $\pm$ 0.00 \\

Binary AllowSet
& 7.18 $\pm$ 0.10
& 1.86 $\pm$ 0.17
& 90.96 $\pm$ 0.19 \\

Single-channel success learner
& 50.00 $\pm$ 0.00
& 1.86 $\pm$ 0.17
& 48.14 $\pm$ 0.17 \\

Cross-evidence without provenance independence
& 16.80 $\pm$ 0.33
& 1.86 $\pm$ 0.17
& 81.34 $\pm$ 0.32 \\

Cross-evidence without quarantine
& 20.27 $\pm$ 0.29
& 1.86 $\pm$ 0.17
& 77.87 $\pm$ 0.25 \\

Proposed cross-evidence method
& 43.74 $\pm$ 0.60
& 1.86 $\pm$ 0.17
& 54.39 $\pm$ 0.50 \\
\bottomrule
\end{tabular}
\end{table*}

The proposed method assigned 43.74\% of requests to the \textit{Review} state, while 54.39\% were allowed and 1.86\% were blocked.

The high review rate is primarily associated with legitimate relations for which the available evidence does not satisfy the confirmation requirements. Instead of treating insufficient evidence as proof of unauthorized access, the method retains an intermediate decision state.

This behavior is consistent with the intended distinction between authorization confidence and runtime enforcement. A relation that has not accumulated sufficient independent evidence is not automatically promoted to the confirmed state. Consequently, some legitimate requests remain subject to review until additional corroborating evidence becomes available.

The results also indicate that the proposed method does not automatically block every request associated with an authorization risk. Most attack requests are identified through the review mechanism rather than immediate rejection. This distinction is important when interpreting the 100.00\% recall reported in Table~\ref{tab:overall-performance}.

\subsection{Evidence Contamination and Ablation Analysis}\label{subsec:contamination-analysis}

The evidence contamination experiment examines whether successful but unauthorized object operations can cause incorrect authorization relations to enter the \textit{Confirmed} state.

The pollution rate is calculated using the initially unauthorized relation set, excluding relations that become unauthorized only after an explicit mid-stream revocation.

Table~\ref{tab:pollution-results} presents the results.

\begin{table}[t]
\centering
\caption{Unauthorized relation pollution rate across five independent runs.}
\label{tab:pollution-results}
\small
\begin{tabular}{lr}
\toprule
Method & Pollution rate (\%) \\
\midrule
Binary AllowSet
& 100.00 $\pm$ 0.00 \\

Single-channel success learner
& 100.00 $\pm$ 0.00 \\

Cross-evidence without provenance independence
& 100.00 $\pm$ 0.00 \\

Cross-evidence without quarantine
& 100.00 $\pm$ 0.00 \\

Proposed cross-evidence method
& 0.00 $\pm$ 0.00 \\
\bottomrule
\end{tabular}
\end{table}

The binary AllowSet and single-channel success learner both reached a pollution rate of 100.00\%. This result reflects their reliance on successful object operations or repeated positive observations without sufficient independent authorization corroboration.

The two ablation methods also reached a pollution rate of 100.00\%. Removing provenance independence allows correlated evidence events to contribute as separate observations. Removing quarantine allows direct-operation evidence to enter the positive authorization support without the corresponding isolation mechanism.

In contrast, the proposed method achieved a pollution rate of 0.00\% in all five runs. None of the initially unauthorized relation keys was promoted to the \textit{Confirmed} state.

These results support the importance of the two mechanisms in the proposed framework. Provenance independence limits the contribution of repeated observations derived from the same source, while quarantine prevents successful object operations from independently establishing authorization relations.

The combined results also show that the mechanisms address different aspects of evidence reliability. Provenance independence constrains the effective number of independent observations, whereas quarantine controls when potentially contaminated positive evidence can contribute to authorization-state promotion.

\begin{figure}[t]
\centering
\includegraphics[width=0.96\textwidth]{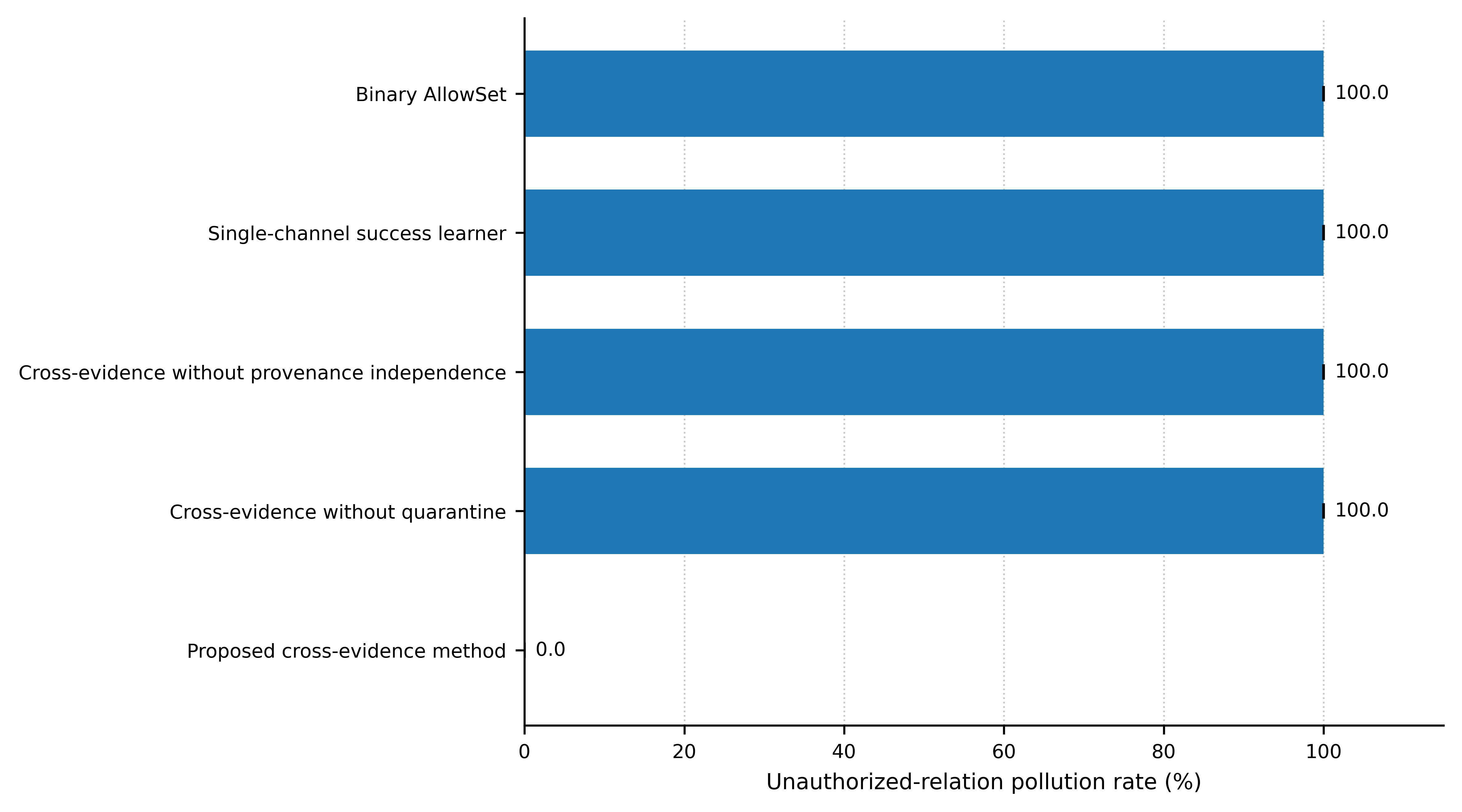}
\caption{Unauthorized-relation pollution rate across the baseline, ablation, and proposed methods. Values are means across five independent seeds; error bars denote sample standard deviations.}
\label{fig:pollution-comparison}
\end{figure}

The 0.00\% pollution rate is specific to the evidence sources and attack patterns represented in this benchmark. Figure~\ref{fig:pollution-comparison} provides a visual comparison of the pollution rates across the baseline, ablation, and proposed methods.

\subsection{Scenario-Level Performance Analysis}\label{subsec:scenario-analysis}

To examine the behavior of the proposed method under different authorization conditions, scenario-level metrics were calculated independently for each seed and subsequently aggregated.

Table~\ref{tab:scenario-results} reports the results for the proposed method. For legitimate-only scenarios, the false-positive rate is reported. For unauthorized-only scenarios, recall is reported. Review and block rates are included to distinguish risk identification from direct enforcement.

\begin{table*}[t]
\centering
\caption{Scenario-level results for the proposed cross-evidence method. Request counts are averages per seed.}
\label{tab:scenario-results}
\scriptsize
\begin{tabular}{lrrrrrr}
\toprule
Scenario
& Requests
& Unauthorized
& Legitimate
& FPR / Recall
& Review
& Block \\
\midrule
Cold start
& 951.2 & 0.0 & 951.2
& FPR 50.00\% & 50.00\% & 0.00\% \\

Correlated sources
& 971.2 & 0.0 & 971.2
& FPR 100.00\% & 100.00\% & 0.00\% \\

Cross-tenant object reference
& 240.0 & 240.0 & 0.0
& Recall 100.00\% & 100.00\% & 0.00\% \\

Delegated two-source access
& 1200.0 & 0.0 & 1200.0
& FPR 0.00\% & 0.00\% & 0.00\% \\

High-volume enumeration
& 640.0 & 640.0 & 0.0
& Recall 100.00\% & 100.00\% & 0.00\% \\

Horizontal BOLA
& 480.0 & 480.0 & 0.0
& Recall 100.00\% & 100.00\% & 0.00\% \\

Low-frequency enumeration
& 240.0 & 240.0 & 0.0
& Recall 100.00\% & 100.00\% & 0.00\% \\

Noisy, then confirmed
& 665.6 & 0.0 & 665.6
& FPR 50.00\% & 50.00\% & 0.00\% \\

Post-revocation access
& 240.0 & 240.0 & 0.0
& Recall 100.00\% & 0.00\% & 100.00\% \\

Public scope
& 301.6 & 0.0 & 301.6
& FPR 100.00\% & 100.00\% & 0.00\% \\

Mid-stream revocation
& 240.0 & 0.0 & 240.0
& FPR 0.00\% & 0.00\% & 0.00\% \\

Single source
& 957.6 & 0.0 & 957.6
& FPR 100.00\% & 100.00\% & 0.00\% \\

Stale evidence
& 515.2 & 0.0 & 515.2
& FPR 100.00\% & 100.00\% & 0.00\% \\

Two independent sources
& 4757.6 & 0.0 & 4757.6
& FPR 0.00\% & 0.00\% & 0.00\% \\

Vertical action escalation
& 480.0 & 480.0 & 0.0
& Recall 100.00\% & 100.00\% & 0.00\% \\
\bottomrule
\end{tabular}
\end{table*}

\begin{figure}[t]
\centering
\includegraphics[width=0.96\textwidth]{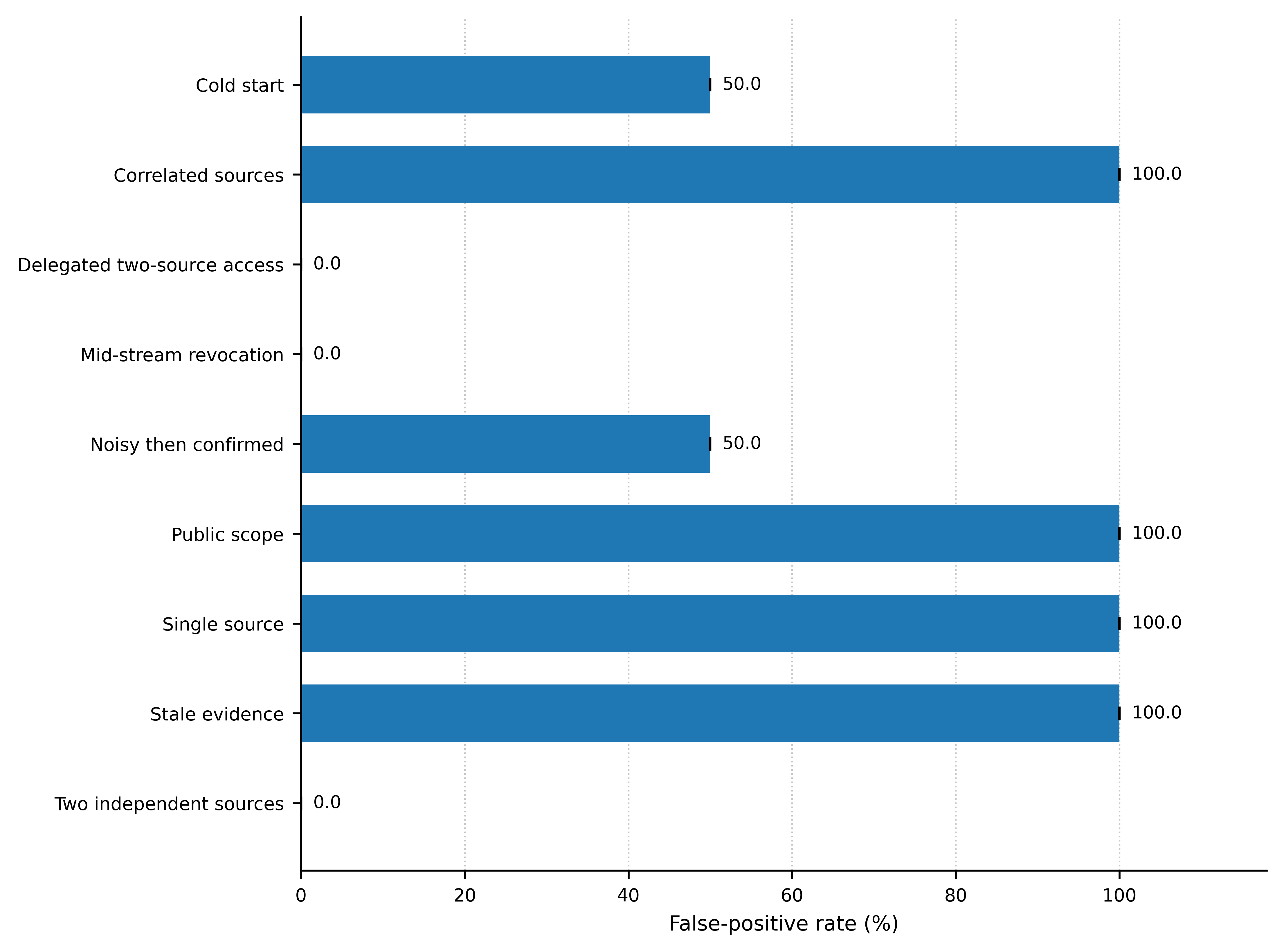}
\caption{False-positive rate of the proposed method on legitimate-only scenarios. Values are means across five independent seeds; error bars denote sample standard deviations.}
\label{fig:legitimate-scenario-fpr}
\end{figure}

\begin{figure}[t]
\centering
\includegraphics[width=0.96\textwidth]{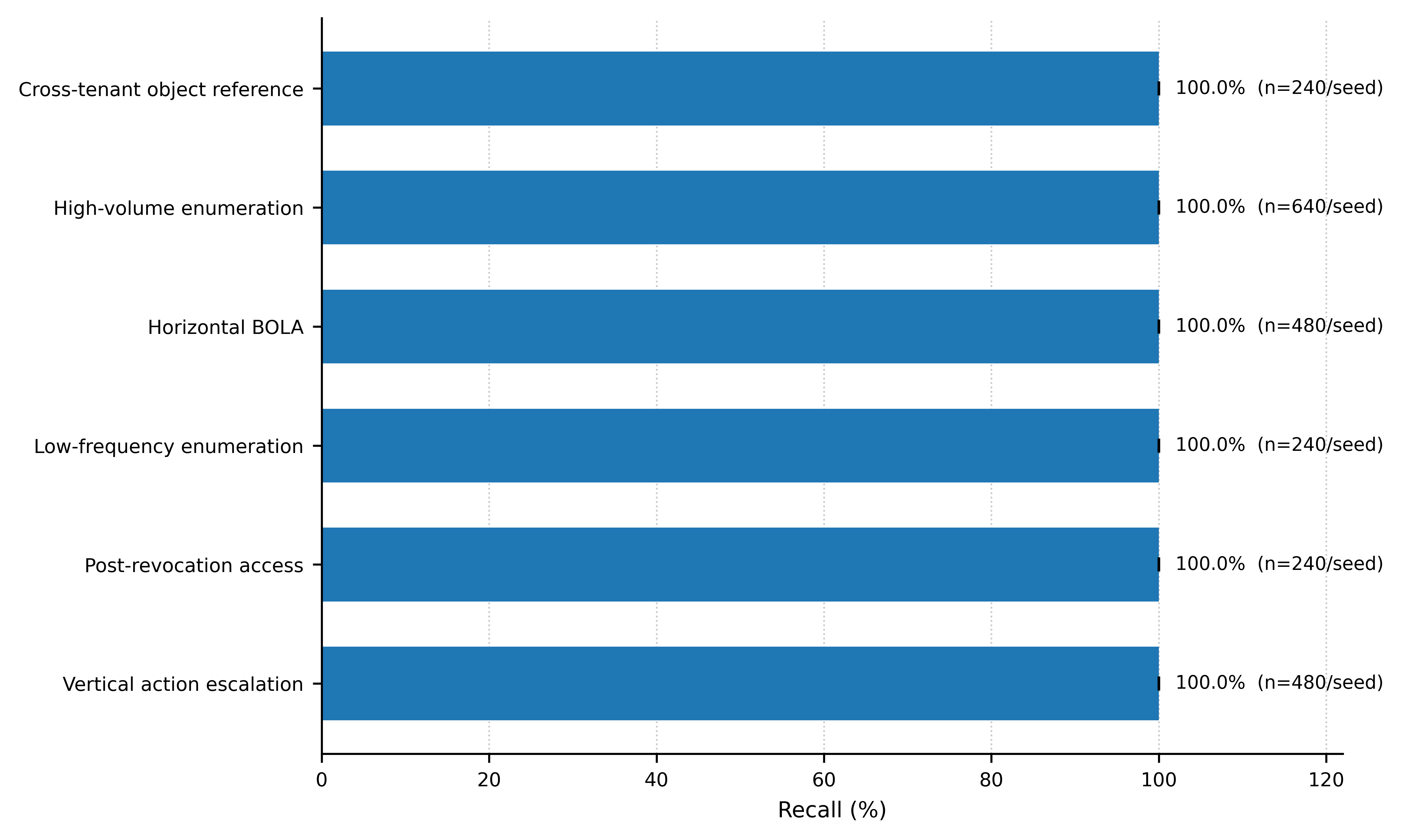}
\caption{Recall of the proposed method on unauthorized-only scenarios. Right-side annotations report the mean number of positive requests per seed.}
\label{fig:attack-scenario-recall}
\end{figure}

Figures~\ref{fig:legitimate-scenario-fpr} and~\ref{fig:attack-scenario-recall} separate the scenario results according to their ground-truth class. The first figure reports false-positive rates for legitimate-only scenarios, while the second reports recall for unauthorized-only scenarios. This separation avoids presenting undefined class-specific metrics as zero.

\FloatBarrier

\subsubsection{Object-Level Unauthorized Access}

The proposed method achieved 100.00\% recall in the horizontal BOLA, cross-tenant object reference, and vertical action escalation scenarios.

The horizontal BOLA scenario evaluates unauthorized access to another subject's object within the same tenant. The cross-tenant scenario evaluates requests that reuse local object identifiers across different tenant namespaces. The vertical action escalation scenario examines requests in which the subject attempts an action that is not supported by the corresponding authorization relation.

The results indicate that the composite authorization relation key provides effective isolation under the evaluated tenant, subject, object, and action combinations.

The low-frequency and high-volume enumeration scenarios also achieved 100.00\% recall. This suggests that the proposed method can identify unauthorized relations without depending exclusively on a high request-volume threshold.

Importantly, the low-frequency enumeration result indicates that object-level authorization evidence can identify suspicious requests even when the request pattern does not exhibit substantial enumeration volume.

\subsubsection{Evidence Sufficiency and Cold-Start Behavior}

The legitimate two-independent-source scenario and delegated two-source access scenario both achieved a false-positive rate of 0.00\%.

This result indicates that the method can recognize legitimate authorization relations when sufficient independent evidence is available. In particular, delegated access is not automatically treated as invalid merely because the subject is not the original object owner.

However, the single-source and correlated-source scenarios produced false-positive rates of 100.00\%. The public-scope scenario and stale-evidence scenario exhibited the same behavior.

The cold-start and noisy-then-confirmed scenarios each produced a false-positive rate of 50.00\%.

These results reveal an important characteristic of the proposed state-learning mechanism. When authorization evidence is insufficient, correlated, stale, or outside the subject-specific scope, the method tends to retain the relation in an intermediate state and request further examination.

This conservative behavior reduces the risk of incorrectly confirming authorization, but it also increases the number of legitimate requests requiring review. Therefore, evidence sufficiency and false-positive reduction remain important areas for further optimization.

\subsubsection{Revocation and Reauthorization Behavior}

The revocation experiments examine the method's response to authorization changes over time.

In the mid-stream revocation scenario, legitimate requests evaluated before the effective revocation produced a false-positive rate of 0.00\%.

For requests explicitly generated after revocation, the method achieved 100.00\% recall and a 100.00\% block rate.

This distinction demonstrates that the method can preserve the validity of previously authorized operations while applying a new enforcement decision after a trusted revocation event.

The reauthorization mechanism is additionally examined by the mechanism-level regression tests. A trusted reauthorization event does not automatically restore superseded positive support. Instead, the relation must obtain new corroborating evidence before it can return to the \textit{Confirmed} state.

These observations support the use of authorization lifecycle management rather than a permanent binary allow-or-deny representation.

\FloatBarrier

\subsection{Parser and Mechanism Validation}\label{subsec:parser-mechanism-results}

The parser benchmark evaluates the configured object-identifier extraction rules independently of the authorization-state learning mechanism.

A total of 3,300 test cases were generated, covering URI path identifiers, query parameters, JSON scalar fields, JSON batch fields, malformed JSON, missing identifiers, generic identifier rejection, and duplicate-identifier deduplication.

Table~\ref{tab:parser-results} summarizes the aggregate results.

\begin{table}[t]
\centering
\caption{Object-identifier extraction results.}
\label{tab:parser-results}
\small
\begin{tabular}{lr}
\toprule
Metric & Result \\
\midrule
Test cases & 3,300 \\
Exact matches & 3,300 \\
Exact-match rate & 100.00\% \\
Precision & 100.00\% \\
Recall & 100.00\% \\
\bottomrule
\end{tabular}
\end{table}

The parser achieved 100.00\% exact-match accuracy under the configured test cases. No false object identifiers or missed expected identifiers were recorded.

This result validates the extraction rules for the supported input formats. It does not imply that the parser can automatically recognize arbitrary object identifiers in previously unseen application schemas. Application-specific field mappings remain necessary when resource identifiers are represented using non-standard names or nested data structures.

The mechanism-level regression suite also passed all 10 tests. These tests covered provenance independence, direct-operation quarantine, quarantine release after independent corroboration, stale evidence handling, explicit revocation, reauthorization, tenant and action isolation, public-scope evidence handling, provenance rollback, and low-quality evidence rejection.

In addition, all 45 dataset-integrity checks passed across the five independent runs. These checks covered request identifier uniqueness, opaque subject identifiers, chronological event ordering, ground-truth isolation, evidence-stream separation, HTTP response consistency, and the absence of externally generated positive authorization evidence for initially unauthorized relations.

Together, these results provide implementation-level evidence that the main components operate according to their configured experimental rules.

\subsection{Discussion of Research Questions}\label{subsec:research-question-discussion}

The experimental results provide different levels of support for the research questions introduced in Section~\ref{subsec:exp-objectives}.

\textbf{RQ1: Detection accuracy under heterogeneous authorization evidence.}

The proposed method achieved 100.00\% recall across the aggregate benchmark and an F1-score of 56.63\%. However, its false-positive rate was 33.65\%. The results demonstrate that the method can identify the unauthorized requests in the controlled benchmark, but they also reveal the operational cost associated with insufficient authorization evidence.

\textbf{RQ2: Comparison with baseline methods.}

The proposed method achieved the highest F1-score and MCC among the seven evaluated methods. In particular, it outperformed the frequency-based and binary AllowSet baselines in recall. Nevertheless, Binary AllowSet achieved higher precision and a lower false-positive rate. The comparison therefore indicates a trade-off between conservative authorization-state assessment and the minimization of risk alerts.

\textbf{RQ3: Effectiveness of provenance independence and quarantine.}

The complete method achieved a pollution rate of 0.00\%, whereas both corresponding ablation methods reached 100.00\%. This result supports the importance of provenance-aware aggregation and evidence quarantine in preventing unauthorized relations from being promoted in the controlled benchmark.

\textbf{RQ4: Behavior under authorization lifecycle and attack scenarios.}

The proposed method achieved 100.00\% recall in the evaluated horizontal BOLA, cross-tenant reference, action escalation, and enumeration scenarios. It also blocked all explicitly generated post-revocation requests. However, legitimate requests with insufficient or unreliable evidence frequently received a Review decision. This behavior confirms that authorization-state uncertainty remains an important factor in the overall detection results.

\textbf{RQ5: Reliability of object-identifier extraction.}

The configured parser achieved 100.00\% exact-match accuracy across 3,300 test cases. This provides evidence that the current extraction rules correctly process the supported URI, query, and JSON formats. Broader validation against independently implemented API schemas is still required.

\subsection{Threats to Validity and Limitations}\label{subsec:results-limitations}

The benchmark is generated by a deterministic simulator. Its authorization graph and ground-truth labels are internally consistent by construction, but they are not derived from independently annotated production traffic. The five random seeds repeat the experiments under different random initializations; they do not represent five independently developed applications. The evaluation therefore measures the behavior and repeatability of the proposed mechanisms within the current benchmark, rather than their generalization across Web applications.

The prototype also uses configured object-extraction rules, evidence-source mappings, and evidence-quality values. Actual applications may expose identifiers through different request structures and response formats, while the subject scope of returned collections may be difficult to determine from traffic alone. The quality of the learned relations consequently depends on the accuracy of these mappings and on the ability to identify dependencies among evidence sources.

The reported classification metrics count both \textit{Review} and \textit{Block} as positive risk alerts. The 100.00\% recall therefore measures detection coverage under this definition; it is not the proportion of unauthorized requests automatically blocked. The false-positive rate of 33.65\% and review rate of 43.74\% also indicate that the current configuration can create a substantial review workload. The evaluation does not include end-to-end proxy latency, throughput, CPU or memory consumption, or multi-instance behavior.

The experiments were conducted on the synthetic benchmark and did not include independently deployed third-party applications such as OWASP crAPI or VAmPI. Evaluating these applications requires controlled test accounts, independently verified permission matrices, and application-side authorization ground truth. These considerations define the scope of the present results and inform the next stage of validation.

\subsection{Summary}\label{subsec:results-summary}

This section evaluated the proposed cross-evidence authorization-state learning method using five deterministic synthetic benchmark runs comprising 64,400 requests.

The proposed method achieved an aggregate recall of 100.00\%, an F1-score of 56.63\%, and an MCC of 0.512. Its unauthorized relation pollution rate was 0.00\%, compared with 100.00\% for the evaluated evidence-promotion ablations.

Scenario-level results demonstrated effective identification of the evaluated object-level authorization violations and correct handling of trusted post-revocation requests. The parser and mechanism regression tests also achieved complete success under their configured test conditions.

At the same time, the false-positive rate of 33.65\% and review rate of 43.74\% indicate that insufficient authorization evidence can generate substantial additional review workload.

Overall, the experiments provide controlled evidence supporting the feasibility of provenance-aware cross-evidence authorization-state learning. They also identify false-positive reduction, evidence calibration, and independent real-application validation as necessary directions for further investigation.

\section{Ablation and Sensitivity Analysis}\label{sec:ablation}

\subsection{Analysis Protocol}\label{subsec:ablation-protocol}

This section further investigates the contribution of individual mechanisms and the sensitivity of the proposed cross-evidence authorization-state learning method to its configuration parameters.

The analysis consists of two parts. First, mechanism ablation examines the effects of removing provenance independence and evidence quarantine. Second, parameter sensitivity analysis evaluates how changes in evidence-quality requirements, authorization-confirmation thresholds, independent-source requirements, and temporal parameters affect detection performance.

The ablation experiments reuse the benchmark results described in Section~\ref{sec:results}. The sensitivity experiments use the same five random seeds, ranging from 20261030 to 20261034. For each parameter setting, the original request and evidence streams are generated once and reused without modification. Only the specified runtime parameter is changed.

A one-factor-at-a-time evaluation strategy is adopted. This design isolates the effect of each parameter while keeping the remaining configuration unchanged. Six parameters are investigated, resulting in 18 parameter settings and 90 seed-setting evaluations.

The baseline configuration follows the implementation in Section~\ref{sec:implementation}. Its principal parameters are a minimum evidence quality of 0.60, a confirmation threshold of 0.75, an independent-group threshold of 0.60, a minimum of two independent provenance groups, a relation time-to-live (TTL) of 100000 simulator time units, and a temporal decay coefficient of $10^{-6}$.

For the temporal decay experiment, the relation TTL is fixed at 150000 simulator time units. This setting allows the effect of exponential decay to be examined independently of hard relation expiration.

All reported sensitivity metrics are calculated as the mean and sample standard deviation across the five independent seeds.

\subsection{Mechanism Ablation Analysis}\label{subsec:mechanism-ablation}

The proposed method incorporates two principal safeguards: provenance independence and positive-evidence quarantine. To investigate their individual contributions, two ablation configurations are evaluated.

The first configuration removes provenance-independence control while retaining the remaining cross-evidence mechanisms. The second disables evidence quarantine while preserving the other components.

Table~\ref{tab:ablation-effects} summarizes the changes relative to the complete proposed method.

\begin{table*}[t]
\centering
\caption{Performance changes caused by removing individual mechanisms.}
\label{tab:ablation-effects}
\small
\begin{tabular}{lrrrrr}
\toprule
Configuration
& $\Delta$ Recall (pp)
& $\Delta$ F1 (pp)
& $\Delta$ FPR (pp)
& $\Delta$ MCC
& Pollution rate (\%) \\
\midrule
Without provenance independence
& -44.83
& -2.43
& -23.02
& -0.073
& 100.00 \\

Without evidence quarantine
& -44.83
& -7.12
& -18.79
& -0.139
& 100.00 \\
\bottomrule
\end{tabular}
\end{table*}

The removal of provenance independence reduced recall by 44.83 percentage points and F1-score by 2.43 percentage points. Although the false-positive rate decreased, this reduction was accompanied by a substantial loss in unauthorized-request detection capability.

The pollution rate reached 100.00\%, indicating that all initially unauthorized relation keys in the evaluated benchmark were incorrectly promoted under this ablation configuration.

This behavior is consistent with the role of provenance grouping. When repeated observations from the same source lineage are counted as separate evidence events, the learner can accumulate apparently sufficient support without obtaining genuinely independent confirmation.

The second ablation produced a larger F1-score reduction of 7.12 percentage points and an MCC reduction of 0.139. Its pollution rate also reached 100.00\%.

Without quarantine, successful direct object operations can contribute to positive authorization support without first satisfying the independent confirmation requirements imposed by the complete mechanism. This weakens the separation between evidence indicating that an operation succeeded and evidence establishing that the operation was authorized.

The two mechanisms therefore provide complementary safeguards. Provenance independence controls the statistical contribution of related observations, whereas quarantine controls the admissibility and release conditions of potentially self-validating evidence.

The results suggest that neither mechanism should be considered a substitute for the other. Their combined use is necessary to maintain the intended authorization-state promotion conditions in the evaluated benchmark.

\subsection{Parameter Sensitivity Analysis}\label{subsec:parameter-sensitivity}

\subsubsection{Minimum Evidence Quality}

The minimum evidence-quality threshold determines whether an incoming observation can participate in authorization-support aggregation.

To investigate its influence, the threshold is evaluated at 0.60, 0.95, and 0.98.

The results show that the proposed method maintains identical aggregate performance at thresholds of 0.60 and 0.95. In both configurations, the F1-score is 56.63\%, with a false-positive rate of 33.65\%.

When the threshold increases to 0.98, the F1-score decreases to 30.53\%, while the false-positive rate reaches 100.00\%.

This change is associated with the credibility values assigned to the configured evidence sources. The object-assignment source has a credibility of approximately 0.9798, while the subject-scoped collection source has a credibility of approximately 0.9519. Consequently, a threshold of 0.98 rejects both sources from positive-support aggregation.

The result demonstrates that excessively restrictive evidence-admission requirements can prevent legitimate authorization relations from reaching the confirmed state. Although the method continues to identify unauthorized requests, legitimate requests are increasingly assigned to the review state.

Therefore, the minimum evidence-quality threshold should be calibrated against the credibility distribution of the available application-specific evidence sources rather than increased indiscriminately.

\subsubsection{Confirmation and Independent-Group Thresholds}

The confirmation threshold and independent-group threshold jointly determine whether accumulated evidence is sufficient to establish an authorization relation.

The confirmation threshold is evaluated at 0.60, 0.75, and 0.9995. The independent-group threshold is evaluated at 0.50, 0.60, and 0.96.

For confirmation thresholds of 0.60 and 0.75, the aggregate metrics remain unchanged. However, increasing the threshold to 0.9995 reduces the F1-score to 30.53\% and increases the false-positive rate to 100.00\%.

This behavior follows from the multiplicative-complement aggregation used by the learner. For the two configured independent evidence sources, the combined positive support is approximately 0.99903. A confirmation threshold above this value cannot be satisfied under the baseline evidence-quality configuration.

A similar effect is observed when the independent-group threshold is increased to 0.96. The credibility of the subject-scoped collection source is approximately 0.9519 and therefore cannot satisfy the group-level requirement. The relation consequently fails to accumulate the required two qualifying independent groups.

These results indicate that the two thresholds control different aspects of authorization confirmation. The confirmation threshold constrains aggregate support, whereas the group threshold constrains the effective contribution of individual provenance groups.

Both parameters must therefore be calibrated jointly with the available evidence-quality distribution.

\subsubsection{Minimum Number of Independent Evidence Groups}

The minimum number of independent provenance groups directly determines how many distinct evidence lineages are required before an authorization relation can be confirmed.

Three configurations are evaluated: one, two, and three independent groups.

When the requirement is reduced to one group, the F1-score increases to 74.06\%, while the false-positive rate decreases to 15.39\%. The review rate also decreases from 43.74\% to 28.77\%.

However, this apparent improvement in classification performance comes at the expense of the intended independent-source confirmation requirement. Single-source and correlated-source observations can now satisfy the confirmation condition.

When the requirement is increased to three groups, the F1-score decreases to 30.53\%, and the false-positive rate reaches 100.00\%. The configured legitimate evidence profiles generally provide only two independent sources, making the confirmation requirement unattainable for these relations.

The baseline requirement of two independent groups therefore represents the configuration consistent with the evidence structure defined in the current benchmark.

It is important to distinguish the observed classification improvement at one group from an improvement in security assurance. The current synthetic attack generator does not inject independently fabricated, non-direct positive authorization evidence into unauthorized relations. Direct-operation evidence remains quarantined. Consequently, the zero pollution rate observed in this particular parameter sweep does not establish that a one-group configuration is resistant to more general evidence-poisoning attacks.

The results instead demonstrate that reducing the number of required independent groups improves legitimate-relation coverage under the current data distribution, while weakening the evidence-independence constraint.

\subsubsection{Relation Time-to-Live}

The relation TTL determines how long previously accumulated positive authorization support remains temporally valid.

Three values are evaluated: 50000, 100000, and 150000 simulator time units.

The results show that TTL values of 50000 and 100000 produce identical aggregate performance. At 150000, the F1-score increases to 60.43\%, and the false-positive rate decreases to 28.78\%.

The difference is primarily associated with the stale-evidence scenario. Under the original configuration, the corresponding positive observations are older than the configured TTL and therefore cannot establish an active confirmed relation.

Increasing the TTL to 150000 places these observations within the configured validity window. As a result, some previously unconfirmed legitimate relations can again satisfy the confirmation condition.

Although this reduces false-positive alerts in the current benchmark, a longer TTL also extends the period during which historical authorization evidence can remain valid.

Consequently, an increase in F1-score does not necessarily imply stronger authorization security. In real deployments, TTL selection must account for permission-change frequency, object ownership transfer, temporary access expiration, and the reliability of historical evidence.

\subsubsection{Temporal Decay Rate}

The temporal decay coefficient controls how rapidly the effective contribution of historical evidence decreases over time.

To isolate exponential decay from hard expiration, the TTL is fixed at 150000 simulator time units while the decay coefficient is varied among 0, $10^{-6}$, and $10^{-5}$.

At decay coefficients of 0 and $10^{-6}$, the F1-score remains 60.43\%. When the coefficient increases to $10^{-5}$, the F1-score decreases to 56.63\%, and the false-positive rate increases to 33.65\%.

The observed change is consistent with the exponential decay function:

\begin{equation}
w(t)=w_0\exp(-\lambda\Delta t),
\label{eq:sensitivity-decay}
\end{equation}

where $w_0$ denotes the original evidence weight, $\lambda$ is the decay coefficient, and $\Delta t$ represents the elapsed simulator time.

For stale observations, a larger decay coefficient reduces the effective group weights. When these weights fall below the independent-group threshold, the corresponding evidence groups no longer contribute to the confirmation requirement.

This result demonstrates that temporal decay and group-level credibility thresholds jointly determine the persistence of authorization support.

The decay coefficient should therefore be selected according to the expected lifetime of authorization evidence rather than treated as an independent optimization parameter.

\subsection{Overall Sensitivity Results}\label{subsec:sensitivity-results}

Table~\ref{tab:parameter-sensitivity} summarizes all 18 parameter settings evaluated in the sensitivity experiments.

\begin{table*}[t]
\centering
\caption{Parameter sensitivity results across five independent seeds. Values are mean $\pm$ standard deviation (\%).}
\label{tab:parameter-sensitivity}
\scriptsize
\setlength{\tabcolsep}{3.5pt}
\begin{tabular}{llrrrrr}
\toprule
Parameter
& Setting
& F1
& Recall
& FPR
& Review
& Pollution \\
\midrule
Minimum evidence quality
& 0.60
& 56.63 $\pm$ 0.72
& 100.00 $\pm$ 0.00
& 33.65 $\pm$ 0.69
& 43.74 $\pm$ 0.60
& 0.00 $\pm$ 0.00 \\

Minimum evidence quality
& 0.95
& 56.63 $\pm$ 0.72
& 100.00 $\pm$ 0.00
& 33.65 $\pm$ 0.69
& 43.74 $\pm$ 0.60
& 0.00 $\pm$ 0.00 \\

Minimum evidence quality
& 0.98
& 30.53 $\pm$ 0.25
& 100.00 $\pm$ 0.00
& 100.00 $\pm$ 0.00
& 98.14 $\pm$ 0.17
& 0.00 $\pm$ 0.00 \\
\midrule
Confirmation threshold
& 0.60
& 56.63 $\pm$ 0.72
& 100.00 $\pm$ 0.00
& 33.65 $\pm$ 0.69
& 43.74 $\pm$ 0.60
& 0.00 $\pm$ 0.00 \\

Confirmation threshold
& 0.75
& 56.63 $\pm$ 0.72
& 100.00 $\pm$ 0.00
& 33.65 $\pm$ 0.69
& 43.74 $\pm$ 0.60
& 0.00 $\pm$ 0.00 \\

Confirmation threshold
& 0.9995
& 30.53 $\pm$ 0.25
& 100.00 $\pm$ 0.00
& 100.00 $\pm$ 0.00
& 98.14 $\pm$ 0.17
& 0.00 $\pm$ 0.00 \\
\midrule
Independent-group threshold
& 0.50
& 56.63 $\pm$ 0.72
& 100.00 $\pm$ 0.00
& 33.65 $\pm$ 0.69
& 43.74 $\pm$ 0.60
& 0.00 $\pm$ 0.00 \\

Independent-group threshold
& 0.60
& 56.63 $\pm$ 0.72
& 100.00 $\pm$ 0.00
& 33.65 $\pm$ 0.69
& 43.74 $\pm$ 0.60
& 0.00 $\pm$ 0.00 \\

Independent-group threshold
& 0.96
& 30.53 $\pm$ 0.25
& 100.00 $\pm$ 0.00
& 100.00 $\pm$ 0.00
& 98.14 $\pm$ 0.17
& 0.00 $\pm$ 0.00 \\
\midrule
Minimum independent groups
& 1
& 74.06 $\pm$ 0.97
& 100.00 $\pm$ 0.00
& 15.39 $\pm$ 0.65
& 28.77 $\pm$ 0.55
& 0.00 $\pm$ 0.00 \\

Minimum independent groups
& 2
& 56.63 $\pm$ 0.72
& 100.00 $\pm$ 0.00
& 33.65 $\pm$ 0.69
& 43.74 $\pm$ 0.60
& 0.00 $\pm$ 0.00 \\

Minimum independent groups
& 3
& 30.53 $\pm$ 0.25
& 100.00 $\pm$ 0.00
& 100.00 $\pm$ 0.00
& 98.14 $\pm$ 0.17
& 0.00 $\pm$ 0.00 \\
\midrule
Relation TTL
& 50000
& 56.63 $\pm$ 0.72
& 100.00 $\pm$ 0.00
& 33.65 $\pm$ 0.69
& 43.74 $\pm$ 0.60
& 0.00 $\pm$ 0.00 \\

Relation TTL
& 100000
& 56.63 $\pm$ 0.72
& 100.00 $\pm$ 0.00
& 33.65 $\pm$ 0.69
& 43.74 $\pm$ 0.60
& 0.00 $\pm$ 0.00 \\

Relation TTL
& 150000
& 60.43 $\pm$ 0.53
& 100.00 $\pm$ 0.00
& 28.78 $\pm$ 0.43
& 39.74 $\pm$ 0.37
& 0.00 $\pm$ 0.00 \\
\midrule
Decay rate
& 0
& 60.43 $\pm$ 0.53
& 100.00 $\pm$ 0.00
& 28.78 $\pm$ 0.43
& 39.74 $\pm$ 0.37
& 0.00 $\pm$ 0.00 \\

Decay rate
& $10^{-6}$
& 60.43 $\pm$ 0.53
& 100.00 $\pm$ 0.00
& 28.78 $\pm$ 0.43
& 39.74 $\pm$ 0.37
& 0.00 $\pm$ 0.00 \\

Decay rate
& $10^{-5}$
& 56.63 $\pm$ 0.72
& 100.00 $\pm$ 0.00
& 33.65 $\pm$ 0.69
& 43.74 $\pm$ 0.60
& 0.00 $\pm$ 0.00 \\
\bottomrule
\end{tabular}
\end{table*}

\begin{figure}[t]
\centering
\includegraphics[width=0.96\textwidth]{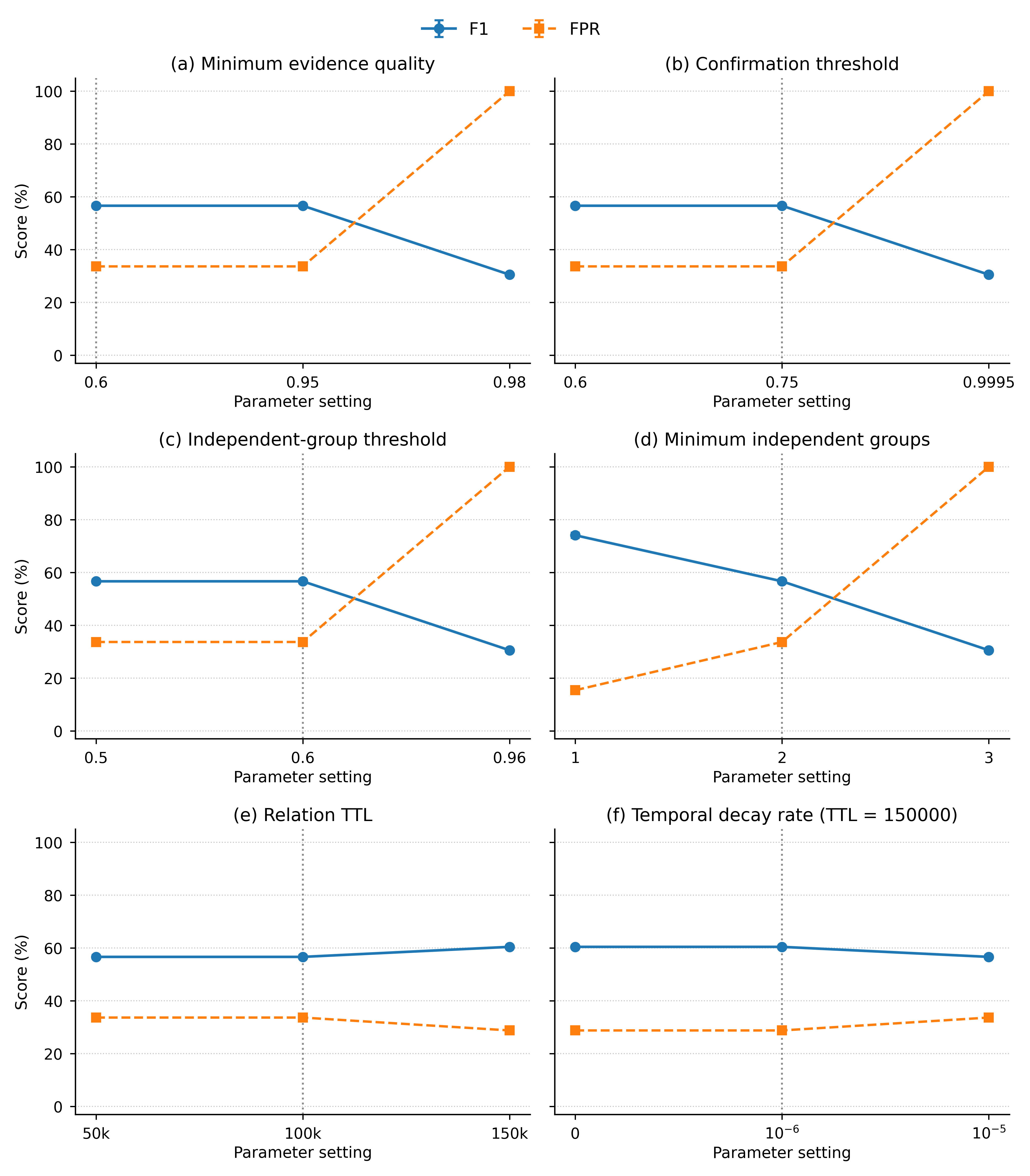}
\caption{F1-score and false-positive rate under six parameter sweeps. Error bars denote sample standard deviations across five independent seeds; dotted vertical lines mark the baseline settings.}
\label{fig:parameter-sensitivity}
\end{figure}

Figure~\ref{fig:parameter-sensitivity} provides a visual comparison of the F1-score and false-positive rate under the different parameter settings.

\FloatBarrier

The results reveal three main characteristics.

First, the proposed method exhibits a relatively stable operating region around its baseline configuration. Moderate changes in minimum evidence quality, confirmation threshold, and independent-group threshold do not alter the aggregate results. However, once the thresholds exceed the credibility or support available from the configured evidence sources, the performance deteriorates sharply.

Second, the minimum number of independent groups has a substantial influence on legitimate authorization-state confirmation. Reducing the requirement improves classification performance, whereas increasing it beyond the number of available independent sources prevents legitimate relations from reaching the confirmed state.

Third, temporal parameters affect the interpretation of historical evidence. A longer TTL can retain previously expired authorization support, while a higher decay coefficient reduces the effective contribution of older observations. These parameters should therefore be configured together with the expected authorization lifecycle of the target application.

The pollution rate remained at 0.00\% across all sensitivity settings. This result is consistent with the current benchmark's attack-evidence construction, in which unauthorized relations do not receive independently fabricated positive evidence and direct-operation observations remain subject to quarantine. It should not be interpreted as proof that every parameter configuration provides equivalent resistance to arbitrary evidence poisoning.

\subsection{Implications for Parameter Configuration}\label{subsec:parameter-implications}

The sensitivity experiments provide several practical implications for configuring the proposed method.

The minimum evidence-quality threshold should remain below the credibility of legitimate evidence sources that are required for authorization confirmation. Excessively high thresholds may reject otherwise valid observations before they can contribute to support aggregation.

The confirmation threshold should be selected according to the maximum support that the expected combination of independent sources can provide. A threshold that cannot be reached under the configured evidence distribution results in persistent candidate states and increased review workload.

The minimum independent-group requirement should reflect the actual independence of the available evidence sources. Although reducing the requirement can improve classification metrics, it weakens the intended resistance to correlated observations. The baseline requirement of two independent groups is retained for the current method.

The relation TTL and decay coefficient should be configured according to the expected lifetime of authorization evidence. Historical observations should not remain authoritative indefinitely, particularly in applications involving temporary permissions, frequently transferred objects, or rapidly changing ownership relations.

Finally, the sensitivity analysis uses a one-factor-at-a-time design. Although it identifies the individual effects of the investigated parameters, it does not fully characterize interactions among all configuration variables. A larger factorial experiment would be required to estimate these interactions and identify an application-independent parameter configuration.

\subsection{Summary}\label{subsec:ablation-summary}

This section investigated the contributions of provenance independence and evidence quarantine, together with the sensitivity of the proposed method to six configuration parameters.

The ablation results showed that removing either provenance independence or evidence quarantine increased unauthorized relation pollution to 100.00\% in the controlled benchmark. Both mechanisms therefore contribute to maintaining reliable authorization-state updates.

The parameter experiments demonstrated that moderate threshold changes can preserve the original detection performance, whereas excessively restrictive confirmation requirements substantially increase false-positive alerts and review workload.

Reducing the minimum number of independent evidence groups improved the classification metrics in the current synthetic benchmark, but weakened the intended evidence-independence constraint. Similarly, increasing relation TTL reduced false-positive alerts in the stale-evidence scenario but extended the validity of historical authorization support.

Overall, the results demonstrate that the effectiveness of cross-evidence authorization-state learning depends not only on the individual learning mechanisms but also on the relationship between evidence quality, source independence, confirmation requirements, and temporal validity.

These findings provide a basis for discussing the applicability, limitations, and deployment considerations of the proposed method in the following section.

\section{Discussion}\label{sec:discussion}

\subsection{Interpretation of the Experimental Findings}\label{subsec:discussion-findings}

The experimental results provide evidence that cross-evidence authorization-state learning can improve the detection of object-level authorization violations under the controlled conditions considered in this study. However, the results also reveal an important trade-off between detection coverage and false-positive control.

The proposed method achieved a mean recall of 100.00\%, an F1-score of 56.63\%, and an MCC of 0.512 across five independent synthetic runs. Its false-positive rate was 33.65\%. Compared with the evaluated baselines, the proposed method achieved the highest F1-score and MCC while maintaining complete recall under the benchmark's alert classification rule.

This result should be interpreted together with the decision distribution. In the evaluation, both \textit{Review} and \textit{Block} are treated as positive risk predictions, whereas \textit{Allow} is treated as a negative prediction. Consequently, a request assigned to \textit{Review} contributes to recall even though it has not been confirmed as an attack and may still be forwarded to the upstream application.

The proposed method generated a mean review rate of 43.74\% and a block rate of 1.86\%. Therefore, its complete recall does not indicate that all unauthorized requests were automatically blocked. Instead, it demonstrates that the method conservatively identifies the evaluated unauthorized requests for either further inspection or enforcement.

The comparison with the Binary AllowSet baseline illustrates this trade-off. Binary AllowSet achieved a higher precision of 65.28\% and a lower false-positive rate of 3.83\%, but its recall was only 32.75\%. It therefore provides a more restrictive decision boundary but misses a substantial proportion of unauthorized requests.

The proposed method adopts a different operating point. It prioritizes identifying potentially unauthorized relations while preserving uncertainty through the \textit{Review} state. This behavior is appropriate for a supplementary runtime monitoring component, particularly when false negatives have a higher security cost than additional review operations.

Nevertheless, the observed false-positive rate indicates that the current confirmation and decision policies require further calibration before the method can be used for automatic enforcement in production environments.

\subsection{Importance of Provenance-Aware Evidence Aggregation}\label{subsec:discussion-provenance}

A central challenge in traffic-side authorization learning is determining whether multiple observations represent genuinely independent evidence.

Repeated successful requests may appear to provide increasingly strong support for an authorization relation. However, these requests may originate from the same session, response, collection operation, or upstream data source. Treating such observations as independent can artificially increase authorization confidence.

The proposed method addresses this problem by grouping observations according to their provenance roots. Within each group, the learner retains the maximum effective positive and negative weights instead of summing repeated observations. This prevents repeated events from the same lineage from independently satisfying the confirmation requirement.

The ablation results support the importance of this mechanism. Removing provenance independence reduced recall to 55.17\%, while the unauthorized relation pollution rate reached 100.00\% in the corresponding controlled experiment.

The result indicates that evidence quantity alone is insufficient for reliable authorization inference. The origin and dependency structure of evidence must also be considered.

However, provenance independence is not equivalent to statistical independence. Two observations with different provenance identifiers may still depend on the same underlying application database, authentication service, or business event. A production implementation must therefore construct provenance roots according to actual data-generation dependencies rather than merely assign different identifiers to different observations.

This distinction is particularly important in distributed applications, where several endpoints may expose the same underlying authorization information through different interfaces.

\subsection{Positive-Evidence Quarantine and State Contamination}\label{subsec:discussion-quarantine}

The second important finding concerns the separation between successful object operations and valid authorization evidence.

A successful HTTP response demonstrates that the application processed a request and returned a response. It does not independently establish that the requester was entitled to perform the operation.

This distinction is particularly important for BOLA detection because the target application may already contain the authorization defect being investigated. If a successful unauthorized request is immediately converted into positive authorization evidence, subsequent requests from the same subject may be incorrectly classified as legitimate.

The proposed quarantine mechanism prevents direct object-operation evidence from immediately promoting an authorization relation. Such observations are retained as held evidence and can only be released after fresh, independent evidence has established the corresponding relation and the release conditions remain satisfied.

The ablation experiment without quarantine resulted in an unauthorized relation pollution rate of 100.00\%. This finding demonstrates the importance of separating operational success from authorization confirmation in the evaluated benchmark.

The quarantine mechanism also preserves the original observation timestamp when held evidence is released. This prevents delayed evidence from being treated as newly generated evidence and avoids unintentionally extending the validity of stale authorization observations.

Importantly, quarantine does not mean that all positive evidence is discarded. Instead, evidence is retained with its provenance, quality, timestamp, and release status. This design provides a traceable mechanism for incorporating previously held observations when sufficient corroboration becomes available.

The current results support this mechanism under the tested synthetic attack conditions. They do not establish resistance to arbitrary poisoning strategies involving compromised independent sources, fabricated application events, or coordinated manipulation of several evidence channels.

\subsection{Authorization Lifecycle and Temporal Validity}\label{subsec:discussion-lifecycle}

Authorization relations are dynamic rather than permanent properties of a subject and an object.

A user may initially own a document, subsequently transfer it to another user, receive temporary delegated access, or lose access following an explicit revocation event. A learner that maintains authorization relations indefinitely may therefore produce incorrect decisions even if its initial inference was accurate.

The proposed lifecycle model addresses these changes through provisional and confirmed states, contradiction handling, revocation, reauthorization, expiration, and temporal decay.

The experimental mechanism tests demonstrate that trusted revocation events take precedence over previously accumulated positive support. Reauthorization does not automatically restore the superseded authorization relation; new independent evidence is required before the relation can return to the confirmed state.

This distinction is important because revocation and reauthorization represent changes in the authorization lifecycle rather than ordinary additions to an evidence stream.

The sensitivity experiments further show that temporal parameters influence the persistence of learned authorization relations. Increasing the relation TTL can retain previously expired evidence, whereas a higher decay coefficient reduces the effective contribution of historical observations.

These parameters should be interpreted in the context of application-specific permission lifetimes. For example, a document-management application with frequent ownership transfers may require shorter authorization validity periods than an application with relatively stable resource ownership.

A single universal TTL or decay coefficient is therefore unlikely to be appropriate for all Web applications.

\subsection{Deployment Considerations}\label{subsec:discussion-deployment}

The proposed method is designed as a supplementary runtime component rather than a replacement for application-side access control.

It can be positioned at a reverse proxy, API gateway, ingress controller, or another observation point where application-layer request and response information is available.

In inline deployment, the component evaluates the authorization state established before the current request is processed. If the request is forwarded, the corresponding response is subsequently used to update the state. This temporal separation prevents the current request from using its own response to establish the authorization state required for its initial decision.

In passive deployment, the same learning mechanism can operate on mirrored traffic or gateway logs. This configuration supports authorization-state learning and retrospective risk analysis without directly affecting request forwarding.

Several deployment conditions must nevertheless be satisfied.

First, subject and tenant identifiers must originate from a trusted authentication or gateway component. Client-supplied identity headers must be removed or overwritten before they are used to construct authorization relation keys. Otherwise, an attacker may manipulate the subject context and invalidate the assumptions of the learning model.

Second, application-specific semantic mappings are required. The current prototype supports configured URI patterns, query parameters, JSON object identifiers, and subject-scoped collection responses. It does not automatically infer the meaning of every endpoint or every field in an arbitrary application.

Third, the observation component must have sufficient application-layer visibility. In deployments where TLS is terminated upstream, the proxy must receive the relevant plaintext request and response information through an authorized interface. If only encrypted traffic metadata is available, the extraction of object identifiers and business outcomes will be limited.

Fourth, the current prototype maintains authorization states in process memory. This implementation is sufficient for controlled experiments but does not provide distributed persistence, multi-instance consistency, or automatic recovery after process termination. A production deployment would require persistent state storage, concurrency control, event ordering, and an appropriate high-availability strategy.

Finally, enforcement policies should distinguish between observation, review, and blocking. A newly encountered or insufficiently supported relation should not automatically be treated as a confirmed security violation. A staged deployment that initially records decisions, evaluates review workloads, and gradually introduces selected blocking rules would reduce operational disruption.

\subsection{Future Research Directions}\label{subsec:future-directions}

The evaluation scope described in Section~\ref{subsec:results-limitations} motivates several directions for further work.

A priority is to evaluate the method on independently developed Web and API applications, including OWASP crAPI and VAmPI. Such experiments should use controlled test accounts and verified permission matrices, with ground truth obtained from application configurations, ownership records, or manually checked test cases. This would allow the method to be assessed against application behavior rather than labels produced by the synthetic generator.

The evidence model should also be examined across resource types and interaction patterns beyond document management. Relevant cases include hierarchical objects, delegated access, nested resources, GraphQL operations, batch requests, and application-specific action semantics. Improving object extraction and semantic mapping for these formats is part of this work.

A further direction is the evaluation of adaptive evidence manipulation. Coordinated poisoning, compromised or correlated sources, delayed attacks, provenance misclassification, and authorization changes during active sessions should be included in the test scenarios. Evidence credibility and promotion thresholds can then be calibrated against observed evidence distributions and the operational costs of false-positive and false-negative decisions.

Runtime evaluation should measure end-to-end latency, throughput, memory consumption, concurrency, and consistency across multiple instances. Persistent state storage, distributed event processing, and failure recovery can be introduced incrementally and assessed under realistic workload conditions.

\section{Conclusion}\label{sec:conclusion}

This paper presented a cross-evidence authorization-state learning method for runtime detection of Broken Object Level Authorization (BOLA) in black-box Web applications.

Unlike approaches that rely primarily on URL patterns, request frequency, binary object allowlists, or individual successful responses, the proposed method models object-level authorization as a dynamic subject--resource--object--action relation. It maintains authorization confidence and lifecycle states using heterogeneous runtime evidence while accounting for evidence quality, provenance, source independence, and temporal validity. The method introduces several complementary mechanisms to improve the reliability of authorization-state learning. Evidence credibility and polarity are evaluated separately, correlated observations are grouped by provenance, and repeated observations from the same lineage cannot independently increase confirmation support. Direct object-operation evidence is quarantined to prevent successful but unauthorized requests from validating their own authorization relations. The lifecycle mechanism further supports evidence rollback, explicit revocation, reauthorization, expiration, and temporal confidence decay.

A lightweight Python reverse-proxy prototype was implemented to validate the proposed approach without modifying the target application's source code or requiring an external database service. The prototype separates request-time authorization decisions from response-time evidence updates and supports configurable extraction of object identifiers from URI paths, query parameters, and JSON request bodies.

The experimental evaluation used a deterministic synthetic benchmark comprising 64,400 requests, 26,858 evidence events, and 16,100 authorization relation records across five independent random seeds. The proposed method achieved a mean precision of 39.50\%, recall of 100.00\%, F1-score of 56.63\%, false-positive rate of 33.65\%, and MCC of 0.512. It also achieved a 0.00\% unauthorized relation pollution rate under the evaluated benchmark conditions. The ablation experiments further demonstrated that removing provenance independence or positive-evidence quarantine resulted in a 100.00\% unauthorized relation pollution rate. The sensitivity analysis showed that authorization-state confirmation depends strongly on the relationship between evidence credibility, independent-source requirements, confirmation thresholds, and temporal validity.

These findings suggest that reliable runtime authorization learning requires more than accumulating successful access observations. It requires explicit control over evidence origin, independence, admissibility, and lifecycle transitions. The results also expose a trade-off between detection coverage and false-positive control. Because both \textit{Review} and \textit{Block} are counted as positive risk alerts, the reported 100.00\% recall measures detection coverage rather than automatic blocking. The false-positive rate and review workload indicate that decision calibration remains important.

The evaluation scope and validity considerations are discussed in Section~\ref{subsec:results-limitations}. Further work will extend the evaluation to independently developed applications and more varied runtime conditions. Overall, this study establishes a reproducible framework for learning dynamic object-level authorization states from runtime traffic and provides an initial experimental basis for investigating provenance-aware, contamination-resistant BOLA detection in black-box Web applications.

\section*{Data availability}

The dataset supporting this study is publicly available on Zenodo at \url{https://doi.org/10.5281/zenodo.23830403}. It includes the synthetic request records, authorization evidence, ground-truth relations, prediction results, and experimental summaries. The dataset is released under the Creative Commons Attribution-NonCommercial 4.0 International License (CC BY-NC 4.0).











\printcredits

\bibliographystyle{cas-model2-names}

\bibliography{references}

@misc{owasp2023bola,
  author       = {{OWASP Foundation}},
  title        = {{API1:2023 Broken Object Level Authorization}},
  year         = {2023},
  howpublished = {OWASP API Security Top 10},
  url          = {https://owasp.org/API-Security/editions/2023/en/0xa1-broken-object-level-authorization/},
  urldate      = {2026-07-19}
}

@article{sandhu1996rbac,
  author  = {Sandhu, Ravi S. and Coyne, Edward J. and Feinstein, Hal L. and Youman, Charles E.},
  title   = {Role-Based Access Control Models},
  journal = {Computer},
  year    = {1996},
  volume  = {29},
  number  = {2},
  pages   = {38--47},
  doi     = {10.1109/2.485845}
}

@techreport{hu2014abac,
  author      = {Hu, Vincent C. and Ferraiolo, David F. and Kuhn, D. Richard and Schnitzer, Adam and Sandlin, Kenneth and Miller, Robert and Scarfone, Karen},
  title       = {Guide to Attribute Based Access Control ({ABAC}) Definition and Considerations},
  institution = {National Institute of Standards and Technology},
  type        = {NIST Special Publication},
  number      = {800-162},
  address     = {Gaithersburg, MD, USA},
  year        = {2014},
  note        = {Includes updates as of August 2, 2019},
  doi         = {10.6028/NIST.SP.800-162}
}

@inproceedings{fong2011rebac,
  author    = {Fong, Philip W. L.},
  title     = {Relationship-Based Access Control: Protection Model and Policy Language},
  booktitle = {Proceedings of the First ACM Conference on Data and Application Security and Privacy},
  series    = {CODASPY '11},
  year      = {2011},
  pages     = {191--202},
  publisher = {Association for Computing Machinery},
  address   = {New York, NY, USA},
  doi       = {10.1145/1943513.1943539}
}

@article{park2004ucon,
  author  = {Park, Jaehong and Sandhu, Ravi},
  title   = {The {UCONABC} Usage Control Model},
  journal = {ACM Transactions on Information and System Security},
  year    = {2004},
  volume  = {7},
  number  = {1},
  pages   = {128--174},
  doi     = {10.1145/984334.984339}
}

@inproceedings{bau2010blackbox,
  author    = {Bau, Jason and Bursztein, Elie and Gupta, Divij and Mitchell, John C.},
  title     = {State of the Art: Automated Black-Box Web Application Vulnerability Testing},
  booktitle = {2010 IEEE Symposium on Security and Privacy},
  year      = {2010},
  pages     = {332--345},
  publisher = {IEEE},
  doi       = {10.1109/SP.2010.27}
}

@inproceedings{doupe2010johnny,
  author    = {Doup{\'e}, Adam and Cova, Marco and Vigna, Giovanni},
  title     = {Why Johnny Can't Pentest: An Analysis of Black-Box Web Vulnerability Scanners},
  booktitle = {Detection of Intrusions and Malware, and Vulnerability Assessment},
  series    = {Lecture Notes in Computer Science},
  volume    = {6201},
  year      = {2010},
  pages     = {111--131},
  publisher = {Springer},
  address   = {Berlin, Heidelberg},
  doi       = {10.1007/978-3-642-14215-4_7}
}

@inproceedings{atlidakis2019restler,
  author    = {Atlidakis, Vaggelis and Godefroid, Patrice and Polishchuk, Marina},
  title     = {{RESTler}: Stateful {REST API} Fuzzing},
  booktitle = {Proceedings of the 41st International Conference on Software Engineering},
  series    = {ICSE '19},
  year      = {2019},
  pages     = {748--758},
  publisher = {IEEE/ACM},
  doi       = {10.1109/ICSE.2019.00083}
}

@inproceedings{martinlopez2021restest,
  author    = {Martin-Lopez, Alberto and Segura, Sergio and Ruiz-Cort{\'e}s, Antonio},
  title     = {{RESTest}: Automated Black-Box Testing of {RESTful Web APIs}},
  booktitle = {Proceedings of the 30th ACM SIGSOFT International Symposium on Software Testing and Analysis},
  series    = {ISSTA 2021},
  year      = {2021},
  pages     = {682--685},
  publisher = {Association for Computing Machinery},
  address   = {New York, NY, USA},
  doi       = {10.1145/3460319.3469082}
}

@article{zhang2023openproblems,
  author    = {Zhang, Man and Arcuri, Andrea},
  title     = {Open Problems in Fuzzing {RESTful APIs}: A Comparison of Tools},
  journal   = {ACM Transactions on Software Engineering and Methodology},
  year      = {2023},
  volume    = {32},
  number    = {6},
  articleno = {144},
  pages     = {1--45},
  doi       = {10.1145/3597205}
}

@inproceedings{cova2007swaddler,
  author    = {Cova, Marco and Balzarotti, Davide and Felmetsger, Viktoria and Vigna, Giovanni},
  title     = {Swaddler: An Approach for the Anomaly-Based Detection of State Violations in Web Applications},
  booktitle = {Recent Advances in Intrusion Detection},
  series    = {Lecture Notes in Computer Science},
  volume    = {4637},
  year      = {2007},
  pages     = {63--86},
  publisher = {Springer},
  address   = {Berlin, Heidelberg},
  doi       = {10.1007/978-3-540-74320-0_4}
}

@inproceedings{felmetsger2010logic,
  author    = {Felmetsger, Viktoria and Cavedon, Ludovico and Kruegel, Christopher and Vigna, Giovanni},
  title     = {Toward Automated Detection of Logic Vulnerabilities in Web Applications},
  booktitle = {Proceedings of the 19th USENIX Security Symposium},
  year      = {2010},
  pages     = {143--160},
  publisher = {USENIX Association},
  address   = {Washington, DC, USA},
  url       = {https://www.usenix.org/legacy/events/sec10/tech/full_papers/Felmetsger.pdf}
}

@inproceedings{zuo2017authscope,
  author    = {Zuo, Chaoshun and Zhao, Qingchuan and Lin, Zhiqiang},
  title     = {{AuthScope}: Towards Automatic Discovery of Vulnerable Authorizations in Online Services},
  booktitle = {Proceedings of the 2017 ACM SIGSAC Conference on Computer and Communications Security},
  series    = {CCS '17},
  year      = {2017},
  pages     = {799--813},
  publisher = {Association for Computing Machinery},
  address   = {New York, NY, USA},
  doi       = {10.1145/3133956.3134089}
}

@inproceedings{huang2024bolaray,
  author    = {Huang, Yongheng and Shi, Chenghang and Lu, Jie and Li, Haofeng and Meng, Haining and Li, Lian},
  title     = {Detecting Broken Object-Level Authorization Vulnerabilities in Database-Backed Applications},
  booktitle = {Proceedings of the 2024 ACM SIGSAC Conference on Computer and Communications Security},
  series    = {CCS '24},
  year      = {2024},
  pages     = {2934--2948},
  publisher = {Association for Computing Machinery},
  address   = {New York, NY, USA},
  doi       = {10.1145/3658644.3690227}
}

@article{santosfilho2025openapi,
  author  = {Santos Filho, Ailton and Rodr{\'i}guez, Ricardo J. and Feitosa, Eduardo L.},
  title   = {Automated Broken Object-Level Authorization Attack Detection in {REST APIs} through {OpenAPI} to Colored Petri Nets Transformation},
  journal = {International Journal of Information Security},
  year    = {2025},
  volume  = {24},
  articleno = {83},
  doi     = {10.1007/s10207-024-00970-5}
}

@unpublished{lin2026framework,
  author = {Lin, Siqi and Luo, Yuchao and Zhu, Zihan and Wang, Yiran and Qiu, Borui},
  title  = {A Non-Intrusive Traffic Analysis Framework for Authorization Risk Detection and Coordinated Response in Web Applications},
  note   = {arXiv preprint},
  year   = {2026},
  doi    = {10.48550/arXiv.2607.16754}
}



\end{document}